\documentclass[twocolumn,aps,prb,
groupedaddress]{revtex4-1}
\usepackage{graphicx}
\usepackage{dcolumn}
\usepackage{bm}
\usepackage{graphics}
\usepackage{epsfig}
\usepackage{epstopdf}
\begin{document}

\title{Effects of Rashba spin-orbit coupling, Zeeman splitting and gyrotropy in two-dimensional cavity polaritons under the influence of the Landau quantization}

\author{S.A. Moskalenko,$^{1}$ I.V. Podlesny,$^{1}$ E.V. Dumanov,$^{1}$
and M.A. Liberman$^{2}$}
%\email[dum@phys.asm.md]
%\thanks{}
\affiliation{$^{1}$Institute of Applied Physics of the Academy of Sciences of Moldova, Academic Str. 5, Chisinau, MD2028, Republic of Moldova\\
$^{2}$Nordita, KTH Royal Institute of Technology and Stockholm University, Roslagstullsbacken 23,
10691 Stockholm, Sweden}

\date{\today}

\begin{abstract}
We consider the energy spectrum of the two-dimensional cavity polaritons under the influence of a strong magnetic and electric fields perpendicular to the surface of the GaAs-type quantum wells (QWs) with p-type valence band embedded into the resonators. As the first step in this direction the Landau quantization (LQ) of the electrons and heavy-holes (hh) was investigated taking into account the Rashba spin-orbit coupling (RSOC) with third-order chirality terms for hh and with nonparabolicity terms in their dispersion low including as well the Zeeman splitting (ZS) effects. The nonparabolicity term is proportional to the strength of the electric field and was introduced to avoid the collapse of the semiconductor energy gap under the influence of the third order chirality terms. The exact solutions for the eigenfunctions and eigenenergies were obtained using the Rashba method [1]. On the second step we derive in the second quantization representation the Hamiltonians describing the Coulomb electron-electron and the electron-radiation interactions. This allow us to determine the magnetoexciton energy branches and to deduce the Hamiltonian of the magnetoexciton-photon interaction. On the third step the fifth order dispersion equation describing the energy spectrum of the cavity magnetoexciton-polariton is investigated. It takes into account the interaction of the cavity photons with two dipole-active and with two quadrupole-active 2D magnetoexciton energy branches. The cavity photons have the circular polarizations $\vec{\sigma }_{\vec{k}}^{\pm }$ oriented along their wave vectors $\vec{k}$, which has the quantized longitudinal component $k_{z}=\pm \pi /L_{c}$, where $L_{c}$ is the resonator length and another small transverse component $\vec{k}_\parallel$ oriented in the plane of the QW. The 2D magnetoexcitons are characterized by the in-plane wave vectors $\vec{k}_{\parallel}$ and by circular polarizations $\vec{\sigma}_{M}$ arising in the p-type valence band with magnetic momentum projection $M=\pm 1$ on the direction of the magnetic field. The selection rules of the exciton-photon interaction have two origins. The first one, of geometrical-type, is expressed through the scalar products of the two-types circular polarizations. They depend on the in-plane wave vectors $\vec{k}_{\parallel}$ even in the case of dipole-active transitions, because the cavity photons have an oblique incidence to the surface of the QW. Another origin is related with the numbers $n_{e}$ and $n_{h}$ of the LQ levels of electrons and heavy-holes taking part in the magnetoexciton formation. So, the dipole-active transitions take place for the condition $n_{e}=n_{h}$, whereas in the quadrupole-active transitions the relation is $n_{e}=n_{h}\pm 1$. It was shown that the Rabi frequency $\Omega_{R}$ of the polariton branches and the magnetoexciton oscillator strength $f_{\text{osc}}$ increase in dependence on the magnetic field strength $B$ as $\Omega_{R}\sim \sqrt{B}$, and $f_{\text{osc}}\sim B$. The optical gyrotropy effects may be revealed if changing the sign of the photon circular polarization at a given sign of the wave vector longitudinal projection $k_{z}$ or equivalently changing the sign of the longitudinal projection $k_{z}$ at the same selected light circular polarization.
\end{abstract}
\pacs{71.35.Lk, 67.85.Jk}
\keywords{polariton, Landau quantization, Rashba spin-orbit coupling, Zeeman splitting, gyrotropy effects}
\maketitle
\tableofcontents


\section{Introduction}
The theoretical description of the electron structure of the magnetoexcitons and of the excitonic spectra in the GaN-based quantum wells in wide range of the perpendicular magnetic fields up to 60T was proposed in Refs [2-4], where the built-in strain and perpendicular electric field were taken into account. The pressure leads to the reordering of the topmost valence subbands having different symmetries. The exciton binding energies at different valence subbands give rise to the reordering of the corresponding optical transitions in the emission and absorption spectra. The reordering in the light - and heavy-hole valence subbands takes place also in dependence on the QW width. The latter is an important feature of the GaN-type semiconductor and it leads also to a stepwise dependence of the excitonic g-factor. The electric field is perpendicular to the layer surface. It was taken into account together with the size quantization. The Landau orbitals in Refs [2, 3] were generated by the ladder operators constructed using the Lamb gauge transformation [5]. They concern only to the states with zero center-of-mass momentum, depend on coordinates of the relative electron-hole motion and act on the wave function of the ground state with the same coordinate dependence.

Contrary to the model of Refs.[2-4], below we will consider the two-dimensional (2D) electron-hole (e-h) system in a GaAs-type structure with only one heavy-hole valence subband with infinitesimal QW width without any reference to the excited levels of the size quantization. We will consider the multilateral phenomena related with Landau quantization accompanied by the Rashba spin-orbit coupling, Zeeman splitting and non-parabolicity of the heavy-hole dispersion law. All of them are induced in the 2D e-h system by the influence of the strong perpendicular magnetic and electric fields. We have followed the Rashba basic paper [1] using the rising and the lowering operators in the space of the single-particle Landau levels. In our case they depend exclusively on the electron and hole coordinates, avoiding their expressing through the relative and center-of-mass coordinates of the e-h pair. The corresponding functions were determined by the exact solutions [6, 7]. The magnetoexciton wave function in coordinate representation when the electron and hole are situated on the lowest Landau levels was obtained in Refs [8, 9]. It reflects the interdependence of the translational and relative motion of electron and hole in the presence of a magnetic field.
Recently much attention was attracted to the influence of the magnetic field on the microcavity polaritons [10-14]. Among many different nonlinear effects induced by the magnetic field one can mention the suppression of the polariton superfluidity and the density dependent Zeeman splitting [12, 14]. The transverse electric-tranverse magnetic (TE-TM) splitting of the cavity modes, the Faraday rotation of the plane polarization of the light passing through the microcavity, and the Zeeman splitting of the exciton eigenstates were observed experimentally in Refs [10, 11].

The magnetic field was used as a tuning parameter for exciton and photon resonances [15]. In this way the change of the exciton energy, of the oscillator strength, of the polariton line width and the redistribution of the polariton density along the dispersion curve owning to the magnetically induced detuning were obtained. The increase of the exciton oscillator strength in dependence of the magnetic field strength $B$ observed experimentally [15] qualitativly agrees with the theoretical result of the Ref [16] (formula (12)), where the Rabi frequency $\Omega_{R}$ of the magnetoexciton polariton is inverse proportional to the magnetic length $l_{0}$ as $\Omega_{R}\sim {l_{0}}^{-1}\sim \sqrt{B}$. It means that the oscillator strength in this case is proportional to $B$. The evolution of the circularly polarized nonequilibrium Bose-Einstein condensates of the spinor-polaritons in different spinor states [13], as well as the influence of the magnetic field on a spinor exciton polariton condensate in rectangular GaAs microcavity pillars were investigated in [17]. The Zeeman splitting and the diamagnetic shift of the exciton-polariton energy spectrum in an external magnetic field for different trap diameters have been studied in Ref.[18]. The new Hopefield [19] coefficients in the presence of the magnetic field and its influence of the relative weights of the photonic and excitonic components in the structure of the polariton branches were determined in Ref. [20]. The combined exciton-cyclotron resonance in GaAs-type QWs was revealed experimentally in Ref. [21] and explained theoretically in Ref.[22].

The paper is organized as follows. In Section 2 the wave functions and the eigenenergies of the 2D electron and heavy holes(hh) undergoing the Landau quantization (LQ), Rashba spin-orbit coupling (RSOC) and Zeeeman splitting(ZS) effects were reviewed. The main results in this field,in general forms were obtained in our previous papers cited below, but their detailed presentation needed for the subsequent applications is given in the present paper. At the same time in this section new results related with the Dresselhaus spin-orbit interaction(DSOI) were included. In the Section 3 on the base of the above mentioned wave functions the Hamiltonian describing in the second quantization representation the electron-electron Coulomb interaction was deduced. The energy spectrum of the 2D magnetoexcitons arising in these conditions was determined. Section 4 is devoted to the electron-radiation interaction, to the magnetoexciton-photon coupling and to the magnetoexciton-polariton formation. The Sections 3 and 4 contain the main original results of our paper. Their novelity is determined by the simultaneous taking into account of four important events such as the LQ, RSOC, ZS and the nonparabolicity of the hh dispersion law. We conclude in Section 5.
\section{2D electrons and holes under the influence of the perpendicular magnetic and electric fields}
The Hamiltonians describing the Landau quantization, Rashba spin-orbit coupling and Zeeman effect with participation of the 2D electrons and holes were derived in Refs.[1, 6, 7, 23-25], and they look as
\begin{eqnarray}
\hat{H}_e &=& \hbar \omega_{ce} \left\lbrace \left( a^{\dag}a + \frac{1}{2}\right)  \hat{I}  + i \alpha \sqrt{2} \left|\begin{array}{cc}
0 & a\\
-a^{\dag} & 0\end{array}\right| +Z_e\hat{\sigma}_z \right\rbrace, \nonumber \\
\hat{H}_h &=& \hbar \omega_{ch} \left\lbrace \left[ \left( a^{\dag}a + \frac{1}{2}\right) + \delta \left( a^{\dag}a + \frac{1}{2}\right)^2 \right] \hat{I} \right. \nonumber \\
&& \left. + i \beta 2\sqrt{2} \left|\begin{array}{cc}
0 & (a^{\dag})^3\\
-a^3 & 0\end{array}\right| -Z_h\hat{\sigma}_z \right\rbrace
\end{eqnarray}
Here the Bose-type operators $a^{\dagger},a$ generating the Fock states $\left| m \right\rangle $ were introduced and the following notation are used [25-28]:
\begin{eqnarray}
Z_{i}&=&\frac{g_{i}\mu _{B}B}{2\hbar {\omega_{ci}}}=\frac{g_{i}m_{i}}{4m_{0}}, \omega_{ci}=\frac{eB}{m_{i}c}, i=e,h, \nonumber\\
 \hat{I}&=&\left| \begin{array}{cc}
   1 & 0  \\
   0 & 1  \\
\end{array} \right|, {{{\hat{\sigma }}}_{z}}=\left| \begin{array}{cc}
   1 & 0  \\
   0 & -1  \\
\end{array} \right|, {{\mu }_{B}}=\frac{e\hbar }{2m_{0}c}, e=\left| e \right|>0, \nonumber\\
 B&=&yT, E_{z}=x kV/cm, \alpha =8\cdot 10^{-3}x/\sqrt{y}, \\
 \beta& =&1.06\cdot 10^{-2}x\sqrt{y}, \delta =10^{-4}Cxy, \nonumber
\end{eqnarray}
where $\omega_{ci}$ are the cyclotron frequencies, $Z_{i}$ are the Zeeman parameters proportional to the $g$-factors $g_{i}$ and to the effective masses $m_{i}$ of the electrons and holes, whereas $m_{0}$ is the bare electron mass. $\delta $ is the nonparabolicity of the heavy-hole dispersion low, $\alpha$ and $\beta$ are the parameters of the chirality terms, which are of the first order in the case of the electrons [1] and of the third order in the case of the heavy-holes [25-28].

For the electron case solutions of these equations in the dimensionless forms are
\begin{eqnarray}
\textit{H}_{e}&=&\frac{H_{e}}{\hbar {\omega _{ce}}}, H_{e}\left| \psi_{e} \right\rangle =\varepsilon \left| \psi_{e} \right\rangle,\left| \psi_{e} \right\rangle =\left| \begin{array}{cc}
   \left| f_{1} \right\rangle   \\
   \left| f_{2} \right\rangle   \\
\end{array} \right| \\
\left| f_{1} \right\rangle &=&\sum\limits_{n}{{a_{n}}}\left| n \right\rangle, \left| f_{2} \right\rangle =\sum\limits_{n}{{b_{n}}\left| n \right\rangle }, {\sum\limits_{n}{\left| a_{n} \right|}}^{2}+{\sum\limits_{n}{\left| b_{n} \right|}}^{2}=1 \nonumber
\end{eqnarray}
The equations, which determine the coefficients $a_{n}$ and $b_{n}$ are [1]:
\begin{eqnarray}
\sum\limits_{n\ge 0}{{{a}_{n}}}\left[ n+\frac{1}{2}+{{Z}_{e}}-\varepsilon  \right]\left| n \right\rangle &=&-i\sqrt{2}\alpha \sum\limits_{n\ge 1}{{{b}_{n}}\sqrt{n}}\left| n-1 \right\rangle   \\
\sum\limits_{n\ge 0}{{{b}_{n}}}\left[ n+\frac{1}{2}-{{Z}_{e}}-\varepsilon  \right]\left| n \right\rangle &=&i\sqrt{2}\alpha \sum\limits_{n\ge 0}{{{a}_{n}}}\sqrt{n+1}\left| n+1 \right\rangle \nonumber
\end{eqnarray}
They are reduced to the equations for the coefficients ${{a}_{n}}$ and ${{b}_{n}}$ at $m\ge 0$
\begin{eqnarray}
\left( m+\frac{1}{2}+{{Z}_{e}}-\varepsilon  \right){{a}_{m}}&=&-i\alpha \sqrt{2}\sqrt{m+1}{{b}_{m+1}}, \nonumber \\
\left( m+\frac{3}{2}-{{Z}_{e}}-\varepsilon  \right){{b}_{m+1}}&=&i\alpha \sqrt{2}\sqrt{m+1}{{a}_{m}}
\end{eqnarray}
Based on these equations we obtain the equalities
\begin{eqnarray}
&\varepsilon _{m}^{\pm }=\left( m+1 \right)\pm \sqrt{{{({1}/{2}\;-{{Z}_{e}})}^{2}}+2{{\left| \alpha  \right|}^{2}}\left( m+1 \right)}, \\
 & {{\left| a_{m}^{-} \right|}^{2}}=\frac{{{[({1}/{2}\;-{{Z}_{e}})+\sqrt{{{({1}/{2}\;-{{Z}_{e}})}^{2}}+2{{\left| \alpha  \right|}^{2}}\left( m+1 \right)}]}^{2}}}{{{[({1}/{2}\;-{{Z}_{e}})+\sqrt{{{({1}/{2}\;-{{Z}_{e}})}^{2}}+2{{\left| \alpha  \right|}^{2}}\left( m+1 \right)}]}^{2}}+2{{\left| \alpha  \right|}^{2}}\left( m+1 \right)}, \nonumber \\
 & {{\left| b_{m+1}^{-} \right|}^{2}}=\frac{2{{\left| \alpha  \right|}^{2}}\left( m+1 \right)}{{{[({1}/{2}\;-{{Z}_{e}})+\sqrt{{{({1}/{2}\;-{{Z}_{e}})}^{2}}+2{{\left| \alpha  \right|}^{2}}\left( m+1 \right)}]}^{2}}+2{{\left| \alpha  \right|}^{2}}\left( m+1 \right)}, \nonumber\\
 & {{\left| a_{m}^{\pm } \right|}^{2}}={{\left| b_{m+1}^{\mp } \right|}^{2}}, {{\left| a_{m}^{\pm } \right|}^{2}}+{{\left| b_{m+1}^{\pm } \right|}^{2}}=1, m\ge 0. \nonumber
\end{eqnarray}
The wave functions in the second quantization representation look as
\begin{eqnarray}
\left| \psi _{m}^{\pm } \right\rangle &=&\left| \begin{array}{cc}
   a_{m}^{\pm }\left| m \right\rangle   \\
   b_{m+1}^{\pm }\left| m+1 \right\rangle   \\
\end{array} \right| \\
\left\langle  \psi _{m}^{\pm } \right|&=&\left| \begin{array}{cc}
{{\left( a_{m}^{\pm } \right)}^{*}}\left\langle  m \right| & {{\left( b_{m+1}^{\pm } \right)}^{*}}\left\langle  m+1 \right|  \\
\end{array} \right| \nonumber
\end{eqnarray}
They satisfy the orthogonality condition
\begin{eqnarray}
\left\langle  \psi _{m}^{+} | \psi _{m}^{-} \right\rangle ={{\left( a_{m}^{+} \right)}^{*}}a_{m}^{-}+{{\left( b_{m+1}^{+} \right)}^{*}}b_{m+1}^{-}=0 \nonumber
\end{eqnarray}
which is fulfilled due the relations
\begin{eqnarray}
&&a_{m}^{+}=\frac{i\alpha \sqrt{2}\sqrt{m+1}b_{m+1}^{+}}{{(1}/{2}\;-{{Z}_{e}})+\sqrt{{{({1}/{2}\;-{{Z}_{e}})}^{2}}+2{{\left| \alpha  \right|}^{2}}\left( m+1 \right)}},   \\
&&b_{m+1}^{-}=\frac{i\alpha \sqrt{2}\sqrt{m+1}a_{m}^{-}}{\left( {1}/{2}\;-{{Z}_{e}} \right)+\sqrt{{{({1}/{2}\;-{{Z}_{e}})}^{2}}+2{{\left| \alpha  \right|}^{2}}\left( m+1 \right)}}.\nonumber
\end{eqnarray}

The normalization condition was mentioned above. The normalized wave functions n the coordinate representation are
\begin{eqnarray}
\left| \psi _{m}^{\pm }\left( {\vec{r}} \right) \right\rangle =\frac{{{e}^{ipx}}}{\sqrt{{{L}_{x}}}}\left| \begin{array}{cc}
   a_{m}^{\pm }{{\varphi }_{m}}\left( \eta  \right)  \\
   b_{m+1}^{\pm }{{\varphi }_{m+1}}\left( \eta  \right)  \\
\end{array} \right|, \eta = \frac{y}{l}-pl
\end{eqnarray}
where $L_{x}$ is the length of the layer. These states were obtained for the first time by Rashba [1] and are reproduced here including the Zeeman coefficient $Z_{e}$.
Side by side with the solutions $\left| \psi _{m}^{\pm } \right\rangle $ with $m\ge 0$ there exist else one solution with $b_{0}=1$ of the type
\begin{eqnarray}
{\varepsilon }_{0}=\frac{1}{2}-{{Z}_{e}}, \left| {{\psi }_{0}} \right\rangle =\left| \begin{array}{cc}
   0  \\
   \left| 0 \right\rangle   \\
\end{array} \right|, \nonumber\\
\left| {{\psi }_{0}}\left( {\vec{r}} \right) \right\rangle =\frac{{{e}^{ipx}}}{\sqrt{{{L}_{x}}}}\left| \begin{array}{cc}
   0  \\
   {{\varphi }_{0}}\left( \eta  \right)  \\
\end{array} \right|,
\end{eqnarray}
which is orthogonal to any solutions (7).
In energy units the energy spectrum is
\begin{eqnarray}
E_{m}^{\pm }=\hbar {{\omega }_{ce}}\varepsilon _{m}^{\pm }, m\ge 0, \nonumber\\
E_{0}=\hbar {{\omega }_{ce}}{{\varepsilon }_{0}}
\end{eqnarray}
Below we will consider only two lowest Landau levels for conduction electrons, namely the state $\left| \psi _{0}^{-} \right\rangle $ (9) and $\left| \psi _{0} \right\rangle $ (10). They will be denoted as $(e{{R}_{1}})$ and $(e{{R}_{2}})$ correspondingly. In the fig.1 the dependences on the magnetic field strength B of the energy levels $E_{0}^{\pm }(B)$ and ${{E}_{0}}(B)$ are denoted by the levels ${{0}^{\pm }}$ and 0 correspondingly. The two lowest energy levels $0^{-}$ and 0 change their positions on the energy scale in dependence on the values of the electron g-factor ${{g}_{e}}$.

\begin{figure}
\resizebox{0.48\textwidth}{!}{%
  \includegraphics{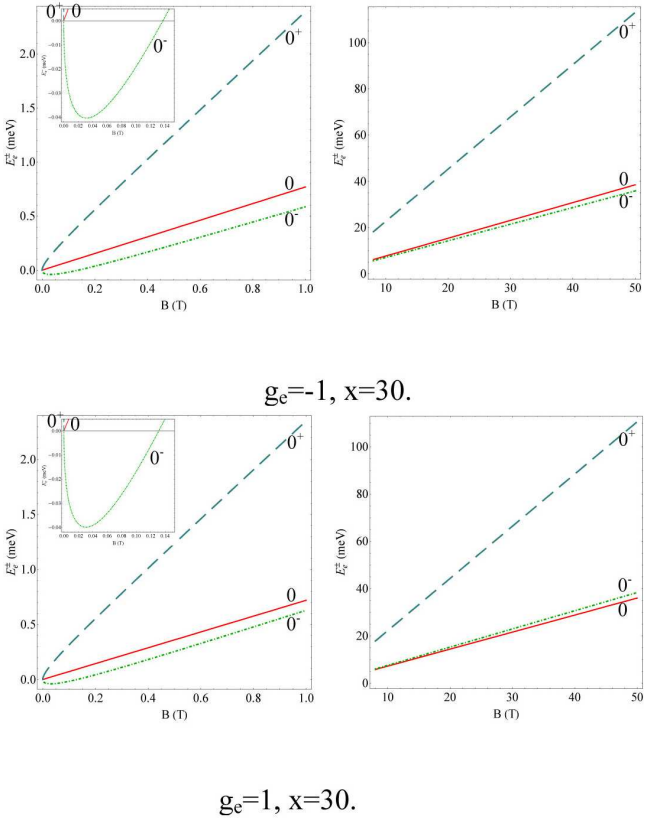}
}
\caption{The electron branches of the Landau quantization levels $E_{e}^{\pm }(0)$ and $E_e(0)$ for two different values of g-factor $g_{e}=\pm 1$ and parameter $x=30$}
\end{figure}
The parameter $\alpha$, which determines the strength of the RSOI in the electron Hamiltonian $H_{e}$ of the formulas (1), can be tuned by changing the external gate voltage applied perpendicular to the interface of the QW. As was mentioned in the Ref.[29] it opens the possibility of an effective spin control in spintronics. Side by side with the RSOI arising due to the structure inversion asymmetry (SIA) there exist another intrinsic mechanism of spin-orbit interaction due to the bulk inversion asymmetry(BIA) revealed by Dresselhaus [30]. Following the Ref.[30] the Dresselhaus SOI is described by the expression $\gamma(\sigma_{x}k_{x}-\sigma_{y}k_{y})$ where $\sigma_{x}$ and $\sigma_{y}$ are the Pauli matrices and $k_{x}$, $k_{y}$ are the components of the electron in-plane wave vector. In the presence of the external magnetic field perpendicular to the layer surface in Landau gauge description the canonical momentum operator, as usual, is substituted by the kinetic momentum operator. Introducing the increasing and decreasing differential operator and passing to second quantization representation on the base of bosonic creation and annihilation operators $a^{\dag}$, $a$, as it is described in-detail in the Ref.[6] we can transcribe the Dresselhaus SOI (DSOI) Hamiltonian in the form $\gamma\left| \begin{array}{cc}
   0 & a^{\dag}  \\
   a & 0   \\
\end{array} \right|$.
The Landau quantization task in the presence of the DSOI, taking into account the Zeeman splitting effects was solved as our previous results using the Rashba method [1]. The dimensionless energy spectrum of the conduction electron in these conditions and the corresponding wave functions in the second quantization representation are:
\begin{eqnarray}
E_{m}^{\pm}&=& m \pm \sqrt{(\frac{1}{2}-Z_{e})^{2}+\beta ^{2}m}, m\geq 1, \nonumber \\
\left| \psi _{m}^{\pm} \right\rangle &=& \left| \begin{array}{cc}
   a_{m}^{\pm}\left| m \right\rangle  \\
   b_{m-1}^{\pm}\left| m-1 \right\rangle   \\
\end{array} \right|, |a_{m}^{\pm}|^{2}+|b_{m-1}^{\pm}|^{2}=1, \nonumber \\
E_{0} &=& \frac{1}{2}-Z_{e}, \left| \psi _{0} \right\rangle=\left| \begin{array}{cc}
   \left| 0 \right\rangle  \\
   0   \\
\end{array} \right|
\end{eqnarray}
They can be used for the subsequent calculations in the same manner as the results concerning the RSOC.

The heavy-hole Hamiltonian in dimensionless form looks as
\begin{eqnarray}
H&=&\frac{{{H}_{h}}}{\hbar {{\omega }_{ch}}}=\{[({{a}^{\dagger}}a+{1}/{2}\;)+\delta {{({{a}^{\dagger}}a+{1}/{2}\;)}^{2}}]\hat{I}+\nonumber \\
&&+i2\sqrt{2}\beta +\left| \begin{array}{cc}
   0 & {{\left( {{a}^{\dagger}} \right)}^{3}}  \\
   -{{(a)}^{3}} & 0  \\
\end{array} \right|-{{Z}_{h}}{{{\hat{\sigma }}}_{z}}\}, \\
H\left| \psi  \right\rangle &=&\varepsilon \left| \psi  \right\rangle ,\left| \psi  \right\rangle =\left| \begin{array}{cc}
   \left| {{f}_{1}} \right\rangle   \\
   \left| {{f}_{2}} \right\rangle   \\
\end{array} \right|,\left| {{f}_{1}} \right\rangle =\sum\limits_{n\ge 0}{{{c}_{n}}\left| n \right\rangle }, \nonumber \\
&&\left| {{f}_{2}} \right\rangle =\sum\limits_{n\ge 0}{{{d}_{n}}\left| n \right\rangle }. \nonumber
\end{eqnarray}
Using the properties of the Fock states
\begin{eqnarray}
{{a}^{3}}\left| n \right\rangle &=&\sqrt{n(n-1)(n-2)}\left| n-3 \right\rangle , \nonumber\\
&& n\ge 3. \nonumber\\
{{({{a}^{\dagger }})}^{3}}\left| n \right\rangle &=&\sqrt{(n+1)(n+2)(n+3)}\left| n+3 \right\rangle , \nonumber\\
&& n\ge 0,
\end{eqnarray}
one can transcribe the Pauli equation (14) in a form of two algebraic equations
\begin{eqnarray}
&&\sum\limits_{n\ge 0}{{{c}_{n}}}[(n+{1}/{2}\;)+\delta {{(n+{1}/{2}\;)}^{2}}-{{Z}_{h}}-\varepsilon ]\left| n \right\rangle =\nonumber\\
&&=-i2\sqrt{2}\beta \sum\limits_{n\ge 0}{{{d}_{n}}\sqrt{(n+1)(n+2)(n+3)}}\left| n+3 \right\rangle , \nonumber\\
&&\sum\limits_{n\ge 0}{{{d}_{n}}}[(n+{1}/{2}\;)+\delta {{(n+{1}/{2}\;)}^{2}}+{{Z}_{h}}-\varepsilon ]\left| n \right\rangle =\nonumber \\
&&i2\sqrt{2}\beta \sum\limits_{n\ge 3}{{{c}_{n}}\sqrt{n(n-1)(n-2)}}\left| n-3 \right\rangle
\end{eqnarray}
Multiplying the first equation (15) form the left hand side by the Fock state $\left\langle  m \right|$ and the second equation (15) by the state $\left\langle  m-3 \right|$ we will obtain two linear algebraic equations determining the coefficients ${{C}_{m}}$ and $dm-3$ at $m\ge 3$:
\begin{eqnarray}
&& [(m+{1}/{2)}\;+\delta (m+{{{1}/{2)}\;}^{2}}-{{Z}_{h}}-\varepsilon ]{{c}_{m}}= \nonumber\\
&& =-i2\sqrt{2}\beta \sqrt{m(m-1)(m-2)}{{d}_{m-3}}, \\
&& [(m-{5}/{2}\;)+\delta {{(m-{5}/{2}\;)}^{2}}+{{Z}_{h}}-\varepsilon ]{{d}_{m-3}}= \nonumber\\
&& =i2\sqrt{2}\beta \sqrt{m(m-1)(m-2)}{{c}_{m}}. \nonumber
\end{eqnarray}
They determine the energy spectrum as follows
\begin{eqnarray}
&& \varepsilon _{m}^{\pm }=\frac{E_{m}^{\pm }}{\hbar {{\omega }_{ch}}}=(m-1)+\delta ({{m}^{2}}-2m+{13}/{4}\;)\pm  \nonumber\\
&& \pm \sqrt{{{[{3}/{2}\;-{{Z}_{h}}+3\delta (m-1)]}^{2}}+8{{\beta }^{2}}m(m-1)(m-2)}, \nonumber\\
&& m\ge 3.
\end{eqnarray}
There are two pairs of the coefficients $c_{m}^{\pm }$, $d_{m-3}^{\pm }$. For $m\ge 3$ they are:
\begin{eqnarray}
&& c_{m}^{-}=-i2\sqrt{2}\beta \sqrt{m\left( m-1 \right)\left( m-2 \right)}d_{m-3}^{-}/ \nonumber\\
&& /\left[ \frac{3}{2}-{{Z}_{h}}+3\delta \left( m-1 \right)+ \right. \nonumber\\
&& +\left. \sqrt{{{\left[ \frac{3}{2}-{{Z}_{h}}+3\delta \left( m-1 \right) \right]}^{2}}+8{{\beta }^{2}}m\left( m-1 \right)\left( m-2 \right)} \right], \nonumber\\
&& d_{m-3}^{+}=-i2\sqrt{2}\beta \sqrt{m\left( m-1 \right)\left( m-2 \right)}c_{m}^{+}/ \nonumber\\
&& /\left[ \frac{3}{2}-{{Z}_{h}}+3\delta \left( m-1 \right)+ \right. \nonumber\\
&& +\left. \sqrt{{{\left[ \frac{3}{2}-{{Z}_{h}}+3\delta \left( m-1 \right) \right]}^{2}}+8{{\beta }^{2}}m\left( m-1 \right)\left( m-2 \right)} \right], \nonumber\\
&& {{\left| d_{m-3}^{-} \right|}^{2}}=\left[ \frac{3}{2}-{{Z}_{h}}+3\delta \left( m-1 \right)+ \right. \nonumber\\
&& {{\left. +\sqrt{{{\left[ \frac{3}{2}-{{Z}_{h}}+3\delta \left( m-1 \right) \right]}^{2}}+8{{\beta }^{2}}m\left( m-1 \right)\left( m-2 \right)} \right]}^{2}}/ \nonumber\\
&& /\left( \left[ \frac{3}{2}-{{Z}_{h}}+3\delta \left( m-1 \right)+ \right. \right. \nonumber\\
&& +{{\left. \sqrt{{{\left[ \frac{3}{2}-{{Z}_{h}}+3\delta \left( m-1 \right) \right]}^{2}}+8{{\beta }^{2}}m\left( m-1 \right)\left( m-2 \right)} \right]}^{2}} \nonumber\\
&& \left. +8{{\beta }^{2}}m\left( m-1 \right)\left( m-2 \right) \right), \\
&& {{\left| c_{m}^{-} \right|}^{2}}=8{{\beta }^{2}}m\left( m-1 \right)\left( m-2 \right)/ \nonumber\\
&& /\left( \left[ \frac{3}{2}-{{Z}_{h}}+3\delta \left( m-1 \right)+ \right. \right. \nonumber\\
&& +{{\left. \sqrt{{{\left[ \frac{3}{2}-{{Z}_{h}}+3\delta \left( m-1 \right) \right]}^{2}}+8{{\beta }^{2}}m\left( m-1 \right)\left( m-2 \right)} \right]}^{2}} \nonumber\\
&& \left. +8{{\beta }^{2}}m\left( m-1 \right)\left( m-2 \right) \right), \nonumber\\
&& {{\left| c_{m}^{+} \right|}^{2}}={{\left| d_{m-3}^{-} \right|}^{2}},{{\left| c_{m}^{-} \right|}^{2}}={{\left| d_{m-3}^{+} \right|}^{2}},m\ge 3. \nonumber
\end{eqnarray}
The corresponding wave functions have the forms
\begin{eqnarray}
&& \left| \psi _{m}^{\pm } \right\rangle =\left| \begin{array}{cc}
   c_{m}^{\pm }\left| m \right\rangle   \\
   d_{m-3}^{\pm }\left| m-3 \right\rangle   \\
\end{array} \right|, \nonumber \\
&&\left| \psi _{m}^{\pm }\left( {\vec{r}} \right) \right\rangle =\frac{{{e}^{iqx}}}{\sqrt{{{L}_{x}}}}\left| \begin{array}{cc}
   c_{m}^{\pm }{{\varphi }_{m}}\left( \eta  \right)  \\
   d_{m-3}^{\pm }{{\varphi }_{m-3}}\left( \eta  \right)  \\
\end{array} \right|, \nonumber\\
&& \eta =\frac{y}{l}+ql
\end{eqnarray}
They obey to the normalization and orthogonality conditions
\begin{eqnarray}
&& \left\langle  \psi _{m}^{\pm } | \psi _{m}^{\pm } \right\rangle ={{\left| c_{m}^{\pm } \right|}^{2}}+\left| d_{m-3}^{\pm } \right|=1, \nonumber\\
&& \left\langle  \psi _{m}^{+} | \psi _{m}^{-} \right\rangle =c_{m}^{+*}c_{m}^{-}+d_{m-3}^{+*}d_{m-3}^{-}=0.
\end{eqnarray}
Side by side with the solutions (19) there are else three solutions with $m=0,1,2.$ they are:
\begin{eqnarray}
&& {{c}_{0}}=1,\left| {{\psi }_{0}} \right\rangle =\left| \begin{array}{cc}
   \left| 0 \right\rangle   \\
   0  \\
\end{array} \right|, \left| {{\psi }_{0}}\left( {\vec{r}} \right) \right\rangle =\frac{{{e}^{iqx}}}{\sqrt{{{L}_{x}}}}\left| \begin{array}{cc}
   {{\varphi }_{0}}\left( \eta  \right)  \\
   0  \\
\end{array} \right|, \nonumber\\
&& {{\varepsilon }_{0}}=\frac{1}{2}+\frac{\delta }{4}-{{Z}_{h}}, \nonumber\\
&& {{c}_{1}}=1, \left| {{\psi }_{1}} \right\rangle =\left| \begin{array}{cc}
   \left| 1 \right\rangle   \\
   0  \\
\end{array} \right|, \left| {{\psi }_{1}}\left( {\vec{r}} \right) \right\rangle =\frac{{{e}^{iqx}}}{\sqrt{{{L}_{x}}}}\left| \begin{array}{cc}
   {{\varphi }_{1}}\left( \eta  \right)  \\
   0  \\
\end{array} \right|, \nonumber\\
&& {{\varepsilon }_{1}}=\frac{3}{2}+\frac{9}{4}\delta -{{Z}_{h}}, \\
&& {{c}_{2}}=1,\left| {{\psi }_{2}} \right\rangle =\left| \begin{array}{cc}
   \left| 2 \right\rangle   \\
   0  \\
\end{array} \right|, {{\psi }_{2}}\left( {\vec{r}} \right)=\frac{{{e}^{iqx}}}{\sqrt{{{L}_{x}}}}\left| \begin{array}{cc}
   {{\varphi }_{2}}\left( \eta  \right)  \\
   0  \\
\end{array} \right|, \nonumber\\
&& {{\varepsilon }_{2}}=\frac{5}{2}+\frac{25}{4}\delta -{{Z}_{h}} \nonumber
\end{eqnarray}
All of them are orthogonal to the previous solutions (19).

Fig. 2 shows the heavy-hole energy levels for two different values of the g-factor $g_{h}=\pm 5$ in two ranges of the magnetic field strength B. One of them concerns the values $B\ge 10$T where the magnetoexcitons in GaAs-type crystals will be formed. There are three groups of energy levels $E_{h}^{\pm }(m)$ with $m=3,4,...,10$ and $E_{h}(m)$ with $m=0,1,2$. The lowest of them are $E_{h}^{-}(3)$, $E_{h}(0)$ and $E_{h}^{-}(4)$. They will be denoted as $(h,R_{j})$ with $j=1,2,3$ correspondingly.
\begin{figure}%[h]
\resizebox{0.48\textwidth}{!}{%
  \includegraphics{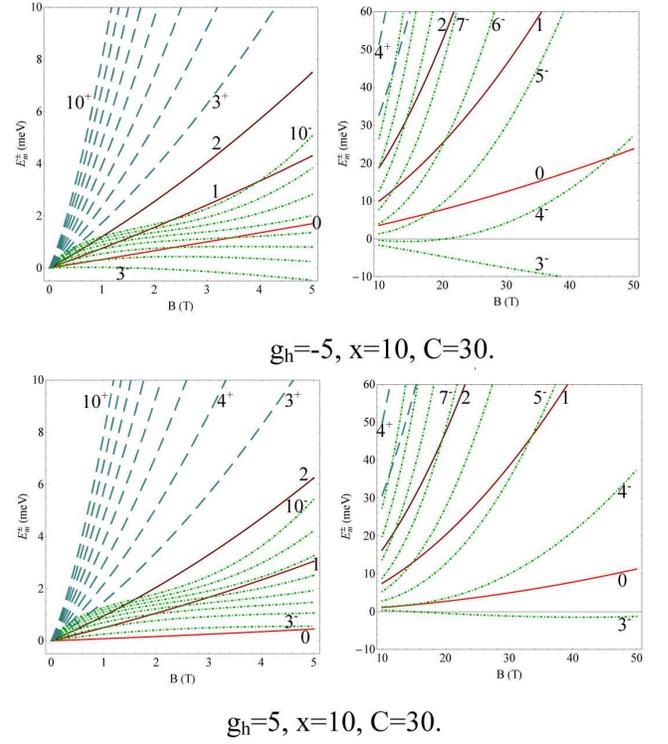}
}
\caption{The heavy-hole Landau quantization levels $E_{h}^{\pm }(0)$, with $m=3,4...10$ and $E_{h}(m)$ with $m=0,1,2$ and two values of the g-factor ${{g}_{h}}=\pm 5$. The parameters $x=10$ and $C=30$ were chosen.}
\end{figure}

The Landau quantization and the energy levels of the electrons and holes determine the energies of the optical band-to-band quantum transitions as well as of the magnetoexciton creation energy.

The band-to-band quantum transitions can be considered for a wider range of magnetic field strength because they Coulomb electron-electron interactions is not taken into account and the knowledge about the ionization potentials of the magnetoexcitons are not needed. The quantum transitions from the ground state of the crystal to the magnetoexciton states and to the states of an electron-hole (e-h) pair without Coulomb e-h interaction will be considered taking into account three lowest Landau levels(LLLs) for holes denoted as $(h,R_{j})$ with $j=1,2,3$ and two LLLs for electrons denoted as $(e,R_{i})$ with $i=1,2.$ These combinations have the energies of the band-to-band transitions ${{E}_{cv}}\left( {{F}_{n}} \right)$. Being accounted from the semiconductor energy band gap $E_{g}^{0}$ they equal to
\begin{eqnarray}
E_{cv}(F_{n})-E_{g}^{0}=E_{e}(R_{i})+E_{h}(R_{j})
\end{eqnarray}
The energies of the band-to-band transitions are presented in Fig. 3, 4 in dependences on the magnetic field strength $B$ and on the $g-$factors $g_{e}$ and $g_{h}$ at different parameters of the Rashba spin-orbit coupling and of the nonparabolic heavy-hole dispersion low.

In the Fig. 3 only the influence of the Landau quantization and of the Zeeman splitting is demonstrated under the influence of the perpendicular magnetic field. In this case we have the parameters $x=0$ and $C=0$. In the presence of a supplementary electric field with the same direction an additional Rashba spin-orbit coupling takes place accompannied by the nonparabolicity of the heavy-hole dispersion law induced also by the electric field. The influence of these four factor, namely of LC, ZS, RSOC and NP is represented in the fig. 4 for some concrete parameters.
\begin{figure}%[h]
\resizebox{0.48\textwidth}{!}{%
  \includegraphics{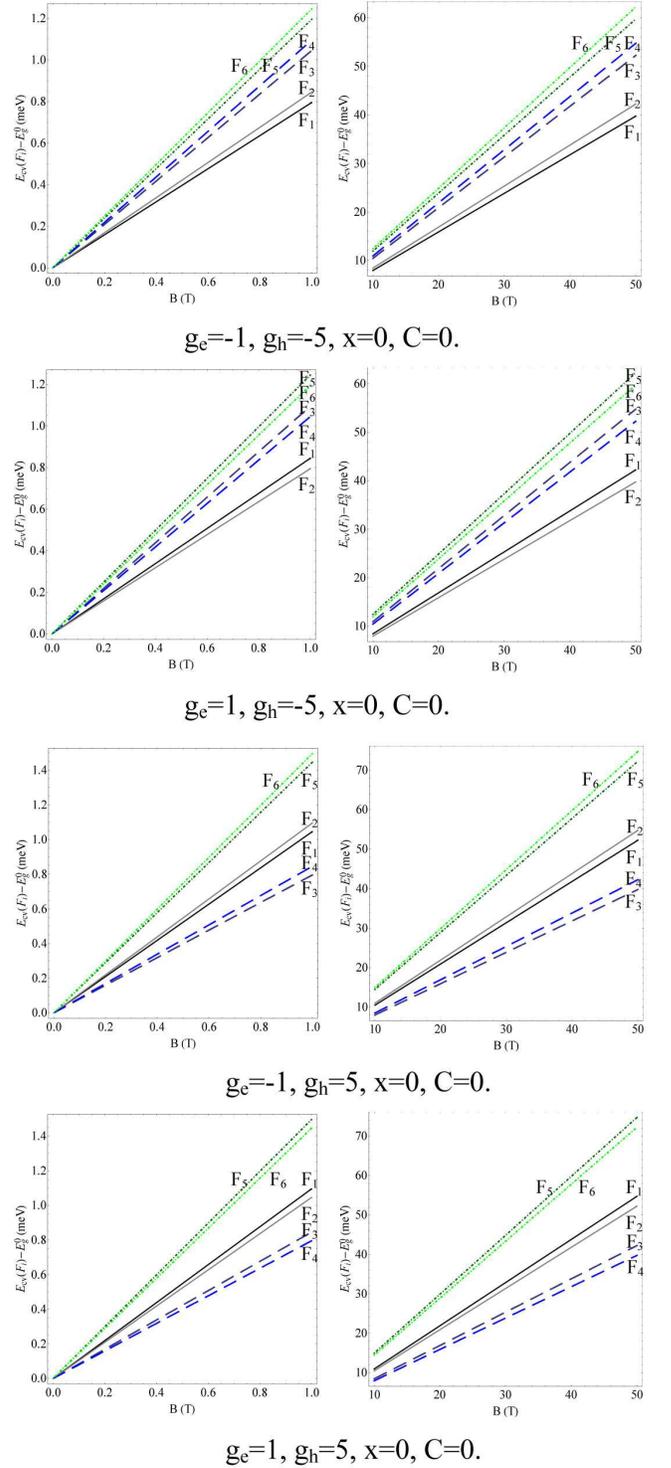}
}
\caption{The band-to-band quantum transitions energies $E_{cv}(F_{i})-E_{g}^{0}$ for six combinations $F_{i}$ of three heavy-hole and of two electron lowest Landau levels in dependence on the magnetic field strength B at different combinations of the electron and of the heavy-hole $g_{e}$ and $g_{h}$. The influence of the electric field is absent what leads to the parameters: $x=0$ and $C=0$.}
\end{figure}

\begin{figure}%[h]
\resizebox{0.48\textwidth}{!}{%
  \includegraphics{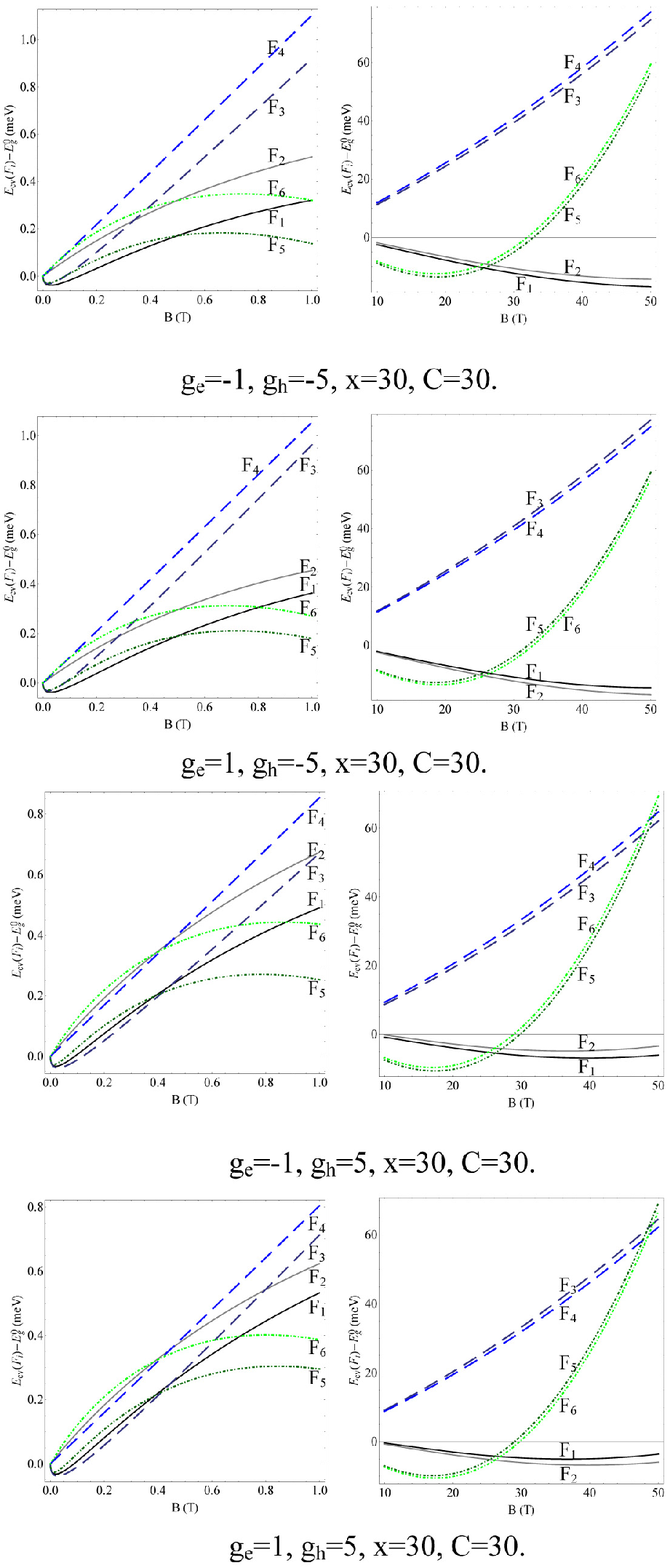}
}
\caption{The band-to-band quantum transitions energies $E_{cv}(F_{i})-E_{g}^{0}$ for six combinations $F_{i}$ of three heavy-hole and of two electron lowest Landau levels in dependence on the magnetic field strength B at different combinations of the electron and of the heavy-hole g-factors $g_{e}$ and $g_{h}$. The influence of the electric field is taken into account choosing the parameters: $x=30$ and $C=30$.}
\end{figure}

\section{The Coulomb electron-electron interaction and the magnetoexcitons}
The three LLLs for 2D heavy-holes $(h,R_{j})$ with $j=1,2,3$ were combined with two LLLs for 2D conduction electrons $(e,R_{i})$ with $i=1,2$ giving rise to six 2D magnetoexciton states ${{F}_{n}}$ with $n=1,2...6$ [23-25]. To calculate their ionization potentials the Hamiltonian of the Coulomb electron-electron interaction in the conditions of the Landau quantization, Rashba spin-orbit coupling and Zeeman splitting is needed. It was obtained in the Refs. [23-25] taking into account only the first two conditions, namely the LQ and RSOC. Below we will generalize these results adding also the ZS effects. Moreover, it can be done because the Pauli Hamiltonians containing the Zeeman effects are represented by the operators $\pm {{g}_{i}}\frac{{{\mu }_{B}}B}{2}{{\hat{\sigma }}_{z}}$, where $\hat{\sigma }_{z}$ has only diagonal matrix elements. For the electron-hole pair in the concrete combination $({{R}_{i}},\varepsilon _{m}^{-})$, the Hamiltonian of the Coulomb interaction is
\begin{eqnarray}
&&{{H}_{coul}}( {{R}_{i}},\varepsilon _{m}^{-} )= \nonumber \\
&&=\frac{1}{2}\sum\limits_{{\vec{a}}}{\{{{W}_{e-e}}( {{{\vec{R}}}_{i}};\vec{Q} )}[{{{\hat{\rho }}}_{e}}( {{R}_{i}};\vec{Q} ){{{\hat{\rho }}}_{e}}( {{R}_{i}},-\vec{Q} )-{{{\hat{N}}}_{e}}( {{R}_{i}} )] \nonumber\\
&& +{{W}_{h-h}}( \varepsilon _{m}^{-};\vec{Q} )[{{{\hat{\rho }}}_{h}}( {{M}_{h}},\varepsilon _{m}^{-};\vec{Q} ){{{\hat{\rho }}}_{h}}( {{M}_{h}},\varepsilon _{m}^{-};\vec{Q} )- \nonumber\\
&&-{{{\hat{N}}}_{h}}( {{M}_{h}},\varepsilon _{m}^{-} )]- \nonumber\\
&& -2{{W}_{e-h}}( {{R}_{i}},\varepsilon _{m}^{-};\vec{Q} ){{{\hat{\rho }}}_{e}}( {{R}_{i}};\vec{Q}){{{\hat{\rho }}}_{h}}( {{M}_{h}},\varepsilon _{m}^{-};-\vec{Q} )\}
\end{eqnarray}
Here the electron and hole density operators are
\begin{eqnarray}
&& {{{\hat{\rho }}}_{e}}({{R}_{i}};\vec{Q})=\sum\limits_{t}{{{e}^{i{{Q}_{y}}tl_{0}^{2}}}}a_{{{R}_{i}},t+\frac{{{Q}_{x}}}{2}}^{\dagger }{{a}_{{{R}_{i}},t-\frac{{{Q}_{x}}}{2}}}, \nonumber\\
&& {{{\hat{\rho }}}_{h}}({{M}_{h}},\varepsilon _{m}^{-};\vec{Q})=\sum\limits_{t}{{{e}^{-i{{Q}_{y}}tl_{0}^{2}}}}b_{{{M}_{h}},\varepsilon _{m}^{-},t+\frac{{{Q}_{x}}}{2}}^{\dagger }{{b}_{{{M}_{h}},\varepsilon _{m}^{-},t-\frac{{{Q}_{x}}}{2}}}, \nonumber\\
&& {{{\hat{N}}}_{e}}({{R}_{i}})={{{\hat{\rho }}}_{e}}({{R}_{i}},0),{{{\hat{N}}}_{h}}({{M}_{h}},\varepsilon _{m}^{-})={{{\hat{\rho }}}_{h}}({{M}_{h}},\varepsilon _{m}^{-};0).
\end{eqnarray}
The coefficients ${{W}_{i-j}}(Q)$ in the case of the electron state ${{R}_{1}}$ described by the formula (9) with coefficients $a_{0}^{-}$ and $b_{1}^{-}$ and of the holes states $( {{M}_{h}},\varepsilon _{m}^{-} )$ were obtained in Ref.[28] and equal to
\begin{eqnarray}
&&{{W}_{e-e}}({{R}_{1}};\vec{Q})=W(\vec{Q}){{({{| a_{0}^{-} |}^{2}}{{A}_{0,0}}(\vec{Q})+{{| b_{1}^{-} |}^{2}}{{A}_{1,1}}(\vec{Q}))}^{2}} \nonumber\\
&& {{W}_{h-h}}(\varepsilon _{m}^{-};\vec{Q})= \nonumber \\
&&=W(\vec{Q}){{({{| d_{m-3}^{-} |}^{2}}{{A}_{m-3,m-3}}(\vec{Q})+{{| c_{m}^{-} |}^{2}}{{A}_{m,m}}(\vec{Q}))}^{2}}, \nonumber\\
&&m\ge 3. \\
&&{{W}_{e-h}}(R1,\varepsilon _{m}^{-};\vec{Q})= \nonumber \\
&&=W(\vec{Q})({{| a_{0}^{-} |}^{2}}{{A}_{0,0}}(\vec{Q})+{{| b_{1}^{-} |}^{2}}{{A}_{1,1}}(\vec{Q}))\times  \nonumber\\
&&\times ({{| d_{m-3}^{-}|}^{2}}{{A}_{m-3,m-3}}(\vec{Q})+{{| c_{m}^{-} |}^{2}}{{A}_{m,m}}(\vec{Q})),m\ge 3 \nonumber
\end{eqnarray}
with the normalization conditions
\begin{eqnarray}
&&{{| a_{0}^{-} |}^{2}}+{{| b_{1}^{-} |}^{2}}=1, \nonumber\\
&&{{| d_{m-3}^{-} |}^{2}}+{{| c_{m}^{-} |}^{2}}=1, \nonumber\\
&&m\ge 3.
\end{eqnarray}
The first five functions ${{A}_{m,m}}(\vec{Q})$ with $m\le 4$ are
\begin{eqnarray}
&&{{A}_{0,0}}(\vec{Q})=1, {{A}_{1,1}}(\vec{Q})=1-\frac{{{Q}^{2}}l_{0}^{2}}{2}, \nonumber\\
&&{{A}_{2,2}}(\vec{Q})=1-{{Q}^{2}}l_{0}^{2}+\frac{{{Q}^{4}}l_{0}^{4}}{8}, \\
&&{{A}_{3,3}}(\vec{Q})=1-\frac{3}{2}{{Q}^{2}}l_{0}^{2}+\frac{3}{8}{{Q}^{4}}l_{0}^{4}-\frac{1}{48}{{Q}^{6}}l_{0}^{6}, \nonumber\\
&&{{A}_{4,4}}(\vec{Q})=1-2{{Q}^{2}}l_{0}^{2}+\frac{3}{4}{{Q}^{4}}l_{0}^{4}-\frac{{{Q}^{6}}l_{0}^{6}}{12}+\frac{{{Q}^{8}}l_{0}^{8}}{384} \nonumber
\end{eqnarray}
In the particular hole state $\varepsilon _{3}^{-}$ the expressions for ${{W}_{i-j}}(\vec{Q})$ were obtained in Ref.[28]:
\begin{eqnarray}
&& {{W}_{e-e}}({{R}_{1}};\vec{Q})=W(\vec{Q}){(1-\frac{|b_{1}^{-}{{|}^{2}}}{2}{{Q}^{2}}l_{0}^{2})^{2}}, \nonumber\\
&& {{W}_{h-h}}(\varepsilon _{3}^{-};\vec{Q})= \nonumber\\
&& =W(\vec{Q}){(1-\frac{|c_{3}^{-}{{|}^{2}}}{2}(3{{Q}^{2}}l_{0}^{2}-\frac{3}{4}{{Q}^{4}}l_{0}^{4}+\frac{1}{24}{{Q}^{6}}l_{0}^{6}))^{2}}, \nonumber\\
&& {{W}_{e-h}}({{R}_{1}},\varepsilon _{3}^{-};\vec{Q})= \\
&& =W(\vec{Q})(1-\frac{|b_{1}^{-}{{|}^{2}}}{2}{{Q}^{2}}l_{0}^{2})\times \nonumber \\
&& \times (1-\frac{|c_{3}^{-}{{|}^{2}}}{2}(3{{Q}^{2}}l_{0}^{2}-\frac{3}{4}{{Q}^{4}}l_{0}^{4}+\frac{1}{24}{{Q}^{6}}l_{0}^{6})), \nonumber\\
&& W(\vec{Q})={{e}^{\frac{-{{Q}^{2}}l_{0}^{2}}{2}}}V(\vec{Q}), V(\vec{Q})=\frac{2\pi {{e}^{2}}}{{{\varepsilon }_{0}}S|\vec{Q}|}\nonumber
\end{eqnarray}
For the case $m=4$ we can add
\begin{eqnarray}
&& {{W}_{h-h}}(\varepsilon _{4}^{-};\vec{Q})= \nonumber\\
&&=W(\vec{Q})(1-\frac{{{Q}^{2}}l_{0}^{2}}{2}- \nonumber \\
&&-\frac{|c_{4}^{-}{{|}^{2}}}{2}(3{{Q}^{2}}l_{0}^{2}-\frac{3}{2}{{Q}^{4}}l_{0}^{4}+\frac{{{Q}^{6}}l_{0}^{6}}{6}-\frac{{{Q}^{8}}l_{0}^{8}}{192}))^{2}, \nonumber\\
&&{{W}_{e-h}}({{R}_{1}},\varepsilon _{4}^{-};\vec{Q})= \\
&&=W(\vec{Q})(1-\frac{|b_{1}^{-}{{|}^{2}}}{2}{{Q}^{2}}l_{0}^{2})\times  \nonumber\\
&&\times (1-\frac{{{Q}^{2}}l_{0}^{2}}{2}-\frac{{{\left| c_{4}^{-} \right|}^{2}}}{2}(3{{Q}^{2}}l_{0}^{2}-\frac{3}{2}{{Q}^{4}}l_{0}^{4}+\frac{{{Q}^{6}}l_{0}^{6}}{6}-\frac{{{Q}^{8}}l_{0}^{8}}{192}))\nonumber
\end{eqnarray}
Side by side with the electron state ${{R}_{1}},$ we will consider also the state ${{R}_{2}}$ described by the formulas (10). The combination of the electron state ${{R}_{2}}$ with the holes states $\varepsilon _{m}^{-}$ gives rise to the e-h states$\left( e{{R}_{2}},h,\varepsilon _{m}^{-} \right)$. The coefficients ${{W}_{i-j}}\left( {{R}_{2}},\varepsilon _{m}^{-};\vec{Q} \right)$ can be obtained from the coefficients ${{W}_{i-j}}\left( {{R}_{1}},\varepsilon _{m}^{-};\vec{Q} \right)$ setting in $b_{1}^{-}=0$
\begin{eqnarray}
&&{{W}_{i-j}}\left( {{R}_{2}},\varepsilon _{m}^{-};\vec{Q} \right)={{W}_{i-j}}\left( {{R}_{1}},\varepsilon _{m}^{-};\vec{Q} \right)\left| \begin{array}{cc}
   {}  \\
   b_{1}^{-}=0  \\
\end{array} \right., \nonumber\\
&& i,j=e,h
\end{eqnarray}
The terms proportional to ${{\hat{N}}_{e}}({{R}_{1}})$ and ${{\hat{N}}_{h}}({{M}_{h}},\varepsilon _{m}^{-})$ in (22) have the coefficients ${{I}_{e}}({{R}_{1}})$ and ${{I}_{h}}(\varepsilon _{m}^{-})$, which describe the Coulomb self-actions of the electrons and holes. They equal to
\begin{eqnarray}
&& {{I}_{e}}\left( {{R}_{1}} \right)=\frac{1}{2}\sum\limits_{{\vec{Q}}}{{{W}_{e-e}}\left( {{R}_{1}},\vec{Q} \right)}, \nonumber\\
&& {{I}_{h}}\left( \varepsilon _{m}^{-} \right)=\frac{1}{2}\sum\limits_{{\vec{Q}}}{{{W}_{h-h}}\left( \varepsilon _{m}^{-};\vec{Q} \right)},\nonumber \\
&& {{I}_{s}}\left( {{R}_{1}},\varepsilon _{m}^{-} \right)={{I}_{e}}\left( {{R}_{1}} \right)+{{I}_{h}}\left( \varepsilon _{m}^{-} \right).
\end{eqnarray}
To determine the binding energy of the magnetoexciton its wave functions $\left| {{\psi }_{ex}}\left( {{F}_{i}},\vec{K} \right) \right\rangle$ are obtained acting on the vacuum state $\left| 0 \right\rangle$ by the magnetoexciton creation operator $\psi _{ex}^{\dagger }({{\vec{k}}_{||}},M_{h},{{R}_{i}},\varepsilon )$ constructed from the electron and hole creation operators $a_{{{R}_{i}},t}^{\dagger }$ and $b_{{{M}_{h}},\varepsilon ,t}^{\dagger }$ correspondingly
\begin{eqnarray}
&& \psi _{ex}^{\dagger }\left( {{{\vec{k}}}_{||}},{{R}_{i}},{{M}_{h}},\varepsilon  \right)=\frac{1}{\sqrt{N}}\sum\limits_{t}{{{e}^{i{{k}_{y}}tl_{0}^{2}}}}a_{{{R}_{i}},t+\frac{{{k}_{x}}}{2}}^{\dagger }b_{{{M}_{h}},\varepsilon ;-t+\frac{{{k}_{x}}}{2}}^{\dagger }, \nonumber\\
&& \left| {{\psi }_{ex}}\left( {{R}_{i}};{{M}_{h}},\varepsilon ;{{{\vec{k}}}_{||}} \right) \right\rangle =\psi _{ex}^{\dagger }\left( {{{\vec{k}}}_{||}},{{R}_{i}},{{M}_{h}},\varepsilon  \right)\left| 0 \right\rangle .
\end{eqnarray}
The vacuum state is determined by the equalities
\begin{eqnarray}
{{a}_{\xi ,t}}\left| 0 \right\rangle ={{b}_{\xi ,t}}\left| 0 \right\rangle =0.
\end{eqnarray}
Below we consider a concrete composition ${{F}_{i}}=\left( {{R}_{1}},{{M}_{h}},\varepsilon _{m}^{-} \right)$ with $m\ge 3$ of the electron and hole states.

The binding energy of the magnetoexciton is determined by the diagonal matrix element of the Hamiltonian (23) calculated with the wave function $\left| {{\psi }_{ex}}\left( {{F}_{i}},\vec{k} \right) \right\rangle $
\begin{eqnarray}
&& \left\langle  {{\psi }_{ex}}\left( {{F}_{i}},\vec{k} \right) \right|{{H}_{coul}}\left| {{\psi }_{ex}}\left( {{F}_{i}},\vec{k} \right) \right\rangle =-{{I}_{ex}}\left( {{F}_{i}} \right)+E\left( {{F}_{i}},\vec{k} \right), \nonumber\\
&& {{I}_{ex}}\left( {{F}_{i}} \right)={{I}_{ex}}\left( {{R}_{1}},\varepsilon _{m}^{-} \right)=\sum\limits_{{\vec{Q}}}{{{W}_{e-h}}}\left( {{R}_{1}};\varepsilon _{m}^{-};\vec{Q} \right), \nonumber\\
&& E\left( {{F}_{i}},\vec{k} \right)=E\left( {{R}_{1}};\varepsilon _{m}^{-};\vec{k} \right)= \nonumber\\
&& =2\sum\limits_{{\vec{Q}}}{{{W}_{e-h}}}\left( {{R}_{1}};\varepsilon _{m}^{-};\vec{Q} \right){{\sin }^{2}}\left( \frac{{{\left[ \vec{k}\times \vec{Q} \right]}_{z}}l_{0}^{2}}{2} \right), \\
&&\lim \limits_{k\rightarrow \infty} E(R_{1};\varepsilon _{m}^{-};\overrightarrow{k})=I_{ex}(R_{1};\varepsilon _{m}^{-}).\nonumber
\end{eqnarray}
The binding energy of the magnetoexciton has the opposite sign as compared with ionization potential. They both tends to zero when the wave vector $\vec{k}$ tends to infinity and the magnetoexciton is transformed into a free e-h pair.
The terms of the Hamiltonian which contain only the operators ${{\hat{N}}_{e}}$ and ${{\hat{N}}_{h}}$ describing the full numbers of electrons and holes including their chemical potentials ${{\mu }_{e}}$ and ${{\mu }_{h}}$ are combined in a form:
\begin{eqnarray}
&& {{H}_{mex,1}}=\nonumber\\
&&=({E_{g}^{0}}/{2}\;+{{E}_{e}}({{R}_{1}})-{{I}_{e}}({{R}_{1}})-{{\mu }_{e}}){{{\hat{N}}}_{e}}({{R}_{1}})+ \\
&& +({E_{g}^{0}}/{2}\;+{{E}_{h}}({{M}_{h}},\varepsilon _{m}^{-})-{{I}_{h}}(\varepsilon _{m}^{-})-{{\mu }_{h}}){{{\hat{N}}}_{h}}({{M}_{h}},\varepsilon _{m}^{-}).\nonumber
\end{eqnarray}
Here we introduced the semiconductor energy band gap $E_{g}^{0}$ in the absence of the Landau quantization.

Below we will introduce the density operators of the optical plasmon denoted as $\hat{\rho }(\vec{Q})$ and of the acoustical plasmon denoted as $\hat{D}(\vec{Q})$ instead of the electron and hole density operators ${{\hat{\rho }}_{e}}(\vec{Q})$ and ${{\hat{\rho }}_{h}}(\vec{Q})$ following the relations
\begin{eqnarray}
&& \hat{\rho }(\vec{Q})={{{\hat{\rho }}}_{e}}(\vec{Q})-{{{\hat{\rho }}}_{h}}(\vec{Q}),\nonumber \\
&& \hat{D}(\vec{Q})={{{\hat{\rho }}}_{e}}(\vec{Q})+{{{\hat{\rho }}}_{h}}(\vec{Q}),\nonumber \\
&& {{{\hat{\rho }}}_{e}}(\vec{Q})=\frac{\hat{\rho }(\vec{Q})+\hat{D}(\vec{Q})}{2}, \nonumber\\
&& {{{\hat{\rho }}}_{h}}(\vec{Q})=\frac{\hat{D}(\vec{Q})-\hat{\rho }(\vec{Q})}{2}.
\end{eqnarray}
For the sake of simplicity, the indexes, which label the electron, hole and plasmon density operators, are omitted in (36). They will be restored in concrete cases.

In the plasmon representation the Hamiltonian ${{H}_{mex,1}}$ will be:
\begin{eqnarray}
&&{{H}_{mex,1}}=({{E}_{mex}}({{R}_{1}};{{M}_{h}},\varepsilon _{m}^{-})-{{\mu }_{mex}})\frac{\hat{D}(0)}{2}+ \nonumber\\
&& +({{G}_{e-h}}({{R}_{1}};{{M}_{h}},\varepsilon _{m}^{-})-{{\mu }_{e}}+{{\mu }_{h}})\frac{\hat{\rho }(0)}{2}
\end{eqnarray}
Here the sums and the differences of the Landau quantization level energies, of the Coulomb self-interaction terms and of the chemical potentials are presented in the following way
\begin{eqnarray}
&& {{E}_{mex}}({{R}_{1}};{{M}_{h}},\varepsilon _{m}^{-})={{E}_{g}}({{R}_{1}};{{M}_{h}},\varepsilon _{m}^{-})-{{I}_{ex}}({{R}_{1}};\varepsilon _{m}^{-}), \nonumber\\
&& {{E}_{g}}({{R}_{1}};{{M}_{h}},\varepsilon _{m}^{-})=E_{g}^{0}+{{E}_{e}}({{R}_{1}})+{{E}_{h}}({{M}_{h}},\varepsilon _{m}^{-}), \nonumber\\
&& {{\mu }_{ex}}={{\mu }_{e}}+{{\mu }_{h}}, \\
&& {{G}_{e-h}}({{R}_{1}};{{M}_{h}},\varepsilon _{m}^{-})={{E}_{e}}({{R}_{1}})-{{E}_{h}}({{M}_{h}},\varepsilon _{m}^{-})- \nonumber\\
&& -{{I}_{e}}({{R}_{1}})+{{I}_{h}}(\varepsilon _{m}^{-}).\nonumber
\end{eqnarray}
The remaining part ${{H}_{mex,2}}$ of the Hamiltonian (23) after the excluding of the linear terms is quadratic in the plasmon density operators. It has the form:
\begin{eqnarray}
&& {{H}_{mex,2}}=\frac{1}{2}\sum\limits_{{\vec{Q}}}{\{}{{W}_{0-0}}(\vec{Q})\hat{\rho }(\vec{Q})\hat{\rho }(-\vec{Q})+\nonumber \\
&&+{{W}_{a-a}}(\vec{Q})\hat{D}(\vec{Q})\hat{D}(-\vec{Q})+ \\
&& +{{W}_{0-a}}(\vec{Q})( \hat{\rho }(\vec{Q})\hat{D}(-\vec{Q})+\hat{D}(\vec{Q})\hat{\rho }(-\vec{Q}))\} \nonumber
\end{eqnarray}
The new coefficients are expressed through the former ones by the formulas:
\begin{eqnarray}
&& {{W}_{0-0}}(\vec{Q})=\frac{1}{4}({{W}_{e-e}}(\vec{Q})+{{W}_{h-h}}(\vec{Q})+2{{W}_{e-h}}(\vec{Q})), \nonumber\\
&& {{W}_{a-a}}(\vec{Q})=\frac{1}{4}({{W}_{e-e}}(\vec{Q})+{{W}_{h-h}}(\vec{Q})-2{{W}_{e-h}}(\vec{Q})), \nonumber\\
&& {{W}_{0-a}}(\vec{Q})=\frac{1}{4}({{W}_{e-e}}(\vec{Q})-{{W}_{h-h}}(\vec{Q})).
\end{eqnarray}
In the case of the e-h pairs of the type $({{R}_{1}};\varepsilon _{m}^{-})$ they take the forms:
\begin{eqnarray}
&& {{W}_{0-0}}({{R}_{1}};\varepsilon _{m}^{-};\vec{Q})= \nonumber\\
&& =\frac{1}{4}({{W}_{e-e}}({{R}_{1}};\vec{Q})+{{W}_{h-h}}(\varepsilon _{m}^{-};\vec{Q})+2{{W}_{e-h}}({{R}_{1}};\varepsilon _{m}^{-};\vec{Q}))= \nonumber\\
&& =\frac{W(\vec{Q})}{4}(|{{a}_{0}}{{|}^{2}}{{A}_{0,0}}(\vec{Q})+|{{b}_{1}}{{|}^{2}}{{A}_{1,1}}(\vec{Q})+ \nonumber\\
&& +|d_{m-3}^{-}{{|}^{2}}{{A}_{m-3,m-3}}(\vec{Q})+|c_{m}^{-}{{|}^{2}}{{A}_{m,m}}(\vec{Q}){{)}^{2}}, \nonumber\\
&& {{W}_{a-a}}({{R}_{1}};\varepsilon _{m}^{-};\vec{Q})= \\
&& =\frac{1}{4}({{W}_{e-e}}({{R}_{1}};\vec{Q})+{{W}_{h-h}}(\varepsilon _{m}^{-};\vec{Q})-2{{W}_{e-h}}({{R}_{1}};\varepsilon _{m}^{-};\vec{Q}))=\nonumber \\
&& =\frac{W(\vec{Q})}{4}(|{{a}_{0}}{{|}^{2}}{{A}_{0,0}}(\vec{Q})+|{{b}_{1}}{{|}^{2}}{{A}_{1,1}}(\vec{Q})- \nonumber\\
&& -|d_{m-3}^{-}{{|}^{2}}{{A}_{m-3,m-3}}(\vec{Q})-|c_{m}^{-}{{|}^{2}}{{A}_{m,m}}(\vec{Q}){{)}^{2}}, \nonumber\\
&& {{W}_{0-a}}({{R}_{1}};\varepsilon _{m}^{-};\vec{Q})=\frac{1}{4}({{W}_{e-e}}({{R}_{1}};\vec{Q})-{{W}_{h-h}}(\varepsilon _{m}^{-};\vec{Q}))=\nonumber \\
&& =\frac{W(\vec{Q})}{4}[{{(|{{a}_{0}}{{|}^{2}}{{A}_{0,0}}(\vec{Q})+|{{b}_{1}}{{|}^{2}}{{A}_{1,1}}(\vec{Q}))}^{2}}- \nonumber\\
&& -{{(|d_{m-3}^{-}{{|}^{2}}{{A}_{m-3,m-3}}(\vec{Q})+|c_{m}^{-}{{|}^{2}}{{A}_{m,m}}(\vec{Q}))}^{2}}].\nonumber
\end{eqnarray}
The energy levels of Landau quantization discussed above can be denoted as $(e,{{R}_{i}})$ for electrons and as $(h,{{R}_{j}})$ for the heavy-holes. The lowest Landau levels for electrons are the state $(e,{{R}_{1}})=\left| \psi _{0}^{-} \right\rangle $ described by the formulas (6)-(9) and the state $(e,{{R}_{2}})=\left| \psi _{0}^{{}} \right\rangle $ determined by the formula (10). For the heavy-holes the lowest Landau levels in the range of the magnetic field strength $B\ge 10$T are the states $\left| \psi _{m}^{-} \right\rangle $ with $m=3,4$ represented by the formulas (17)-(19) and the state $\left| \psi _{0}^{{}} \right\rangle $ described by the formula (21). Just this e-h levels determined the lowest exciton branches, what can be observed looking at the figures 3, 4. This range of the magnetic field strength B is of special interest for the magnetoexciton physics because just in this range the cyclotron energies for the electrons and holes are greater than the binding energy of the 2D Wannier-Mott excitons in GaAs-type crystals. The magnetoexcitons can exist only in these conditions. The range of smaller magnetic field B is also of interest for the band-to-band optical quantum transitions. The renormalized energy band gap ${{E}_{cv}}$ is greater than the value $E_{g}^{0}$ in the absence of the external magnetic field and equals to
\begin{eqnarray}
{{E}_{cv}}({{F}_{n}})=E_{g}^{0}+{{E}_{e}}({{R}_{i}})+{{E}_{h}}({{R}_{j}})
\end{eqnarray}
where ${{F}_{n}}$ is the combination of the states $(e,{{R}_{i}})$ and $(h,{{R}_{j}})$. We will consider six such variants ${{F}_{n}}$ with $n=1,2,...6$ appearing from the combinations of two electron states $(e,{{R}_{i}})$ with $i=1,2$ and of three hole states $(h,{{R}_{j}})$ with $j=1,2,3$, which will be enumerated below. The creation of the magnetoexciton with the wave vector $\vec{k}=0$, as a result of the optical quantum transition in the point $\vec{k}=0$ of the Brillouine zone, requires a photon with energy smaller than the renormalized band gap ${{E}_{cv}}(e,{{R}_{i}};h,{{R}_{j}})$ by the value of the ionization potential ${{I}_{ex}}(e,{{R}_{i}};h,{{R}_{j}})$:
\begin{eqnarray}
&&{{E}_{mex}}(e,{{R}_{i}};h,{{R}_{j}})= \\
&&={{E}_{cv}}(e,{{R}_{i}};h,{{R}_{j}})-{{I}_{ex}}(e,{{R}_{i}};h,{{R}_{j}})\nonumber
\end{eqnarray}
In the case $(e,{{R}_{i}})$ and $(h,\varepsilon _{m}^{-})$ the corresponding ionization potentials equal to:
\begin{eqnarray}
&& {{I}_{ex}}(e,{{R}_{1}};h,\varepsilon _{m}^{-})=\sum\limits_{{\vec{Q}}}{{{W}_{e-h}}({{R}_{1}};\varepsilon _{m}^{-};\vec{Q})}= \nonumber\\
&& ={{\left| a_{0}^{-} \right|}^{2}}{{\left| d_{m-3}^{-} \right|}^{2}}I_{ex}^{(0,m-3)}+{{\left| a_{0}^{-} \right|}^{2}}{{\left| c_{m}^{-} \right|}^{2}}I_{ex}^{0,m}+ \nonumber\\
&& +{{\left| b_{1}^{-} \right|}^{2}}{{\left| d_{m-3}^{-} \right|}^{2}}I_{ex}^{(1,m-3)}+{{\left| b_{1}^{-} \right|}^{2}}{{\left| c_{m}^{-} \right|}^{2}}I_{ex}^{(1,m)}, \nonumber\\
&& {{I}_{ex}}(e,{{R}_{2}};h,\varepsilon _{m}^{-})=\sum\limits_{{\vec{Q}}}{{{W}_{e-h}}({{R}_{2}};\varepsilon _{m}^{-};\vec{Q})}= \nonumber\\
&& ={{\left| d_{m-3}^{-} \right|}^{2}}I_{ex}^{(0,m-3)}+{{\left| c_{m}^{-} \right|}^{2}}I_{ex}^{(0,m)}
\end{eqnarray}
Here we introduced the partial ionization potentials
\begin{eqnarray}
I_{ex}^{(p,q)}=\sum\limits_{{\vec{Q}}}{{{W}_{e-h}}(\vec{Q})}{{A}_{p,p}}(\vec{Q}){{A}_{q,q}}(\vec{Q})
\end{eqnarray}
The ionization potentials in the combinations $(e,{{R}_{i}};h,{{\varepsilon }_{0}})$ are:
\begin{eqnarray}
&& {{I}_{ex}}(e,{{R}_{1}};h,{{\varepsilon }_{0}})={{\left| a_{0}^{-} \right|}^{2}}I_{ex}^{(0,0)}+{{\left| b_{1}^{-} \right|}^{2}}I_{ex}^{(1,0)}, \nonumber\\
&& {{I}_{ex}}(e,{{R}_{2}};h,{{\varepsilon }_{0}})=I_{ex}^{(0,0)}
\end{eqnarray}
The energies of the band-to-band optical quantum transitions accounted from the semiconductor energy band gap $E_{g}^{0}$ in the absence of the external magnetic field, as well as the creation energies of the magnetoexcitons in the point $\vec{k}=0$ in the same reference frame are illustrated in the figures 5-7 for six possible combinations of the electrons and of the holes LLLs.

\begin{figure}%[h]
\resizebox{0.48\textwidth}{!}{%
  \includegraphics{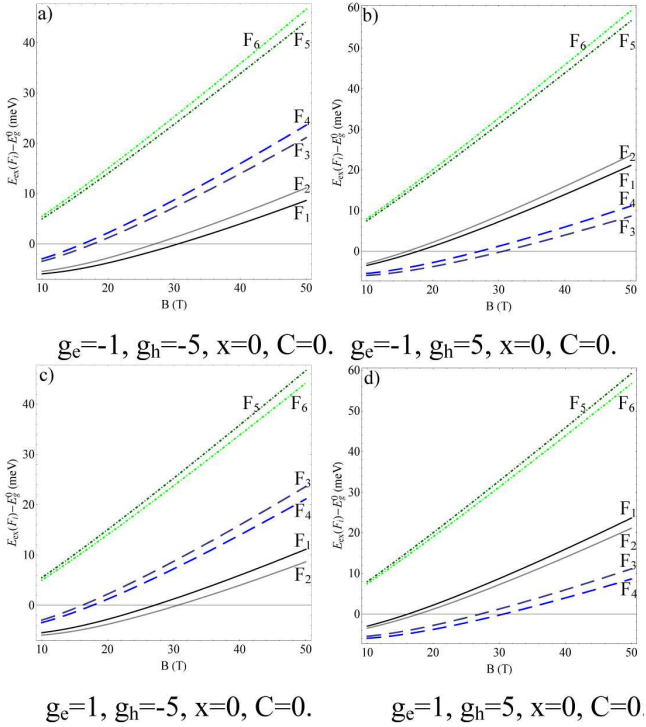}
}
\caption{The magnetoexciton energy levels ${{E}_{ex}}\left( {{F}_{i}} \right)-E_{g}^{0}$ for six combinations ${{F}_{i}}$ of three heavy-hole and of two electron lowest Landau levels in dependence on the magnetic field strength B at four combinations of the heavy-hole and electron g-factors ${{g}_{e}}$ and ${{g}_{h}}$ as follows: a) ${{g}_{e}}=-1$, ${{g}_{h}}=-5$; b) ${{g}_{e}}=-1$, ${{g}_{h}}=5$; c) ${{g}_{e}}=1$, ${{g}_{h}}=-5$; d) ${{g}_{e}}=1$, ${{g}_{h}}=5$. The influence of the external electric field is neglected taking the parameters $x=0$ and $C=0$.}
\end{figure}

\begin{figure}%[h]
\resizebox{0.48\textwidth}{!}{%
  \includegraphics{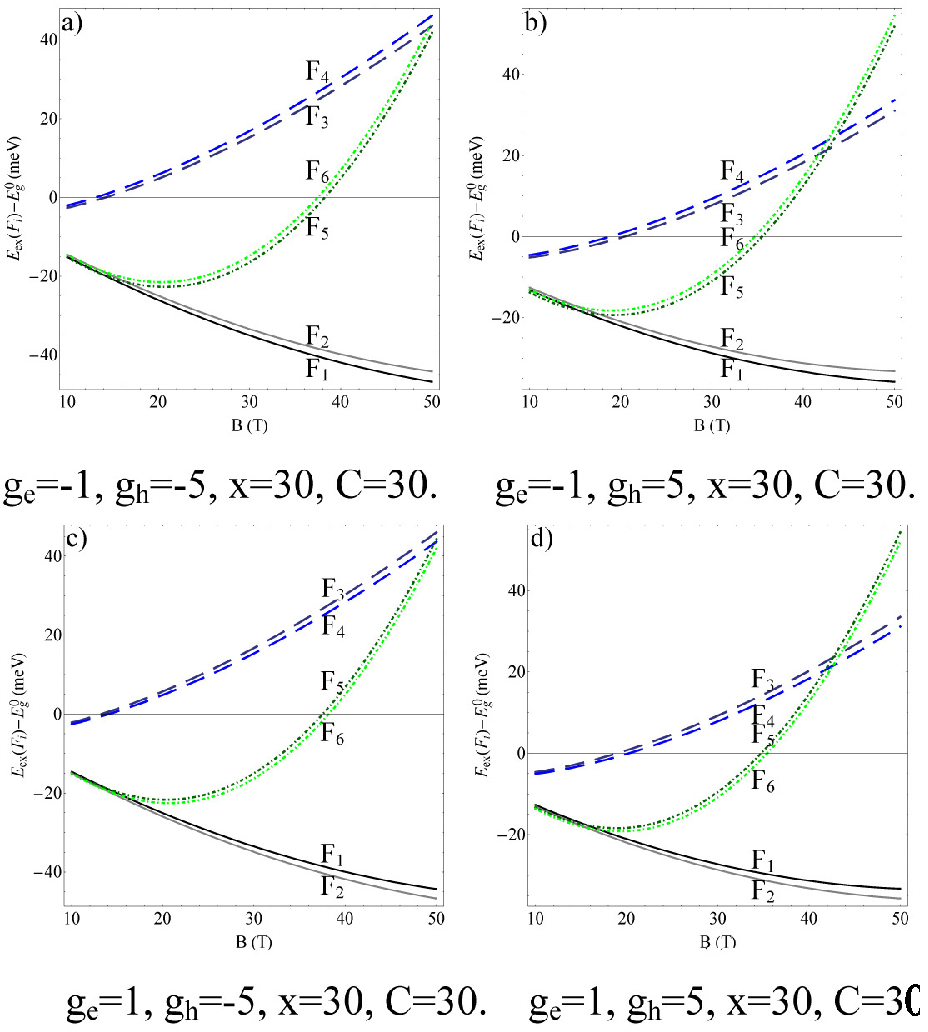}
}
\caption{The magnetoexciton energy levels ${{E}_{ex}}\left( {{F}_{i}} \right)-E_{g}^{0}$ for six combinations ${{F}_{i}}$ of three heavy-hole and of two electron lowest Landau levels in dependence on the magnetic field strength B at four combinations of the heavy-hole and electron g-factors ${{g}_{e}}$ and ${{g}_{h}}$ as follows: a) ${{g}_{e}}=-1$, ${{g}_{h}}=-5$; b) ${{g}_{e}}=-1$, ${{g}_{h}}=5$; c) ${{g}_{e}}=1$, ${{g}_{h}}=-5$; d) ${{g}_{e}}=1$, ${{g}_{h}}=5$. The influence of the external electric field is taken into account using the parameters $x=30$ and $C=30$.}
\end{figure}

\begin{figure}%[h]
\resizebox{0.48\textwidth}{!}{%
  \includegraphics{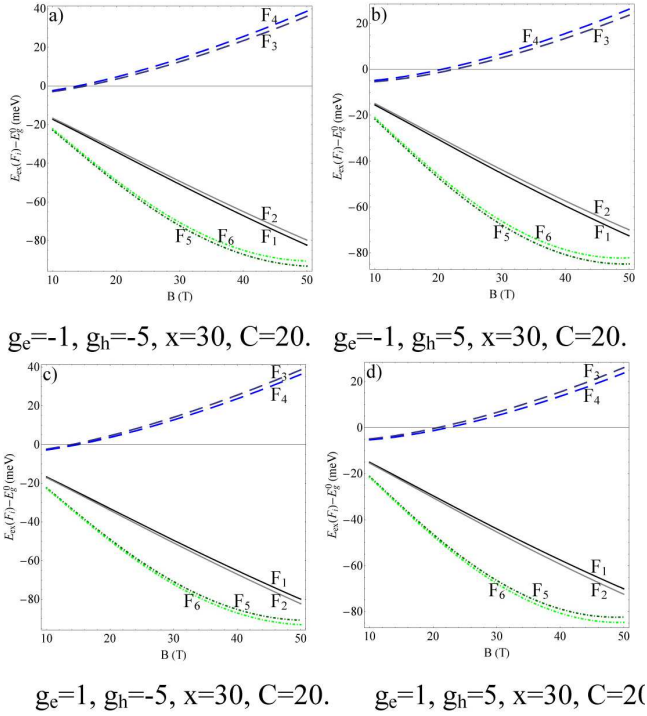}
}
\caption{The magnetoexciton energy levels ${{E}_{ex}}\left( {{F}_{i}} \right)-E_{g}^{0}$ for six combinations ${{F}_{i}}$ of three heavy-hole and of two electron lowest Landau levels in dependence on the magnetic field strength B at four combinations of the heavy-hole and electron g-factors ${{g}_{e}}$ and ${{g}_{h}}$ as follows: a) ${{g}_{e}}=-1$, ${{g}_{h}}=-5$; b) ${{g}_{e}}=-1$, ${{g}_{h}}=5$; c) ${{g}_{e}}=1$, ${{g}_{h}}=-5$; d) ${{g}_{e}}=1$, ${{g}_{h}}=5$. The influence of the external electric field is taken into account using the parameters $x=30$ and $C=20$.}
\end{figure}
The dependences ${{E}_{cv}}({{F}_{n}})-E_{g}^{0}$ shown in the figures 3, 4 and the dependences ${{E}_{ex}}({{F}_{n}},0)-E_{g}^{0}$ represented in the figures 5-7 differ by the values of the magnetoexciton ionization potentials ${{I}_{ex}}({{F}_{n}},0)$ calculated in the point $\vec{k}=0$, where the direct optical quantum transitions take place. For simplicity they were denoted as ${{I}_{ex}}({{F}_{n}})$. The fig. 5 describes the action of a strong perpendicular magnetic field giving rise to the Landau quantization and to the Zeeman splitting of the 2D electrons and holes and to the change of the Coulomb electron-hole interaction. The additional action of a perpendicular electric field giving rise to the Rashba spin-orbit coupling as well as to the nonparabolicity of the heavy-hole dispersion law side by side with the LQ and with the ZS induced by the magnetic field is represented in the figures 6 and 7. They are characterized by the parameter x and C. One can remember that the introduction of the nonparabolicity of the dispersion law is needed to compensate the action of the third order chirality terms in the Hamiltonian (13), otherwise the usual picture of the semiconductor energy band gap will be destroyed due to the unlimited penetration of the heavy-hole energy levels inside the band gap. Comparing the figures 6 and 7 one can observe that the decreasing of the nonparabolicity parameter C leads to strong penetration of the energy levels in the range of negative values.

In general case $\vec{k}\ne 0$ these combinations are:
\begin{eqnarray}
&& {{E}_{ex}}({{F}_{n}},{{{\vec{k}}}_{||}})={{E}_{cv}}({{F}_{n}})-{{I}_{ex}}({{F}_{n}},{{{\vec{k}}}_{||}}), \nonumber\\
&& {{I}_{ex}}({{F}_{n}};0)={{I}_{ex}}({{F}_{n}}), \nonumber\\
&& {{E}_{cv}}({{F}_{n}})-E_{g}^{0}={{E}_{e}}({{R}_{i}})+{{E}_{h}}({{R}_{j}}),\nonumber \\
&& n=1,...6; i=1,2; j=1,2,3
\end{eqnarray}
We present the obtained explicit expression for the ionization potentials of magnetoexcitons at arbitrary values of the wave vectors ${{\vec{k}}_{\parallel }}$, which will be used in the next section.
\begin{eqnarray}
&& {{F}_{1}}=(e,{{R}_{1}};h,\varepsilon _{3}^{-}), {{E}_{cv}}({{F}_{1}})-E_{g}^{0}={{E}_{e}}({{R}_{1}})+{{E}_{h}}(\varepsilon _{3}^{-}), \nonumber\\
&& {{I}_{ex}}({{F}_{1}};{{{\vec{k}}}_{\parallel }})={{I}_{ex}}(e,{{R}_{1}};h,\varepsilon _{3}^{-};{{{\vec{k}}}_{\parallel }})= \nonumber\\
&& ={{\left| a_{0}^{-} \right|}^{2}}{{\left| d_{0}^{-} \right|}^{2}}I_{ex}^{(0,0)}({{{\vec{k}}}_{\parallel }})+{{\left| a_{0}^{-} \right|}^{2}}{{\left| c_{3}^{-} \right|}^{2}}I_{ex}^{(0,3)}({{{\vec{k}}}_{\parallel }})+ \nonumber\\
&& +{{\left| b_{1}^{-} \right|}^{2}}{{\left| d_{0}^{-} \right|}^{2}}I_{ex}^{(1,0)}({{{\vec{k}}}_{\parallel }})+{{\left| b_{1}^{-} \right|}^{2}}{{\left| c_{3}^{-} \right|}^{2}}I_{ex}^{(1,3)}({{{\vec{k}}}_{\parallel }}), \nonumber\\
&& {{F}_{2}}=(e,{{R}_{2}};h,\varepsilon _{3}^{-}), {{E}_{cv}}({{F}_{2}})-E_{g}^{0}={{E}_{e}}({{R}_{2}})+{{E}_{h}}(\varepsilon _{3}^{-}), \nonumber\\
&& {{I}_{ex}}({{F}_{2}};{{{\vec{k}}}_{\parallel }})={{I}_{ex}}(e,{{R}_{2}};h,\varepsilon _{3}^{-};{{{\vec{k}}}_{\parallel }})= \nonumber\\
&& ={{\left| d_{0}^{-} \right|}^{2}}I_{ex}^{(0,0)}({{{\vec{k}}}_{\parallel }})+{{\left| c_{3}^{-} \right|}^{2}}I_{ex}^{(0,3)}({{{\vec{k}}}_{\parallel }}), \nonumber\\
&& {{F}_{3}}=(e,{{R}_{1}};h,\varepsilon _{0}^{{}}), {{E}_{cv}}({{F}_{3}})-E_{g}^{0}={{E}_{e}}({{R}_{1}})+{{E}_{h}}(\varepsilon _{0}^{{}}), \nonumber\\
&& {{I}_{ex}}({{F}_{3}};{{{\vec{k}}}_{\parallel }})={{I}_{ex}}(e,{{R}_{1}};h,\varepsilon _{0}^{{}};{{{\vec{k}}}_{\parallel }})= \nonumber\\
&& ={{\left| a_{0}^{-} \right|}^{2}}I_{ex}^{(0,0)}({{{\vec{k}}}_{\parallel }})+{{\left| b_{1}^{-} \right|}^{2}}I_{ex}^{(1,0)}({{{\vec{k}}}_{\parallel }}), \nonumber\\
&& {{F}_{4}}=(e,{{R}_{2}};h,\varepsilon _{0}^{{}}), {{E}_{cv}}({{F}_{4}})-E_{g}^{0}={{E}_{e}}({{R}_{2}})+{{E}_{h}}(\varepsilon _{0}^{{}}), \nonumber\\
&& {{I}_{ex}}({{F}_{4}};{{{\vec{k}}}_{\parallel }})={{I}_{ex}}(e,{{R}_{2}};h,\varepsilon _{0}^{{}};{{{\vec{k}}}_{\parallel }})=I_{ex}^{(0,0)}({{{\vec{k}}}_{\parallel }}), \\
&& {{F}_{5}}=(e,{{R}_{1}};h,\varepsilon _{4}^{-}), {{E}_{cv}}({{F}_{5}})-E_{g}^{0}={{E}_{e}}({{R}_{1}})+{{E}_{h}}(\varepsilon _{4}^{-}), \nonumber\\
&& {{I}_{ex}}({{F}_{5}};{{{\vec{k}}}_{\parallel }})={{I}_{ex}}(e,{{R}_{1}};h,\varepsilon _{4}^{-};{{{\vec{k}}}_{\parallel }})= \nonumber\\
&& ={{\left| a_{0}^{-} \right|}^{2}}{{\left| d_{1}^{-} \right|}^{2}}I_{ex}^{(0,1)}({{{\vec{k}}}_{\parallel }})+{{\left| a_{0}^{-} \right|}^{2}}{{\left| c_{4}^{-} \right|}^{2}}I_{ex}^{(0,4)}({{{\vec{k}}}_{\parallel }})+ \nonumber\\
&& +{{\left| b_{1}^{-} \right|}^{2}}{{\left| d_{1}^{-} \right|}^{2}}I_{ex}^{(1,1)}({{{\vec{k}}}_{\parallel }})+{{\left| b_{1}^{-} \right|}^{2}}{{\left| c_{4}^{-} \right|}^{2}}I_{ex}^{(1,4)}({{{\vec{k}}}_{\parallel }}), \nonumber\\
&& {{F}_{6}}=(e,{{R}_{2}};h,\varepsilon _{4}^{-}), {{E}_{cv}}({{F}_{6}})-E_{g}^{0}={{E}_{e}}({{R}_{2}})+{{E}_{h}}(\varepsilon _{4}^{-}), \nonumber\\
&& {{I}_{ex}}({{F}_{6}};{{{\vec{k}}}_{\parallel }})={{I}_{ex}}(e,{{R}_{2}};h,\varepsilon _{4}^{-};{{{\vec{k}}}_{\parallel }})= \nonumber\\
&& ={{\left| d_{1}^{-} \right|}^{2}}I_{ex}^{(0,1)}({{{\vec{k}}}_{\parallel }})+{{\left| c_{4}^{-} \right|}^{2}}I_{ex}^{(0,4)}({{{\vec{k}}}_{\parallel }}).\nonumber
\end{eqnarray}
These expressions contains a partial ionization potentials $I_{ex}^{(n,m)}\left( {{{\vec{k}}}_{\parallel }} \right)$, which are listed below together with their series expansions on the small variable $\vec{k}_{\parallel }^{2}l_{0}^{2}<1$
\begin{eqnarray}
&& I_{ex}^{(0,0)}({{{\vec{k}}}_{||}})={{I}_{l}}{{e}^{-\frac{\vec{k}_{\parallel }^{2}l_{0}^{2}}{2}}}{}_{1}{{F}_{1}}\left( \frac{1}{2},1,\frac{\vec{k}_{\parallel }^{2}l_{0}^{2}}{2} \right)\approx {{I}_{l}}\left( 1-\frac{\vec{k}_{\parallel }^{2}l_{0}^{2}}{4} \right), \nonumber\\
&& I_{ex}^{\left( 1,0 \right)}({{{\vec{k}}}_{||}})={{I}_{l}}{{e}^{-\frac{\vec{k}_{\parallel }^{2}l_{0}^{2}}{2}}}[{}_{1}{{F}_{1}}\left( \frac{1}{2},1,\frac{\vec{k}_{\parallel }^{2}l_{0}^{2}}{2} \right)- \nonumber\\
&& -\frac{1}{2}{}_{1}{{F}_{1}}\left( -\frac{1}{2},1,\frac{\vec{k}_{\parallel }^{2}l_{0}^{2}}{2} \right)]\approx {{I}_{l}}\left( \frac{1}{2}+\frac{1}{8}\vec{k}_{\parallel }^{2}l_{0}^{2} \right), \nonumber\\
&& I_{ex}^{\left( 0,3 \right)}({{{\vec{k}}}_{||}})={{I}_{l}}{{e}^{-\frac{\vec{k}_{\parallel }^{2}l_{0}^{2}}{2}}}[{}_{1}{{F}_{1}}\left( \frac{1}{2},1,\frac{\vec{k}_{\parallel }^{2}l_{0}^{2}}{2} \right)- \nonumber\\
&& -\frac{3}{2}{}_{1}{{F}_{1}}\left( -\frac{1}{2},1,\frac{\vec{k}_{\parallel }^{2}l_{0}^{2}}{2} \right)+\frac{9}{8}{}_{1}{{F}_{1}}\left( -\frac{3}{2},1,\frac{\vec{k}_{\parallel }^{2}l_{0}^{2}}{2} \right)- \nonumber\\
&& -\frac{5}{16}{}_{1}{{F}_{1}}\left( -\frac{5}{2},1,\frac{\vec{k}_{\parallel }^{2}l_{0}^{2}}{2} \right)]\approx {{I}_{l}}\left( \frac{5}{16}+\frac{1}{64}\vec{k}_{\parallel }^{2}l_{0}^{2} \right), \nonumber\\
&& I_{ex}^{\left( 1,3 \right)}({{{\vec{k}}}_{||}})={{I}_{l}}{{e}^{-\frac{\vec{k}_{\parallel }^{2}l_{0}^{2}}{2}}}[{}_{1}{{F}_{1}}\left( \frac{1}{2},1,\frac{\vec{k}_{\parallel }^{{}}l_{0}^{2}}{2} \right)- \nonumber\\
&& -2{}_{1}{{F}_{1}}\left( -\frac{1}{2},1,\frac{\vec{k}_{\parallel }^{{}}l_{0}^{2}}{2} \right)+\frac{27}{8}{}_{1}{{F}_{1}}\left( -\frac{3}{2},1,\frac{\vec{k}_{\parallel }^{2}l_{0}^{2}}{2} \right)- \nonumber\\
&& -\frac{25}{8}{}_{1}{{F}_{1}}\left( -\frac{5}{2},1,\frac{\vec{k}_{\parallel }^{2}l_{0}^{2}}{2} \right)+\frac{35}{32}{}_{1}{{F}_{1}}\left( -\frac{7}{2},1,\frac{\vec{k}_{\parallel }^{2}l_{0}^{2}}{2} \right)]\approx  \\
&& \approx {{I}_{l}}\left( \frac{11}{32}+\frac{5}{128}\vec{k}_{\parallel }^{2}l_{0}^{2} \right).\nonumber
\end{eqnarray}
These series expansions, including the terms proportional to $\vec{k}_{\parallel }^{2}l_{0}^{2}<1$, make it possible to determine the magnetic mass $M({{F}_{1}},B)$ of the magnetoexciton in the state ${{F}_{1}}$ as follows
\begin{eqnarray}
&& {{I}_{ex}}\left( {{F}_{1}},B,{{{\vec{k}}}_{\parallel }} \right)= \nonumber\\
&& ={{I}_{l}}A\left( {{F}_{1}},B \right)-{{I}_{l}}\frac{\vec{k}_{\parallel }^{2}l_{0}^{2}}{2}G\left( {{F}_{1}},B \right)= \\
&& ={{I}_{ex}}\left( {{F}_{1}},B,0 \right)-\frac{{{\hbar }^{2}}\vec{k}_{\parallel }^{2}}{2M({{F}_{1}},B)}.\nonumber
\end{eqnarray}
Here the denotations $A\left( {{F}_{1}},B \right)$ and $G\left( {{F}_{1}},B \right)$ were introduced
\begin{eqnarray}
&& {{I}_{ex}}\left( {{F}_{1}},B,0 \right)={{I}_{l}}A\left( {{F}_{1}},B \right), \nonumber\\
&& M\left( {{F}_{1}},B \right)=\frac{{{\hbar }^{2}}}{{{I}_{l}}l_{0}^{2}G\left( {{F}_{1}},B \right)}, \nonumber\\
&& A\left( {{F}_{1}},B \right)= \nonumber\\
&& ={{\left| d_{0}^{-} \right|}^{2}}({{\left| a_{0}^{-} \right|}^{2}}+\frac{1}{2}{{\left| b_{1}^{-} \right|}^{2}})+\frac{1}{16}\left| c_{3}^{-} \right|(5{{\left| a_{0}^{-} \right|}^{2}}+\frac{11}{2}{{\left| b_{1}^{-} \right|}^{2}}), \nonumber\\
&& G\left( {{F}_{1}},B \right)= \\
&& =\frac{1}{2}{{\left| a_{0}^{-} \right|}^{2}}({{\left| d_{0}^{-} \right|}^{2}}-\frac{1}{16}{{\left| c_{3}^{-} \right|}^{2}})-\frac{1}{4}{{\left| b_{1}^{-} \right|}^{2}}({{\left| d_{0}^{-} \right|}^{2}}+\frac{5}{16}{{\left| c_{3}^{-} \right|}^{2}}). \nonumber
\end{eqnarray}
In the case of the dipole-active state ${{F}_{4}}$ the ionization potential equals to $I_{ex}^{\left( 0,0 \right)}\left( {{{\vec{k}}}_{\parallel }} \right)$ and the magnetic mass $M\left( {{F}_{4}},B \right)$ was determined in Ref. [9]. Both masses depend on the magnetic field strength
\begin{eqnarray}
&& M\left( {{F}_{4}},B \right)=\frac{2{{\hbar }^{2}}}{{{I}_{l}}l_{0}^{2}}, \nonumber\\
&& M\left( {{F}_{1}},B \right)=\frac{\sqrt{2}{{\hbar }^{2}}{{\varepsilon }_{0}}}{\sqrt{\pi }{{e}^{2}}{{l}_{0}}G\left( {{F}_{1}},B \right)}, \nonumber\\
&& {{I}_{l}}=\frac{{{e}^{2}}}{{{\varepsilon }_{0}}{{l}_{0}}}\sqrt{\frac{\pi }{2}}.
\end{eqnarray}
Their expressions normalized to the free electron mass ${{m}_{0}}$, $M\left( {{F}_{4}},B \right)/{{m}_{0}}$, $M\left( {{F}_{1}},B \right)/{{m}_{0}}$ are presented in the fig. 8.
\begin{figure}%[h]
\resizebox{0.48\textwidth}{!}{%
  \includegraphics{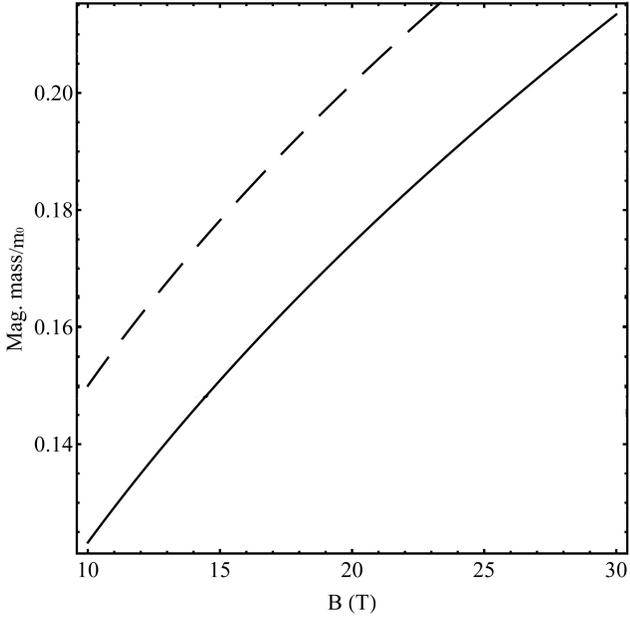}
}
\caption{The dependences on the magnetic field strength of the magnetic masses in two dipole-active states ${{F}_{1}}$(dashed line) and ${{F}_{4}}$(solid line).}
\end{figure}

In the next section we will take into account also the interaction of 2D magnetoexcitons with the photons confined in microcavity giving rise to the formation of the polaritons.
\section{Interaction of the magnetoexcitons with the electromagnetic radiation and the formation of the cavity polaritons}
In the Refs.[16, 22, 25, 28] the Hamiltonians describing in different approximations the electron-radiation interaction in the system of two-dimensional coplanar electrons and holes accumulated in the semiconductor quantum well and subjected to the action of a strong perpendicular magnetic field giving rise to the Landau quantization of their energy levels were obtained. However, only the case of the inter-band optical quantum transitions with the creation or annihilation of one electron-hole pair was considered in Refs.[16, 22, 25, 28]. The intra-band quantum transitions were discussed in Ref.[20]. In Ref.[22] the exciton-cyclotron resonance and the optical orientation phenomena [31] arising under the influence of the circularly polarized laser radiation were studied but without taking into account the RSOC. The Hamiltonian derived in Ref.[22] in the e-h representation was transcribed in Ref.[16] with the purpose to described the magnetoexciton-photon interaction for the light propagating arbitrarily in the three-dimensional (3D) space as well as being confined in the microcavity. The dispersion law of the magnetoexciton-polariton in microcavity was deduced. The dependence of the Rabi frequency on the magnetic field strength as well as the selection rule for the numbers of the LQ levels of the e-h pair engaged in the dipole-active and in the quadrupole-active transitions were determined [16]. The influence of the RSOC on the optical properties of the magnetoexcitons and on the band-to-band quantum transitions arising due to the action of a supplementary electric field perpendicular to the plane of the QW was investigated in the Refs.[23, 24, 25, 28]. The third order chirality terms in the Hamiltonian (1) of the heavy-hole induced by the electric field obliged one to introduce an additional nonparabolicity (NP) term of the same origin in the heavy-hole dispersion law. The NP term together with the chirality terms change essentially the dependence of the energy levels on the magnetic field strength, as one can see looking at the figures 2, 4, 6, 7. The role of the NP term is to prevent the unlimited deep penetration of the energy levels of the 2D heavy-hole, as well as of the e-h pair and of the magnetoexciton inside the energy band gap and furthers to the stability of the semiconductor band structure. In the mentioned Refs. [23, 24, 25, 28] the effects related with the Zeeman splitting were not discussed. This lack is removed below. We will discussed the properties of the magnetoexcitons and of the magnetoexciton-polaritons taking into account the influence of a full set of four factors such as LQ, RSOC, ZS and NP. The Hamiltonian describing the magnetoexciton-photon interaction including the ZS effects looks exactly the same as in Ref. [25] with only one difference, that the coefficients $a_{0}^{-}$, $b_{1}^{-}$, $c_{m}^{-}$ and $d_{m-3}^{-}$ with $m\ge 3$ must be determined by the expressions (8) and (17) containing the Zeeman coefficients ${{Z}_{e}}$ and ${{Z}_{h}}$ different from zero. In the previous papers [23, 24, 25, 28] these coefficients were absent.

Using formula (42) of Ref.[25], the Hamiltonian of the magnetoexciton-photon interaction can be written as
\begin{eqnarray}
&& {{{\hat{H}}}_{e-rad}}=(-\frac{e}{{{m}_{0}}{{l}_{0}}})\sum\limits_{\vec{k}({{{\vec{k}}}_{||}},{{k}_{z}})}{\sum\limits_{{{M}_{h}}=\pm 1}{\sum\limits_{i=1,2}^{{}}{\sum\limits_{\varepsilon ={{\varepsilon }_{m}},\varepsilon _{m}^{-}}{\sqrt{\frac{\hbar }{{{L}_{z}}{{\omega }_{{\vec{k}}}}}}}}}}\times  \nonumber\\
&& \times \{{{P}_{cv}}(0)T({{R}_{i}},\varepsilon ,{{{\vec{k}}}_{||}})[{{C}_{\vec{k},-}}(\vec{\sigma }_{{\vec{k}}}^{+}\cdot \vec{\sigma }_{{{M}_{h}}}^{*})+ \nonumber\\
&& +{{C}_{\vec{k},+}}(\vec{\sigma }_{{\vec{k}}}^{-}\cdot \vec{\sigma }_{{{M}_{h}}}^{*})]\hat{\Psi }_{ex}^{\dagger }({{{\vec{k}}}_{||}},{{M}_{h}},{{R}_{i}},\varepsilon )+ \nonumber\\
&& +{{P}^{*}}_{cv}(0){{T}^{*}}({{R}_{i}},\varepsilon ,{{{\vec{k}}}_{||}})[{{({{C}_{\vec{k},-}})}^{\dagger }}(\vec{\sigma }_{{\vec{k}}}^{-}\cdot \vec{\sigma }_{{{M}_{h}}}^{{}})+ \nonumber\\
&& +{{({{C}_{\vec{k},+}})}^{\dagger }}(\vec{\sigma }_{{\vec{k}}}^{+}\cdot \vec{\sigma }_{{{M}_{h}}}^{{}})]\hat{\Psi }_{ex}^{{}}({{{\vec{k}}}_{||}},{{M}_{h}},{{R}_{i}},\varepsilon )+ \\
&& +{{P}_{cv}}(0)T({{R}_{i}},\varepsilon ,-{{{\vec{k}}}_{||}})[{{({{C}_{\vec{k},-}})}^{\dagger }}(\vec{\sigma }_{{\vec{k}}}^{-}\cdot \vec{\sigma }_{{{M}_{h}}}^{*})+ \nonumber\\
&& +{{({{C}_{\vec{k},+}})}^{\dagger }}(\vec{\sigma }_{{\vec{k}}}^{+}\cdot \vec{\sigma }_{{{M}_{h}}}^{*})]\hat{\Psi }_{ex}^{\dagger }(-{{{\vec{k}}}_{||}},{{M}_{h}},{{R}_{i}},\varepsilon )+ \nonumber\\
&& +{{P}^{*}}_{cv}(0){{T}^{*}}({{R}_{i}},\varepsilon ,-{{{\vec{k}}}_{||}})[{{C}_{\vec{k},-}}(\vec{\sigma }_{{\vec{k}}}^{+}\cdot \vec{\sigma }_{{{M}_{h}}}^{{}})+ \nonumber\\
&& +{{C}_{\vec{k},+}}(\vec{\sigma }_{{\vec{k}}}^{-}\cdot \vec{\sigma }_{{{M}_{h}}}^{{}})]\hat{\Psi }_{ex}^{{}}(-{{{\vec{k}}}_{||}},{{M}_{h}},{{R}_{i}},\varepsilon )\}\nonumber
\end{eqnarray}
It differs from the Hamiltonian (9) in Ref. [16] by the more complicate coefficients $T\left( {{R}_{i}},\varepsilon ,{{{\vec{k}}}_{||}} \right)$, which contain the generalized coefficients $a_{0}^{-}$, $b_{1}^{-}$, $c_{m}^{-}$ and $d_{m-3}^{-}$ given by the expressions (8) and (18).
The Hamiltonian (49) contains the creation and annihilation operators of the magnetoexcitons $\hat{\Psi }_{ex}^{\dagger }({{\vec{k}}_{||}},{{M}_{h}},{{R}_{i}},\varepsilon )$, $\hat{\Psi }_{ex}^{{}}({{\vec{k}}_{||}},{{M}_{h}},{{R}_{i}},\varepsilon )$ and of the photons $C_{\vec{k},\xi }^{\dagger }$, $C_{\vec{k},\xi }^{{}}$. First of them were determined by the formula (32) and are characterized by the in-plane wave vectors ${{\vec{k}}_{||}}$, by the orbital projection ${{M}_{h}}$ of the hole state in the frame of the p-type valence band, and by the quantum states ${{R}_{i}}$ and $\varepsilon $ of the electron and hole in the conditions of the LQ accompanied by the RSOC, ZS and NP. Instead of the quantum number ${{M}_{h}}$ we will use the circular polarization vector $\vec{\sigma }_{{{M}_{h}}}^{{}}$. The photon operators depend on the arbitrary oriented in the 3D space wave vectors $\vec{k}={{\vec{a}}_{3}}{{k}_{z}}+{{\vec{k}}_{||}}$, where ${{\vec{a}}_{3}}$ is the unit vector perpendicular to the layer and on the polarization label $\xi $, which takes two values 1 and 2 in the case of the light with linear polarizations ${{\vec{e}}_{\vec{k},i}}$ or the signs $\pm $ in the case of circular polarizations $\vec{\sigma }_{{\vec{k}}}^{\pm }$. Below we will use the following denotations:
\begin{eqnarray}
&& {{C}_{\vec{k},\pm }}=\frac{1}{\sqrt{2}}({{C}_{\vec{k},1}}\pm i{{C}_{\vec{k},2}}),{{({{C}_{\vec{k},\pm }})}^{\dagger }}=\frac{1}{\sqrt{2}}(C_{\vec{k},1}^{\dagger }\mp iC_{\vec{k},2}^{\dagger }), \nonumber\\
&& \vec{\sigma }_{{\vec{k}}}^{\pm }=\frac{1}{\sqrt{2}}({{{\vec{e}}}_{\vec{k},1}}\pm i{{{\vec{e}}}_{\vec{k},2}}),({{{\vec{e}}}_{\vec{k},1}}\cdot \vec{k})=0,i=1,2, \nonumber\\
&& \sum\limits_{i=1}^{2}{{{C}_{\vec{k},i}}{{{\vec{e}}}_{\vec{k},i}}}={{C}_{\vec{k},-}}\vec{\sigma }_{{\vec{k}}}^{+}+{{C}_{\vec{k},+}}\vec{\sigma }_{{\vec{k}}}^{-},\nonumber \\
&& \sum\limits_{i=1}^{2}{{{({{C}_{\vec{k},\pm }})}^{\dagger }}{{{\vec{e}}}_{\vec{k},i}}}={{({{C}_{\vec{k},-}})}^{\dagger }}\vec{\sigma }_{{\vec{k}}}^{-}+{{({{C}_{\vec{k},+}})}^{\dagger }}\vec{\sigma }_{{\vec{k}}}^{+}, \nonumber\\
&& {{{\vec{\sigma }}}_{{{M}_{h}}}}=\frac{1}{\sqrt{2}}({{{\vec{a}}}_{1}}\pm i{{{\vec{a}}}_{2}}), {{{\vec{k}}}_{||}}={{k}_{x}}{{{\vec{a}}}_{1}}+{{k}_{y}}{{{\vec{a}}}_{2}}, \nonumber\\
&& ({{{\vec{\sigma }}}_{{{M}_{h}}}}\cdot {{{\vec{a}}}_{3}})=0, {{M}_{h}}=\pm 1,
\end{eqnarray}
Here ${{\vec{a}}_{1}}$ and ${{\vec{a}}_{2}}$ are the in-plane orthogonal unit vectors. The scalar products $\left( \vec{\sigma }_{{\vec{k}}}^{\pm }\cdot \vec{\sigma }_{{{M}_{h}}}^{*} \right)$ appearing in the Hamiltonian (52) determine the ability of the photon circular polarization $\vec{\sigma }_{{\vec{k}}}^{\pm }$ to create a circular polarization ${{\vec{\sigma }}_{{{M}_{h}}}}$ with the probability ${{\left| \left( \vec{\sigma }_{{\vec{k}}}^{\pm }\cdot \vec{\sigma }_{{{M}_{h}}}^{*} \right) \right|}^{2}}$, which can be considered as a geometrical selection rule. This probability is the same for any in-plane wave vector $\vec{k}={{\vec{k}}_{||}}$ and equals to 1/4. In the case of the incident wave vector $\vec{k}$ perpendicular to the layer $\vec{k}={{\vec{a}}_{3}}{{k}_{z}}$ the light with the circular polarization $\vec{\sigma }_{{\vec{k}}}^{\pm }$ excites the magnetoexciton with the same circular polarization ${{\vec{\sigma }}_{{{M}_{h}}}}=\vec{\sigma }_{{\vec{k}}}^{\pm }$ with the probability equal to unity. Another spin-orbital selection rules are determined by the coefficients $T\left( {{R}_{i}},\varepsilon ,{{{\vec{k}}}_{||}} \right)$, which are expressed by the formulas (36) of the Ref.[25] as follows:
\begin{eqnarray}
&& T\left( {{R}_{1}},\varepsilon _{m}^{-};{{{\vec{k}}}_{||}} \right)=a_{0}^{-*}d_{m-3}^{-*}\tilde{\phi }\left( 0,m-3;{{{\vec{k}}}_{||}} \right)- \nonumber\\
&& -b_{1}^{-*}c_{m}^{-*}\tilde{\phi }\left( 1,m;{{{\vec{k}}}_{||}} \right), m\ge 3, \nonumber\\
&& T\left( {{R}_{1}},\varepsilon _{m}^{{}};{{{\vec{k}}}_{||}} \right)=-b_{1}^{-*}\tilde{\phi }\left( 1,m;{{{\vec{k}}}_{||}} \right), m=0,1,2, \nonumber\\
&& T\left( {{R}_{2}},\varepsilon _{m}^{-};{{{\vec{k}}}_{||}} \right)=-c_{m}^{-*}\tilde{\phi }\left( 0,m;{{{\vec{k}}}_{||}} \right), m\ge 3, \nonumber\\
&& T\left( {{R}_{2}},\varepsilon _{m}^{{}};{{{\vec{k}}}_{||}} \right)=-\tilde{\phi }\left( 0,m;{{{\vec{k}}}_{||}} \right), m=0,1,2.
\end{eqnarray}
The integrals $\tilde{\phi }\left( {{n}_{e}},{{n}_{h}};{{{\vec{k}}}_{||}} \right)$ were introduces by the formulas (32) of the Ref.[25]. They have a general form and some concrete values given below
\begin{eqnarray}
&&\tilde{\phi }( {{n}_{e}},{{n}_{h}};{{{\vec{k}}}_{||}} )=\int\limits_{-\infty }^{\infty }{dy}{{\varphi }_{{{n}_{e}}}}( y-\frac{{{k}_{x}}l_{0}^{2}}{2} ){{\varphi }_{{{n}_{h}}}}( y+\frac{{{k}_{x}}l_{0}^{2}}{2}){{e}^{i{{k}_{y}}y}}, \nonumber\\
&& \tilde{\phi }( 0,0;{{{\vec{k}}}_{||}} )=1-( \frac{k_{x}^{2}}{2}+k_{y}^{2} )\frac{l_{0}^{2}}{4}, \nonumber\\
&& \tilde{\phi }( 0,1;{{{\vec{k}}}_{||}})=\frac{( {{k}_{x}}+i{{k}_{y}} ){{l}_{0}}}{\sqrt{2}}.
\end{eqnarray}
In the point ${{\vec{k}}_{||}}=0$ they coincide with the normalization and with the orthogonality conditions for the wave functions ${{\varphi }_{n}}(y)$, which have real values.

The integrals (56) play the role of the orbital selection rules for the quantum transitions from the ground state of the crystal to the magnetoexciton states as well as for the band-to-band optical transitions. Following them in the case of the dipole-active transitions with ${{\vec{k}}_{||}}=0$ the selection rules is ${{n}_{e}}={{n}_{h}}$. In the case of the quadrupole-active quantum transitions, their amplitudes are proportional to $\left( {{k}_{x}}+i{{k}_{y}} \right){{l}_{0}}$ and their probabilities are proportional to $\vec{k}_{||}^{2}l_{0}^{2}$, which is a small factor of the order ${{\left( {{l}_{0}}/\lambda  \right)}^{2}}<<1$ where $\lambda $ is the light wavelength. In this case following (56) the difference between the quantum numbers ${{n}_{e}}$ and ${{n}_{h}}$ satisfies the requirements ${{n}_{e}}={{n}_{h}}\pm 1$. The selection rules (56) allow also higher order multi-pole quantum transitions with probabilities of the order ${{\left( {{l}_{0}}/\lambda  \right)}^{2n}}$ with $n>1$, but these infinitesimal values will be neglected and the corresponding quantum transitions will be considered for simplicity forbidden. We will discuss two types of the allowed inter-band quantum transitions. For the dipole-active transitions with ${{\vec{k}}_{||}}=0$, the quantum numbers ${{n}_{e}}$ and ${{n}_{h}}$ of the electron and heavy-hole Landau quantization states coincide i.e. ${{n}_{e}}={{n}_{h}}$.
For the quadrupole-active quantum transitions the probabilities are proportional to a small factor ${{\left( \vec{k}_{||}^{{}}\cdot l_{0}^{{}} \right)}^{2}}$ whereas the quantum numbers ${{n}_{e}}$ and ${{n}_{h}}$ satisfy to the requirements ${{n}_{e}}={{n}_{h}}\pm 1$ as it follows from  the expressions (56). The quantum transitions in the magnetoexciton state ${{F}_{1}}$ composed from the electron state ${{R}_{1}}$ and from the heavy-hole state $\varepsilon _{3}^{-}$, what means ${{F}_{1}}=\left( {{R}_{1}},\varepsilon _{3}^{-} \right)$, is determined by the coefficient $T\left( {{R}_{1}},\varepsilon _{3}^{-},{{{\vec{k}}}_{||}} \right)$ written in the list (55). It has a concrete expression
\begin{eqnarray}
&& T({{F}_{1}},{{{\vec{k}}}_{||}})=T({{R}_{1}},\varepsilon _{3}^{-},{{{\vec{k}}}_{||}})= \\
&& =a_{0}^{-*}d_{0}^{-*}\tilde{\phi }(0,0;{{{\vec{k}}}_{||}})-b_{1}^{-*}c_{3}^{-*}\tilde{\phi }(1,3;{{{\vec{k}}}_{||}})\approx a_{0}^{-*}d_{0}^{-*}. \nonumber
\end{eqnarray}
The magnetoexciton state ${{F}_{1}}$ is dipole-active with the probability of the quantum transition proportional to $|a_{0}^{-}{{|}^{2}}|d_{0}^{-}{{|}^{2}}$. The state ${{F}_{1}}$ described above in the presence of the RSOC, ZS and NP term is a successor of a state with ${{M}_{h}}=-1$ discussed in the Refs.[16, 22] in the absence of the RSOC. To verify this statement one can analyze the energy spectra for electron state $\varepsilon _{0}^{-}$ (6) and for heavy-hole state $\varepsilon _{3}^{-}$ (16) directing their chirality terms parameters $\alpha$ and $\beta$ including the NP term $\delta$ to zero.
In this case we will obtain the expressions$\begin{array}{cc}
   {\varepsilon _{0}^{-}}  \\
   {\alpha \to 0}  \\
\end{array} =\frac{1}{2}+{{Z}_{e}}$
and  $\begin{array}{cc}
   {\varepsilon _{3}^{-}}  \\
   {\beta \to 0,\delta \to 0}  \\
\end{array} = \frac{1}{2}+{{Z}_{h}}$.
In this limit we can see that the electron spin projection is $S_{z}^{e}=1/2$, whereas the heavy-hole effective spin projection $\tilde{S}_{z}^{h}$ equals $-1/2$, what is equivalent to the full angular momentum projection $j_{z}^{h}=-3/2$. It should be noticed that the heavy-hole spinor-type wave function with the quantum number $j_{z}^{h}=-3/2$ can be constructed only with the orbital periodic part of the p-type valence band wave function with quantum number ${{M}_{h}}=-1$.
As a result the state ${{F}_{1}}=({{R}_{1}},\varepsilon _{3}^{-})$ in the limit of the small chirality terms is characterized by the summary angular projection $j_{z}^{e-h}=S_{z}^{e}+j_{z}^{h}$ equal to -1 and at the same time by the quantum number ${{M}_{h}}=-1$. Indeed the state ${{F}_{1}}=({{R}_{1}},\varepsilon _{3}^{-})$ is a successor of the state with ${{M}_{h}}=-1$ discussed in Refs.[9, 10] in the absence of the RSOC. This state is dipole-active. Another dipole-active state has the composition ${{F}_{4}}=({{R}_{2}},\varepsilon _{0}^{{}})$.
In the case of vanishing chirality term parameters $\alpha ,\beta $ and $\delta $ the energy spectra are $\begin{array}{cc}
   {\varepsilon _{0}^{e}}  \\
   {\alpha \to 0}  \\
\end{array} =\frac{1}{2}-{{Z}_{e}}$ and $\begin{array}{cc}
   {\varepsilon _{0}^{h}}  \\
   {\beta \to 0,\delta \to 0}  \\
\end{array} = \frac{1}{2}-{{Z}_{h}}$.
They are characterized by the electron spin projection $S_{z}^{e}=-1/2$ and by the heavy-hole quantum numbers $\tilde{S}_{z}^{h}=1/2$, $j_{z}^{h}=3/2$ and ${{M}_{h}}=1$. This state in the limit of the small chirality terms is characterized by the quantum numbers $j_{z}^{e-h}=S_{z}^{e}+j_{z}^{h}=1$ and ${{M}_{h}}=1$. it is a successor of the state with ${{M}_{h}}=1$ discussed in Refs.[16, 22] in the absence of the RSOC. Its coefficient $T\left( {{R}_{2}},{{\varepsilon }_{0}};{{{\vec{k}}}_{||}} \right)$ is:
\begin{eqnarray}
T({{F}_{4}},{{\vec{k}}_{||}})=T({{R}_{2}},{{\varepsilon }_{0}};{{\vec{k}}_{||}})=-\tilde{\phi }(0,0;{{\vec{k}}_{||}})\approx -1
\end{eqnarray}
Between six initially selected lowest magnetoexciton states two of them are quadrupole-active. They have the e-h combinations ${{F}_{3}}=\left( {{R}_{1}},{{\varepsilon }_{0}} \right)$ and ${{F}_{5}}=\left( {{R}_{1}},\varepsilon _{4}^{-} \right)$. In both cases the electron energy spectrum in the limit $\alpha \to 0$ looks as$\begin{array}{cc}
   {\varepsilon _{0}}  \\
   {\alpha \to 0}  \\
\end{array} =\frac{1}{2}+{{Z}_{e}}$ and has the quantum number $S_{z}^{e}=1/2$. The two heavy-hole states ${{\varepsilon }_{0}}$ and $\varepsilon _{4}^{-}$ in the limit $\beta \to 0$, $\delta \to 0$ have the form $\begin{array}{cc}
   {\varepsilon _{0}}  \\
   {\beta \to 0,\delta \to 0}  \\
\end{array} = \frac{1}{2}-{{Z}_{h}}$ and $\begin{array}{cc}
   {\varepsilon _{4}^{-}}  \\
   {\beta \to 0,\delta \to 0}  \\
\end{array} = \frac{3}{2}+{{Z}_{h}}$. The state ${{F}_{3}}=\left( {{R}_{1}},{{\varepsilon }_{0}} \right)$ is characterized by the quantum numbers $S_{z}^{e}=-1/2$, $\tilde{S}_{z}^{h}=1/2$, $j_{z}^{h}=3/2$, ${{M}_{h}}=1$ and $J_{z}^{e-h}=2$. Its coefficient $T\left( {{R}_{1}},{{\varepsilon }_{0}};{{{\vec{k}}}_{||}} \right)$ equals to
\begin{eqnarray}
&&T({{F}_{3}},{{\vec{k}}_{||}})=T({{R}_{1}},{{\varepsilon }_{0}};{{\vec{k}}_{||}})=\nonumber \\
&&=-b_{1}^{-*}\tilde{\phi }(1,0;{{\vec{k}}_{||}})=-b_{1}^{-*}(\frac{-{{k}_{x}}+i{{k}_{y}}}{\sqrt{2}}){{l}_{0}}
\end{eqnarray}
It is a small amplitude of the quadrupole-active transition due to essential difference between the values $J_{z}^{e-h}=2$ and ${{M}_{h}}=1$, and due to the small value of the coefficient $b_{1}^{-}$. Contrary to the state ${{F}_{3}}=\left( {{R}_{1}},{{\varepsilon }_{0}} \right)$ the state ${{F}_{5}}=\left( {{R}_{1}},\varepsilon _{4}^{-} \right)$ has the quantum numbers $S_{z}^{e}=1/2$, $\tilde{S}_{z}^{h}=-1/2$, $j_{z}^{h}=-3/2$, ${{M}_{h}}=-1$ and $J_{z}^{e-h}=-1$. In this state the coefficient $T\left( {{R}_{1}},\varepsilon _{4}^{-};{{{\vec{k}}}_{||}} \right)$ looks as:
\begin{eqnarray}
&& T({{F}_{5}},{{{\vec{k}}}_{||}})=T({{R}_{1}},\varepsilon _{4}^{-};{{{\vec{k}}}_{||}})= \nonumber\\
&& =a_{0}^{-*}d_{0}^{-*}\tilde{\phi }(0,1;{{{\vec{k}}}_{||}})-b_{1}^{-*}c_{4}^{-*}\tilde{\phi }(1,4;{{{\vec{k}}}_{||}})\approx \nonumber \\
&& \approx a_{0}^{-*}d_{0}^{-*}(\frac{{{k}_{x}}+i{{k}_{y}}}{\sqrt{2}}){{l}_{0}}.
\end{eqnarray}
In the case of a small chirality terms the coefficients $a_{0}^{-}$ and $d_{0}^{-}$ can approach unity and the amplitude of this quadrupole-transitions is larger than the previous one.
The last two states ${{F}_{2}}=\left( {{R}_{2}},\varepsilon _{3}^{-} \right)$ and ${{F}_{6}}=\left( {{R}_{2}},\varepsilon _{4}^{-} \right)$ are forbidden. Both of them have the electron state $\begin{array}{cc}
   {\varepsilon _{0}}  \\
   {\beta \to 0,\delta \to 0}  \\
\end{array} = \frac{1}{2}-{{Z}_{e}}$ and the hole states $\begin{array}{cc}
   {\varepsilon _{m}^{-}}  \\
   {\beta \to 0,\delta \to 0}  \\
\end{array} =(m-1)-\frac{3}{2}+{{Z}_{h}}$ with $m=3,4$. They are characterized by the quantum numbers $S_{z}^{e}=-1/2$, $\tilde{S}_{z}^{h}=-1/2$, $j_{z}^{h}=-3/2$, ${{M}_{h}}=-1$ and $J_{z}^{e-h}=-2$. Their coefficients $T\left( {{R}_{2}},\varepsilon _{m}^{-};{{{\vec{k}}}_{||}} \right)$ with $m=3,4$ are:
\begin{eqnarray}
&& T({{R}_{2}},\varepsilon _{m}^{-};{{{\vec{k}}}_{||}})=-c_{m}^{-*}\tilde{\phi }(0,m;{{{\vec{k}}}_{||}})\approx 0, \nonumber\\
&& m=3,4
\end{eqnarray}
The inter-band matrix element ${{P}_{cv}}({{\vec{k}}_{||}},g)$ determined by the formula (30) of Ref.[25]:
\begin{eqnarray}
&&{{P}_{cv}}({{\vec{k}}_{||}},g)=\nonumber\\
&&=\frac{1}{{{v}_{0}}}\int\limits_{{{v}_{0}}}{d\vec{\rho }}U_{c,s,g}^{*}(\vec{\rho }){{e}^{i{{k}_{y}}{{\rho }_{y}}}}\frac{\partial }{\partial {{\rho }_{i}}}{{U}_{v,p,i,g-{{k}_{x}}}}(\vec{\rho }).
\end{eqnarray}
It involves the canonical momentum operator $\hat{P}=-i\hbar \vec{\nabla }$ and the periodic parts of the electron Bloch wave functions integrated on the volume ${{v}_{0}}$ of the elementary lattice cell. The periodic part of the s-type conduction band is denoted as $U_{c,s,g}^{{}}(\vec{\rho })$, whereas the periodic parts of the p-type valence band have the structures $\left( U_{v,p,x,q}^{{}}(\vec{\rho })\pm U_{v,p,y,q}^{{}}(\vec{\rho }) \right)/\sqrt{2}$ similar to the expressions $(x\pm iy)/\sqrt{2}$ characterized by the projections $M$ of the orbital momentum on the magnetic field direction equal to $\pm 1$. For the c-v bands of different parities, as is the present case, and takes place in the GaAs-type crystals, this matrix element in the zeroth approximation on the small wave vector ${{\vec{k}}_{||}}$ and g is different from zero. In the definition of Elliott [32, 33] it is considered to be of the allowed type and is denoted as ${{P}_{cv}}({{\vec{k}}_{||}},g)\approx {{P}_{cv}}(0)$. Its value cannot be changed by the magnetic field strength. Even if a magnetic field is considered to be strong with the electron cyclotron energy greater than the exciton binding energy, nevertheless this energy is much smaller than the semiconductor energy band gap $E_{g}^{0}$.

The two selected dipole-active magnetoexciton states ${{F}_{1}}$ and ${{F}_{4}}$ are characterized by the coefficients $\varphi \left( {{F}_{1}},B,\vec{k} \right)$ and $\varphi \left( {{F}_{4}},B,\vec{k} \right)$determining the magnetoexciton-photon interaction. They have the expressions:
\begin{eqnarray}
&& \varphi \left( {{F}_{1}},B,\vec{k} \right)=\left( \frac{-e}{{{m}_{0}}{{l}_{0}}} \right)\sqrt{\frac{\hbar }{{{L}_{c}}{{\omega }_{{\vec{k}}}}}}{{P}_{cv}}\left( 0 \right)a_{0}^{-*}d_{0}^{-*}, \nonumber\\
&& \varphi \left( {{F}_{4}},B,\vec{k} \right)=\left( \frac{e}{{{m}_{0}}{{l}_{0}}} \right)\sqrt{\frac{\hbar }{{{L}_{c}}{{\omega }_{{\vec{k}}}}}}{{P}_{cv}}\left( 0 \right),
\end{eqnarray}
where the frequency ${{\omega }_{{\vec{k}}}}$ of the photon confined in the microcavity can be written in the form:
\begin{eqnarray}
&& \hbar {{\omega }_{{\vec{k}}}}=\frac{\hbar c}{{{n}_{c}}}\sqrt{{{\left( \frac{\pi }{{{L}_{c}}} \right)}^{2}}+\vec{k}_{||}^{2}}\approx  \nonumber\\
&& \approx \hbar {{\omega }_{c}}+\frac{{{\hbar }^{2}}\vec{k}_{\parallel }^{2}}{2{{m}_{c}}}= \nonumber\\
&& =\hbar {{\omega }_{c}}\left( 1+\frac{{{x}^{2}}}{2} \right),x<1, \\
&& \vec{k}=\pm \frac{\pi }{{{L}_{c}}}{{{\vec{a}}}_{3}}+\vec{k}_{||}^{{}},\left| \vec{k}_{||}^{{}} \right|=x\frac{\pi }{{{L}_{c}}}, \nonumber\\
&& {{\omega }_{c}}=\frac{c}{{{n}_{c}}}\frac{\pi }{{{L}_{c}}},{{\omega }_{c}}{{L}_{c}}=\frac{c\pi }{{{n}_{c}}},{{m}_{c}}=\frac{\hbar \pi {{n}_{c}}}{c{{L}_{c}}} \nonumber
\end{eqnarray}
In the limit $x\to 0$ the Rabi splitting energies are determined by the doubled absolute values of the interaction coefficients
\begin{eqnarray}
&& \hbar {{\Omega }_{R}}({{F}_{1}},B)=2\left| \varphi ({{F}_{1}},B,\pi /{{L}_{c}}) \right|= \nonumber\\
&& =\frac{2e}{{{m}_{0}}{{l}_{0}}}\sqrt{\frac{\hbar {{n}_{c}}}{\pi c}}\left| {{P}_{cv}}(0) \right|\left| a_{0}^{-} \right|\left| d_{0}^{-} \right|, \nonumber\\
&& \hbar {{\Omega }_{R}}({{F}_{4}},B)=2\left| \varphi ({{F}_{4}},B,\pi /{{L}_{c}}) \right|= \nonumber\\
&& =\frac{2e}{{{m}_{0}}{{l}_{0}}}\sqrt{\frac{\hbar {{n}_{c}}}{\pi c}}\left| {{P}_{cv}}(0) \right|
\end{eqnarray}
In this approximation the Rabi splitting frequencies do not depend on the cavity frequency ${{\omega }_{c}}$ or on the cavity length ${{L}_{c}}$. They were estimated in the Ref.[16] in the case of GaAs-type quantum well with $\left| {{P}_{cv}}(0) \right|\approx 2\cdot {{10}^{-20}} g \cdot$ cm/sec. Their dependences on the magnetic field strength is determined by the magnetic length ${{l}_{0}}$. The Rabi splitting increases with the increasing of the magnetic field strength B as $\sqrt{B}$: $\hbar {{\Omega }_{R}}\sim \sqrt{B}$.
	
The creation energy of a magnetoexciton in the state ${{F}_{n}}$ and with the wave vector ${{\vec{k}}_{||}}$ is determined by the formulas (47) and (50) and looks as
\begin{eqnarray}
&&{{E}_{mex}}({{F}_{n}},{{{\vec{k}}}_{||}})=E_{g}^{0}+{{E}_{e}}({{R}_{i}})+{{E}_{h}}({{R}_{j}})- \nonumber\\
&& -{{I}_{ex}}({{F}_{n}},B,0)+\frac{{{\hbar }^{2}}\vec{k}_{||}^{2}}{2M({{F}_{n}},B)}
\end{eqnarray}
Introducing the energy detuning $\Delta $ between the semiconductor energy band gap $E_{g}^{0}$ and the cavity mode energy $\hbar {{\omega }_{c}}$ in the way $\Delta =-E_{g}^{0}+\hbar {{\omega }_{c}}$ and using the dimensionless wave number x related with the cavity length ${{L}_{c}}$ as is shown in the formula (64), the expression (66) can be transcribed in the form:
\begin{eqnarray}
&& {{E}_{mex}}({{F}_{n}},x)=\hbar {{\omega }_{c}}-\Delta +L({{F}_{n}},B)+\hbar g({{F}_{n}},B)\frac{{{x}^{2}}}{2}, \nonumber\\
&& L({{F}_{n}},B)={{E}_{e}}({{R}_{i}})+{{E}_{h}}({{R}_{j}})-{{I}_{ex}}({{F}_{n}},B,0), \nonumber\\
&& \hbar g({{F}_{n}},B)=\frac{{{\hbar }^{2}}{{\pi }^{2}}}{M({{F}_{n}},B)L_{c}^{2}}, \nonumber\\
&& x=\frac{\left| {{{\vec{k}}}_{||}} \right|{{L}_{c}}}{\pi }<1
\end{eqnarray}
On the base of the expressions (63)-(67) one can write the dispersion law of the cavity magnetoexciton polaritons as follows [32, 33]:
\begin{eqnarray}
&&{{E}_{\begin{array}{cc}
 up \\
 lp
\end{array}}}=\hbar {{\omega }_{c}}+\frac{1}{2}[-\Delta +L({{F}_{n}},B)+\hbar g({{F}_{n}},B)\frac{{{x}^{2}}}{2}+\nonumber \\
&&+\hbar {{\omega }_{c}}\frac{{{x}^{2}}}{2}\pm  \\
&&\pm ({(-\Delta +L({{F}_{n}},B)+\hbar g({{F}_{n}},B)\frac{{{x}^2}}{2}-\hbar {{\omega }_{c}}\frac{{{x}^{2}}}{2})^2}+ \nonumber\\
&&+|\hbar {{\Omega }_{R}}({{F}_{n}},B){{|}^2})^{1/2}] \nonumber
\end{eqnarray}
In difference on the Ref.[16], where the Rabi energy $\hbar {{\omega }_{R}}=\left| \varphi ({{F}_{n}},B) \right|$ was introduced above the Rabi splitting energy $\hbar {{\Omega }_{R}}=2\hbar {{\omega }_{R}}$ following the Ref.[34, 35] is used.

The zeroth order Hamiltonian describing free 2D magnetoexcitons, the cavity photons and their interaction has a quadratic form and consists of three parts
\begin{eqnarray}
{{H}_{2}}=H_{mex}^{0}+H_{ph}^{0}+{{H}_{mex-ph}}
\end{eqnarray}
For the sake of simplicity the denotation (31) of the magnetoexciton creation operators will be shortened as follows
\begin{eqnarray}
&& \psi _{ex}^{\dagger }\left( {{F}_{1}},{{{\vec{k}}}_{\parallel }} \right)=\psi _{ex}^{\dagger }\left( {{{\vec{k}}}_{\parallel }},{{R}_{1}},-1,\varepsilon _{3}^{-} \right) \nonumber\\
&& \psi _{ex}^{\dagger }\left( {{F}_{2}},{{{\vec{k}}}_{\parallel }} \right)=\psi _{ex}^{\dagger }\left( {{{\vec{k}}}_{\parallel }},{{R}_{2}},-1,\varepsilon _{3}^{-} \right) \nonumber\\
&& \psi _{ex}^{\dagger }\left( {{F}_{3}},{{{\vec{k}}}_{\parallel }} \right)=\psi _{ex}^{\dagger }\left( {{{\vec{k}}}_{\parallel }},{{R}_{1}},1,{{\varepsilon }_{0}} \right) \nonumber\\
&& \psi _{ex}^{\dagger }\left( {{F}_{4}},{{{\vec{k}}}_{\parallel }} \right)=\psi _{ex}^{\dagger }\left( {{{\vec{k}}}_{\parallel }},{{R}_{2}},1,{{\varepsilon }_{0}} \right) \nonumber\\
&& \psi _{ex}^{\dagger }\left( {{F}_{5}},{{{\vec{k}}}_{\parallel }} \right)=\psi _{ex}^{\dagger }\left( {{{\vec{k}}}_{\parallel }},{{R}_{1}},-1,\varepsilon _{4}^{-} \right) \nonumber\\
&& \psi _{ex}^{\dagger }\left( {{F}_{6}},{{{\vec{k}}}_{\parallel }} \right)=\psi _{ex}^{\dagger }\left( {{{\vec{k}}}_{\parallel }},{{R}_{2}},-1,\varepsilon _{4}^{-} \right)
\end{eqnarray}
In the Hamiltonian ${{H}_{2}}$ only the dipole-active magnetoexciton states ${{F}_{1}}$ and ${{F}_{4}}$ as well as the quadrupole-active states ${{F}_{3}}$ and ${{F}_{5}}$ were included
\begin{eqnarray}
&&H_{mex}^{0}=\sum\limits_{n=1,3,4,5}{E}_{ex}({{F}_{n}},{{{\vec{k}}}_{\parallel }})\psi_{ex}^{\dagger }({{F}_{n}},{{\vec{k}}_{\parallel }}){{\psi }_{ex}}({{F}_{n}},{{\vec{k}}_{\parallel }})\nonumber \\
\end{eqnarray}
The remained two states ${{F}_{2}}$ and ${{F}_{6}}$ were excluded because they are forbidden in both approximations.

The cavity photons have the wave vectors $\vec{k}={{\vec{a}}_{3}}{{\vec{k}}_{z}}+{{\vec{k}}_{\parallel }}$ consisting from two parts. The longitudinal component is oriented along the axis of the resonator determined by the unit vector ${{\vec{a}}_{3}}$ perpendicular to the surface of the quantum well embedded inside the microcavity. It has a well definite values ${{k}_{z}}=\pm \frac{\pi }{{{L}_{c}}}$, where ${{L}_{c}}$ is the cavity length. The transverse component ${{\vec{k}}_{\parallel }}={{\vec{a}}_{1}}{{k}_{x}}+{{\vec{a}}_{2}}{{k}_{y}}$ is a 2D vector oriented in-plane of the QW and is determined by two in-plane unit vectors ${{\vec{a}}_{1}}$ and ${{\vec{a}}_{2}}$. Vectors of the light circular polarizations $\vec{\sigma }_{{\vec{k}}}^{\pm }$ can be constructed introducing two unit vectors $\vec{s}$ and $\vec{t}$ perpendicular to the light wave vector $\vec{k}$ and to each other in the way
\begin{eqnarray}
&& \vec{\sigma }_{{\vec{k}}}^{\pm }=\frac{1}{\sqrt{2}}\left( \vec{s}\pm i\vec{t} \right),\vec{k}={{{\vec{a}}}_{3}}{{k}_{z}}+{{{\vec{k}}}_{\parallel }},{{k}_{z}}=\pm \frac{\pi }{{{L}_{c}}}, \nonumber\\
&& {{{\vec{k}}}_{\uparrow }}={{{\vec{a}}}_{3}}\frac{\pi }{{{L}_{c}}}+{{{\vec{k}}}_{\parallel }},{{{\vec{k}}}_{\downarrow }}=-{{{\vec{a}}}_{3}}\frac{\pi }{{{L}_{c}}}+{{{\vec{k}}}_{\parallel }}, \nonumber\\
&& \vec{s}={{{\vec{a}}}_{3}}\frac{\left| {{{\vec{k}}}_{\parallel }} \right|}{\left| {\vec{k}} \right|}-\frac{{{{\vec{k}}}_{\parallel }}\cdot {{k}_{z}}}{\left| {\vec{k}} \right|\left| {{{\vec{k}}}_{\parallel }} \right|},\vec{t}=\frac{{{{\vec{a}}}_{1}}{{k}_{y}}-{{{\vec{a}}}_{2}}{{k}_{x}}}{\left| {{{\vec{k}}}_{\parallel }} \right|}, \nonumber\\
&& \left| {{{\vec{k}}}_{\uparrow }} \right|=\left| {{{\vec{k}}}_{\downarrow }} \right|=\left| {\vec{k}} \right|.
\end{eqnarray}
They obey to the orthogonality and normalization conditions
\begin{eqnarray}
&& \left( \vec{k}\cdot \vec{t} \right)=\left( \vec{s}\cdot \vec{k} \right)=\left( \vec{t}\cdot \vec{s} \right)=0, \nonumber\\
&& \left| {\vec{s}} \right|=\left| {\vec{t}} \right|=1,{{(\vec{\sigma }_{{\vec{k}}}^{+})}^{*}}=\vec{\sigma }_{{\vec{k}}}^{-}, \left| \vec{\sigma }_{{\vec{k}}}^{\pm } \right|=1 \nonumber\\
&& (\vec{\sigma }_{{\vec{k}}}^{\pm }\cdot \vec{\sigma }_{{\vec{k}}}^{{{\mp }^{*}}})=0,(\vec{\sigma }_{{\vec{k}}}^{\pm }\cdot \vec{\sigma }_{{\vec{k}}}^{{{\pm }^{*}}})=1,
\end{eqnarray}
and look as
\begin{eqnarray}
&& \vec{\sigma }_{{\vec{k}}}^{\pm }=\frac{1}{\sqrt{2}\left| {\vec{k}} \right|\left| {{{\vec{k}}}_{\parallel }} \right|}\{{{{\vec{a}}}_{3}}{{\left| {{{\vec{k}}}_{\parallel }} \right|}^{2}}+{{{\vec{a}}}_{1}}(-{{k}_{x}}{{k}_{z}}\pm i{{k}_{y}}\left| {\vec{k}} \right|)+ \nonumber\\
&& +{{{\vec{a}}}_{2}}(-{{k}_{y}}{{k}_{z}}\mp i{{k}_{x}}\left| {\vec{k}} \right|)\}={{(\vec{\sigma }_{{\vec{k}}}^{\mp })}^{*}} \\
&& {{{\vec{\sigma }}}_{\pm 1}}=\frac{1}{\sqrt{2}}\left( {{{\vec{a}}}_{1}}\pm i{{{\vec{a}}}_{2}} \right)=\vec{\sigma }_{\mp 1}^{*}\nonumber
\end{eqnarray}
Here the magnetoexciton circular polarization vectors ${{\vec{\sigma }}_{\pm 1}}$ determined by the formula (54) are remembered. The needed scalar products of the photon and magnetoexciton circular polarization vectors are listed below
\begin{eqnarray}
&& |(\vec{\sigma }_{{\vec{k}}}^{\pm }\cdot \vec{\sigma }_{1}^{*}){{|}^{2}}=\frac{1}{4|\vec{k}{{|}^{2}}}{{({{k}_{z}}\pm |\vec{k}|)}^{2}}\cong  \nonumber\\
&& \cong \frac{1}{2}(1\pm \frac{{{k}_{z}}}{\left| {{k}_{z}} \right|})(1-\frac{{{x}^{2}}}{2}+\frac{{{x}^{4}}}{2})\mp \frac{{{x}^{4}}}{16}\cdot \frac{{{k}_{z}}}{\left| {{k}_{z}} \right|}, \nonumber\\
&& {{\left| \left( \vec{\sigma }_{{\vec{k}}}^{\pm }\cdot \vec{\sigma }_{-1}^{*} \right) \right|}^{2}}=\frac{1}{4|\vec{k}{{|}^{2}}}{{({{k}_{z}}\mp |\vec{k}|)}^{2}}\cong  \nonumber\\
&& \cong \frac{1}{2}(1\mp \frac{{{k}_{z}}}{\left| {{k}_{z}} \right|})(1-\frac{{{x}^{2}}}{2}+\frac{{{x}^{4}}}{2})\pm \frac{{{x}^{4}}}{16}\cdot \frac{{{k}_{z}}}{\left| {{k}_{z}} \right|}, \nonumber\\
&& |\vec{k}|=\sqrt{{{\left| {{k}_{z}} \right|}^{2}}+{{\left| {{{\vec{k}}}_{\parallel }} \right|}^{2}}}=\left| {{{\vec{k}}}_{z}} \right|\sqrt{1+{{x}^{2}}}\cong  \\
&& \cong \left| {{k}_{z}} \right|(1+\frac{{{x}^{2}}}{2}-\frac{{{x}^{4}}}{8}),{{x}^{2}}=\frac{|{{{\vec{k}}}_{\parallel }}{{|}^{2}}L_{c}^{2}}{{{\pi }^{2}}}<1,{{k}_{z}}=\pm \frac{\pi }{{{L}_{c}}}.\nonumber
\end{eqnarray}
The geometric rules (75) depend not only on the signs of the circular polarization vectors $\vec{\sigma }_{{\vec{k}}}^{\pm }$, but also on the signs of the projections ${{k}_{z}}$ of the photon wave vector $\vec{k}$. We will use new denotations for the circular polarization vectors $\vec{\sigma }_{{{{\vec{k}}}_{\uparrow }}}^{\pm }$ and $\vec{\sigma }_{{{{\vec{k}}}_{\downarrow }}}^{\pm }$ to take these details into account, so that the geometric selection rules (75) can be transcribed as follows
\begin{eqnarray}
&& {{\left| \left( \vec{\sigma }_{{{{\vec{k}}}_{\uparrow }}}^{+}\cdot \vec{\sigma }_{1}^{*} \right) \right|}^{2}}={{\left| \left( \vec{\sigma }_{{{k}_{\uparrow }}}^{-}\cdot \vec{\sigma }_{-1}^{*} \right) \right|}^{2}}={{\left| \left( \vec{\sigma }_{{{{\vec{k}}}_{\downarrow }}}^{+}\cdot \vec{\sigma }_{-1}^{*} \right) \right|}^{2}}= \nonumber\\
&& ={{\left| \left( \vec{\sigma }_{{{{\vec{k}}}_{\downarrow }}}^{-}\cdot \vec{\sigma }_{1}^{*} \right) \right|}^{2}}=\left( 1-\frac{{{x}^{2}}}{2}+\frac{7}{16}{{x}^{4}} \right), \nonumber\\
&& {{\left| \left( \vec{\sigma }_{{{{\vec{k}}}_{\uparrow }}}^{+}\cdot \vec{\sigma }_{-1}^{*} \right) \right|}^{2}}={{\left| \left( \vec{\sigma }_{{{{\vec{k}}}_{\uparrow }}}^{-}\cdot \vec{\sigma }_{1}^{*} \right) \right|}^{2}}= \nonumber\\
&& ={{\left| \left( \vec{\sigma }_{{{k}_{\downarrow }}}^{+}\cdot \vec{\sigma }_{1}^{*} \right) \right|}^{2}}={{\left| \left( \vec{\sigma }_{{{k}_{\downarrow }}}^{-}\cdot \vec{\sigma }_{-1}^{*} \right) \right|}^{2}}=\frac{{{x}^{4}}}{16}.
\end{eqnarray}
The zeroth order Hamiltonian of the cavity photons with the wave vectors $\vec{k}$  and circular polarizations $\vec{\sigma }_{{\vec{k}}}^{\pm }$ described in (72) look as
\begin{eqnarray}
&&H_{ph}^{0}=\sum\limits_{{{{\vec{k}}}_{\parallel }},{{k}_{z}}=\pm \frac{\pi }{{{L}_{C}}}}{\hbar {{\omega }_{{\vec{k}}}}}[ {{( {{C}_{\vec{k},+}} )}^{+}}{{C}_{\vec{k},+}}+{{( {{C}_{\vec{k},-}} )}^{+}}{{C}_{\vec{k},-}} ]\nonumber\\
\end{eqnarray}
The creation and annihilation operators of the photons with circular polarizations $\vec{\sigma }_{{\vec{k}}}^{\pm }$  and with the linear polarizations $\vec{s}$ and $\vec{t}$ are related as follows
\begin{eqnarray}
&& {{C}_{\vec{k},\pm }}=\frac{1}{\sqrt{2}}\left( {{C}_{\vec{k},s}}\pm i{{C}_{\vec{k},t}} \right) \nonumber\\
&& {{\left( {{C}_{\vec{k},\pm }} \right)}^{+}}=\frac{1}{\sqrt{2}}\left( C_{\vec{k},s}^{+}\mp iC_{\vec{k},t}^{+} \right) \nonumber\\
&& \vec{s}{{C}_{\vec{k},s}}+\vec{t}{{C}_{\vec{k},t}}={{C}_{\vec{k},-}}\vec{\sigma }_{{\vec{k}}}^{+}+{{C}_{\vec{k},+}}\vec{\sigma }_{{\vec{k}}}^{-} \nonumber\\
&& \vec{s}C_{\vec{k},s}^{+}+\vec{t}C_{\vec{k},t}^{+}={{\left( {{C}_{\vec{k},-}} \right)}^{+}}\vec{\sigma }_{k}^{-}+{{\left( {{C}_{\vec{k},+}} \right)}^{+}}\vec{\sigma }_{{\vec{k}}}^{+}
\end{eqnarray}
The Hamiltonian describing the magnetoexciton-photon interaction including only the resonance terms and taking into account the two photon circular polarizations has the form
\begin{eqnarray}
&& {{H}_{mex-ph}}=\sum\limits_{{{{\vec{k}}}_{\parallel }},{{k}_{z}}=\pm \frac{\pi }{{{L}_{c}}}}{{}}\{\varphi ({{F}_{1}},{{{\vec{k}}}_{\parallel }})[{{C}_{\vec{k},-}}\left( \vec{\sigma }_{{\vec{k}}}^{+}\cdot \vec{\sigma }_{-1}^{*} \right)+ \nonumber\\
&& +{{C}_{\vec{k},+}}\left( \vec{\sigma }_{{\vec{k}}}^{-}\cdot \vec{\sigma }_{-1}^{*} \right)]\psi _{ex}^{+}({{F}_{1}},{{F}_{\parallel }})+ \nonumber\\
&& +\varphi ({{F}_{4}},{{{\vec{k}}}_{\parallel }})[{{C}_{\vec{k},-}}\left( \vec{\sigma }_{{\vec{k}}}^{+}\cdot \vec{\sigma }_{1}^{*} \right)+\nonumber \\
&&+{{C}_{\vec{k},+}}\left( \vec{\sigma }_{{\vec{k}}}^{-}\cdot {{{\vec{\sigma }}}_{1}} \right)]\psi _{ex}^{+}({{F}_{4}},{{{\vec{k}}}_{\parallel }})+ \nonumber\\
&& +\varphi ({{F}_{3}},{{{\vec{k}}}_{\parallel }})[{{C}_{\vec{k},-}}\left( \vec{\sigma }_{{\vec{k}}}^{+}\cdot \vec{\sigma }_{1}^{*} \right)+ \nonumber \\
&&+{{C}_{\vec{k},+}}\left( \vec{\sigma }_{{\vec{k}}}^{-}\cdot {{{\vec{\sigma }}}_{1}} \right)]\psi _{ex}^{+}({{F}_{3}},{{{\vec{k}}}_{\parallel }})+ \nonumber\\
&& +\varphi ({{F}_{5}},{{{\vec{k}}}_{\parallel }})[{{C}_{\vec{k},-}}\left( \vec{\sigma }_{{\vec{k}}}^{+}\cdot \vec{\sigma }_{-1}^{*} \right)+ \nonumber \\
&&+{{C}_{\vec{k},+}}\left( \vec{\sigma }_{{\vec{k}}}^{-}\cdot \vec{\sigma }_{-1}^{*} \right)]\psi _{ex}^{+}({{F}_{5}},{{{\vec{k}}}_{\parallel }})+ \nonumber\\
&& +\left. h.c. \right\}
\end{eqnarray}
The coefficients $\varphi \left( {{F}_{n}},{{{\vec{k}}}_{\parallel }} \right)$  and their square moduli are
\begin{eqnarray}
&& \varphi \left( {{F}_{1}},{{{\vec{k}}}_{\parallel }} \right)=-{{\phi }_{cv}}a{{_{0}^{-}}^{*}}d{{_{0}^{-}}^{*}}, \nonumber\\
&& {{\left| \varphi \left( {{F}_{1}},{{{\vec{k}}}_{\parallel }} \right) \right|}^{2}}={{\left| {{\phi }_{cv}} \right|}^{2}}{{\left| a_{0}^{-} \right|}^{2}}{{\left| d_{0}^{-} \right|}^{2}}, \nonumber\\
&& \varphi \left( {{F}_{4}},{{{\vec{k}}}_{\parallel }} \right)={{\phi }_{cv}}, {{\left| \varphi \left( {{F}_{4}},{{{\vec{k}}}_{\parallel }} \right) \right|}^{2}}={{\left| {{\phi }_{cv}} \right|}^{2}}, \nonumber\\
&& \varphi \left( {{F}_{3}},{{{\vec{k}}}_{\parallel }} \right)={{\phi }_{cv}}b{{_{1}^{-}}^{*}}\left( -\frac{{{k}_{x}}+i{{k}_{y}}}{\sqrt{2}} \right){{l}_{0}}, \nonumber\\
&& {{\left| \varphi \left( {{F}_{3}},{{{\vec{k}}}_{\parallel }} \right) \right|}^{2}}={{\left| {{\phi }_{cv}} \right|}^{2}}{{\left| b_{1}^{-} \right|}^{2}}\frac{{{\left| {{{\vec{k}}}_{\parallel }} \right|}^{2}}l_{0}^{2}}{2}, \nonumber\\
&& \varphi \left( {{F}_{5}},{{{\vec{k}}}_{\parallel }} \right)=-{{\phi }_{cv}}a{{_{0}^{-}}^{*}}d{{_{0}^{-}}^{*}}\left( \frac{{{k}_{x}}+i{{k}_{y}}}{\sqrt{2}} \right){{l}_{0}}, \nonumber\\
&& {{\left| \varphi \left( {{F}_{5}},{{{\vec{k}}}_{\parallel }} \right) \right|}^{2}}={{\left| {{\phi }_{cv}} \right|}^{2}}{{\left| a_{0}^{-} \right|}^{2}}{{\left| d_{0}^{-} \right|}^{2}}\cdot \frac{{{\left| {{{\vec{k}}}_{\parallel }} \right|}^{2}}l_{0}^{2}}{2}, \nonumber\\
&& {{\phi }_{cv}}=\frac{e}{{{m}_{0}}{{l}_{0}}}\sqrt{\frac{\hbar {{n}_{c}}}{\pi c}}{{P}_{cv}}\left( 0 \right),{{\left| {{\phi }_{cv}} \right|}^{2}}=\left( \frac{\hbar {{n}_{c}}}{\pi c} \right){{\left| \frac{e{{P}_{cv}}\left( 0 \right)}{{{m}_{0}}{{l}_{0}}} \right|}^{2}}, \nonumber\\
&& {{f}_{osc}}={{\left| \frac{{{\phi }_{cv}}}{\hbar {{\omega }_{c}}} \right|}^{2}}=\frac{\hbar {{n}_{c}}}{\pi c}\cdot {{\left| \frac{e{{P}_{cv}}\left( 0 \right)}{{{m}_{0}}{{l}_{0}}\hbar {{\omega }_{c}}} \right|}^{2}}.
\end{eqnarray}
Below we will use the dimensionless value ${{f}_{osc}}={{\left| \frac{{{\phi }_{ev}}}{\hbar {{\omega }_{c}}} \right|}^{2}}$ playing the role of the oscillator strength. One can observe that the probabilities of the dipole-active quantum transitions in the magnetoexciton states ${{F}_{1}}$ and ${{F}_{4}}$ determined by the expressions ${{\left| \varphi \left( {{F}_{1}},{{{\vec{k}}}_{\parallel }} \right) \right|}^{2}}$ and ${{\left| \varphi \left( {{F}_{4}},{{{\vec{k}}}_{\parallel }} \right) \right|}^{2}}$ are proportional to ${{\left| {{\phi }_{cv}} \right|}^{2}}\approx l_{0}^{-2}\approx B$ and have a linear dependence on the magnetic field strength B. On the contrary, the probabilities of the quadrupole-active quantum transitions in the magnetoexciton states ${{F}_{3}}$ and ${{F}_{5}}$ are proportional to the expression ${{\left| {{\phi }_{cv}} \right|}^{2}}\cdot l_{0}^{2}$, which does not depend on the magnetic field strength $B$ at all.

The equations of motion for four magnetoexciton annihilation operators ${{\psi }_{ex}}\left( {{F}_{n}},{{{\vec{k}}}_{\parallel }} \right)$ with $n=1,3,4,5$ as well as for the similar photon operators ${{C}_{\vec{k},\pm }}$ in the stationary conditions look as
\begin{eqnarray}
&& i\hbar \frac{d}{dt}{{\psi }_{ex}}\left( {{F}_{n}},{{{\vec{k}}}_{\parallel }} \right)=[{{\psi }_{ex}}\left( {{F}_{n}},{{{\vec{k}}}_{\parallel }} \right),{{{\hat{H}}}_{2}}]=\nonumber \\
&&=\hbar \omega {{\psi }_{ex}}\left( {{F}_{n}},{{{\vec{k}}}_{\parallel }} \right), \nonumber\\
&& i\hbar \frac{d}{dt}{{C}_{\vec{k},\pm }}=[{{C}_{\vec{k},\pm }},{{{\hat{H}}}_{2}}]=\hbar \omega {{C}_{\vec{k},\pm }}.
\end{eqnarray}
Their concrete expressions are
\begin{eqnarray}
&& \left( \hbar \omega -{{E}_{ex}}\left( {{F}_{i}},{{{\vec{k}}}_{\parallel }} \right) \right){{\psi }_{ex}}\left( {{F}_{i}},{{{\vec{k}}}_{\parallel }} \right)= \nonumber\\
&& =\left[ {{C}_{\vec{k},-}}\left( \vec{\sigma }_{{\vec{k}}}^{+}\cdot \vec{\sigma }_{-1}^{*} \right)+{{C}_{\vec{k},+}}\left( \vec{\sigma }_{{\vec{k}}}^{-}\cdot \vec{\sigma }_{-1}^{*} \right) \right]\varphi \left( {{F}_{i}},{{{\vec{k}}}_{\parallel }} \right), \nonumber\\
&& i=1,5, \nonumber\\
&& \left( \hbar \omega -{{E}_{ex}}\left( {{F}_{j}},{{{\vec{k}}}_{\parallel }} \right) \right){{\psi }_{ex}}\left( {{F}_{j}},{{{\vec{k}}}_{\parallel }} \right)= \nonumber\\
&& =\left[ {{C}_{\vec{k},-}}\left( \vec{\sigma }_{{\vec{k}}}^{+}\cdot \vec{\sigma }_{1}^{*} \right)+{{C}_{\vec{k},+}}\left( \vec{\sigma }_{{\vec{k}}}^{-}\cdot \vec{\sigma }_{1}^{*} \right) \right]\varphi \left( {{F}_{j}},{{{\vec{k}}}_{\parallel }} \right), \nonumber\\
&& j=3,4, \nonumber\\
&& \left( \hbar \omega -\hbar {{\omega }_{{\vec{k}}}} \right){{C}_{\vec{k},\pm }}=\nonumber \\
&&=\left( \vec{\sigma }_{{\vec{k}}}^{\pm }\cdot {{{\vec{\sigma }}}_{-1}} \right){{\varphi }^{*}}\left( {{F}_{1}},{{{\vec{k}}}_{\parallel }} \right){{\psi }_{ex}}\left( {{F}_{1}},{{{\vec{k}}}_{\parallel }} \right)+ \nonumber\\
&& +\left( \vec{\sigma }_{{\vec{k}}}^{\pm }\cdot {{{\vec{\sigma }}}_{1}} \right){{\varphi }^{*}}\left( {{F}_{4}},{{{\vec{k}}}_{\parallel }} \right)\psi \left( {{F}_{4}},{{{\vec{k}}}_{\parallel }} \right)+ \nonumber\\
&& +\left( \vec{\sigma }_{{\vec{k}}}^{\pm }\cdot {{{\vec{\sigma }}}_{1}} \right){{\varphi }^{*}}\left( {{F}_{3}},{{{\vec{k}}}_{\parallel }} \right){{\psi }_{ex}}\left( {{F}_{3}},{{{\vec{k}}}_{\parallel }} \right)+ \nonumber\\
&& +\left( \vec{\sigma }_{{\vec{k}}}^{\pm }\cdot {{{\vec{\sigma }}}_{-1}} \right){{\varphi }^{*}}\left( {{F}_{5}},{{{\vec{k}}}_{\parallel }} \right){{\psi }_{ex}}\left( {{F}_{5}},{{{\vec{k}}}_{\parallel }} \right)
\end{eqnarray}
The dipole-active state ${{F}_{1}}$ and the quadrupole-active state ${{F}_{5}}$ can be preferentially excited by the light with the circular polarization $\vec{\sigma }_{{\vec{k}}}^{-}$ propagating with ${{k}_{z}}=\frac{\pi }{{{L}_{c}}}>0$, whereas the dipole-active state ${{F}_{4}}$ and the quadrupole-active state ${{F}_{3}}$ mainly react to the light with circular polarization $\vec{\sigma }_{{\vec{k}}}^{+}$ propagating in the same direction with ${{k}_{z}}=\frac{\pi }{{{L}_{c}}}>0$. In the case when the Hamiltonians (77) and (79) contain the photons only with one circular polarization either $\vec{\sigma }_{{\vec{k}}}^{+}$ or $\vec{\sigma }_{{\vec{k}}}^{-}$ in this cases the equations (82) give rise to the relations
\begin{eqnarray}
&& \psi \left( {{F}_{i}},{{{\vec{k}}}_{\parallel }} \right)={{C}_{\vec{k},\pm }}\frac{\left( \vec{\sigma }_{{\vec{k}}}^{\mp }\cdot \vec{\sigma }_{-1}^{*} \right)\varphi \left( {{F}_{i}},{{{\vec{k}}}_{\parallel }} \right)}{\hbar \omega -{{E}_{ex}}\left( {{F}_{i}},{{{\vec{k}}}_{\parallel }} \right)}, \nonumber\\
&& i=1,5, \nonumber\\
&& \psi \left( {{F}_{j}},{{{\vec{k}}}_{\parallel }} \right)={{C}_{\vec{k},\pm }}\frac{\left( \vec{\sigma }_{{\vec{k}}}^{\mp }\cdot \vec{\sigma }_{1}^{*} \right)\varphi \left( {{F}_{j}},{{{\vec{k}}}_{\parallel }} \right)}{\hbar \omega -{{E}_{ex}}\left( {{F}_{j}},{{{\vec{k}}}_{\parallel }} \right)}, \nonumber\\
&& j=3,4.
\end{eqnarray}
Substituting these relations in the equations of motion (82) for the photon operators ${{C}_{\vec{k},\pm }}$ we will find two dispersion equations describing five branches of the energy spectrum of two different systems. One of them concerns the photons with circular polarization $\vec{\sigma }_{{\vec{k}}}^{-}$ propagating with ${{k}_{z}}=\frac{\pi }{{{L}_{c}}}>0$ and exciting mainly the magnetoexciton state ${{F}_{1}}$ as well as another states${{F}_{3}},{{F}_{4}},{{F}_{5}}$ with smaller oscillator strengths. The second system consist from the magnetoexcitons in the same four states ${{F}_{1}},{{F}_{3}},{{F}_{4}},{{F}_{5}}$ existing in the frame of the microcavity fulfilled by the photons with circular polarization $\vec{\sigma }_{{\vec{k}}}^{+}$ propagating with ${{k}_{z}}=\frac{\pi }{{{L}_{c}}}>0$. These five order algebraic dispersion equations are
\begin{eqnarray}
&& \left( \hbar \omega -\hbar {{\omega }_{{\vec{k}}}} \right)= \nonumber\\
&& =\frac{{{\left| \left( \vec{\sigma }_{{\vec{k}}}^{\mp }\cdot \vec{\sigma }_{-1}^{*} \right) \right|}^{2}}{{\left| \varphi \left( {{F}_{1}}{{{\vec{k}}}_{\parallel }} \right) \right|}^{2}}}{\hbar \omega -{{E}_{ex}}\left( {{F}_{1}},{{{\vec{k}}}_{\parallel }} \right)}+\nonumber \\
&&+\frac{{{\left| \left( \vec{\sigma }_{{\vec{k}}}^{\mp }\cdot \vec{\sigma }_{1}^{*} \right) \right|}^{2}}{{\left| \varphi \left( {{F}_{4}},{{{\vec{k}}}_{\parallel }} \right) \right|}^{2}}}{\hbar \omega -{{E}_{ex}}\left( {{F}_{4}},{{{\vec{k}}}_{\parallel }} \right)}+ \nonumber\\
&& +\frac{{{\left| \left( \vec{\sigma }_{{\vec{k}}}^{\mp }\cdot \vec{\sigma }_{1}^{*} \right) \right|}^{2}}{{\left| \varphi \left( {{F}_{3}}{{{\vec{k}}}_{\parallel }} \right) \right|}^{2}}}{\hbar \omega -{{E}_{ex}}\left( {{F}_{3}},{{{\vec{k}}}_{\parallel }} \right)}+ \nonumber \\
&&+\frac{{{\left| \left( \vec{\sigma }_{{\vec{k}}}^{\mp }\cdot \vec{\sigma }_{-1}^{*} \right) \right|}^{2}}{{\left| \varphi \left( {{F}_{5}},{{{\vec{k}}}_{\parallel }} \right) \right|}^{2}}}{\hbar \omega -{{E}_{ex}}\left( {{F}_{5}},{{{\vec{k}}}_{\parallel }} \right)}.
\end{eqnarray}
They describe four magnetoexciton branches and one photon branch with a given circular polarization either $\vec{\sigma }_{{\vec{k}}}^{-}$ or $\sigma _{{\vec{k}}}^{+}$ when the photons are propagating with ${{k}_{z}}=\pm \frac{\pi }{{{L}_{c}}}$.

Below we will consider the case, when the energy of the cavity mode $\hbar {{\omega }_{c}}$ may be tuned to the energy of the selected magnetoexciton level, in the vicinity of which the maximal development of the polariton branches is expected. There are two dipole-active magnetoexciton energy levels ${{F}_{1}}$ and ${{F}_{4}}$ but in different circular polarizations. Their energies are essentially changed in dependence on the magnetic field strength B and on another parameters of the theory. To obtain the needed detuning the cavity-mode energy $\hbar {{\omega }_{c}}$ must be changed adjusting it to new values of the selected magnetoexciton energy level.

It is useful to introduce the dimensionless values and the detunings ${{\Delta }_{1}}$ and ${{\Delta }_{4}}$ as follows:
\begin{eqnarray}
&& \hbar \omega =\hbar {{\omega }_{c}}+E,\frac{\hbar \omega }{\hbar {{\omega }_{c}}}=1+\varepsilon , \nonumber\\
&& \varepsilon =\frac{E}{\hbar {{\omega }_{c}}},{{\Delta }_{n}}=\hbar {{\omega }_{c}}-{{E}_{ex}}({{F}_{n}},B,0),\nonumber \\
&& \frac{{{\Delta }_{n}}}{\hbar {{\omega }_{c}}}={{\delta }_{n}}=1-\frac{{{E}_{ex}}({{F}_{n}},B,0)}{\hbar {{\omega }_{c}}}, \nonumber\\
&& \hbar {{\omega }_{c}}=\frac{{{E}_{ex}}({{F}_{n}},B,0)}{1-{{\delta }_{n}}},\left| {{\delta }_{n}} \right|<1, \nonumber\\
&& \frac{\hbar \omega -{{E}_{ex}}({{F}_{n}},B,0)}{\hbar {{\omega }_{c}}}=\varepsilon +{{\delta }_{n}},n=1,4.
\end{eqnarray}
In these denotations the dispersion equation (83) can be transcribed for two different cases. One of them concerns the cavity with the mode in resonance with the magnetoexciton energy level ${{E}_{ex}}({{F}_{1}},B,0)$ with the detuning ${{\delta }_{1}}$, when the cavity photons have the circular polarization $\vec{\sigma }_{{{{\vec{k}}}_{\uparrow }}}^{-}$ or $\vec{\sigma }_{{{{\vec{k}}}_{\downarrow }}}^{+}$ and interact mainly with the magnetoexcitons in the state ${{F}_{1}}$ with circular polarization ${{\vec{\sigma }}_{-1}}$ and wave vectors ${{\vec{k}}_{||}}$. These photons interact also with another magnetoexcitons such as the dipole-active state ${{F}_{4}}$ but in opposite circular polarization ${{\vec{\sigma }}_{+1}}$ as well as with the quadrupole-active magnetoexcitons in the states ${{F}_{3}}$ and ${{F}_{5}}$. The corresponding dispersion equation has the form:
\begin{eqnarray}
&& \left( \varepsilon -\frac{{{x}^{2}}}{2} \right)={{f}_{osc}}\cdot \left\{ \frac{{{\left| a_{0}^{-} \right|}^{2}}{{\left| d_{0}^{-} \right|}^{2}}\left( 1-\frac{{{x}^{2}}}{2}+\frac{7}{16}{{x}^{4}} \right)}{\left( \varepsilon +{{\delta }_{1}}-\frac{n_{c}^{2}\hbar {{\omega }_{c}}}{M\left( {{F}_{1}},B \right){{c}^{2}}}\cdot \frac{{{x}^{2}}}{2} \right)}+ \right. \nonumber\\
&& +\frac{\frac{{{x}^{4}}}{16}}{\left( \varepsilon +1-\frac{{{E}_{ex}}\left( {{F}_{4}},B,0 \right)}{{{E}_{ex}}\left( {{F}_{1}},B,0 \right)}\left( 1-{{\delta }_{1}} \right)-\frac{n_{c}^{2}\hbar {{\omega }_{c}}}{M\left( {{F}_{4}},B \right){{c}^{2}}}\cdot \frac{{{x}^{2}}}{2} \right)}+ \nonumber\\
&& +\frac{{{\left| b_{1}^{-} \right|}^{2}}{{\left( \frac{\pi {{l}_{0}}}{{{L}_{c}}} \right)}^{2}}\frac{{{x}^{6}}}{32}}{\left( \varepsilon +1-\frac{{{E}_{ex}}\left( {{F}_{3}},B,0 \right)}{{{E}_{ex}}\left( {{F}_{1}},B,0 \right)}\left( 1-{{\delta }_{1}} \right)-\frac{n_{c}^{2}\hbar {{\omega }_{c}}}{M\left( {{F}_{3}},B \right){{c}^{2}}}\cdot \frac{{{x}^{2}}}{2} \right)}+ \nonumber\\
&& \left. +\frac{{{\left| a_{0}^{-} \right|}^{2}}{{\left| d_{0}^{-} \right|}^{2}}{{\left( \frac{\pi {{l}_{0}}}{{{L}_{c}}} \right)}^{2}}\frac{{{x}^{2}}}{2}\left( 1-\frac{{{x}^{2}}}{2}+\frac{7}{16}{{x}^{4}} \right)}{\left( \varepsilon +1-\frac{{{E}_{ex}}\left( {{F}_{5}},B \right)}{{{E}_{ex}}\left( {{F}_{1}},B \right)}\left( 1-{{\delta }_{1}} \right)-\frac{n_{c}^{2}\hbar {{\omega }_{c}}}{M\left( {{F}_{5}},B \right){{c}^{2}}}\cdot \frac{{{x}^{2}}}{2} \right)} \right\}
\end{eqnarray}
The second case concerns the cavity mode in resonance with the magnetoexciton energy level ${{E}_{ex}}({{F}_{4}},B,0)$ with the detuning ${{\delta }_{4}}$, when the cavity photon have the circular polarization $\vec{\sigma }_{{{{\vec{k}}}_{\uparrow }}}^{+}$or $\vec{\sigma }_{{{{\vec{k}}}_{\downarrow }}}^{-}$. They interact mainly with the magnetoexcitons in the state ${{F}_{4}}$ with circular polarization ${{\vec{\sigma }}_{+1}}$. There are also weak interactions with the dipole-active magnetoexciton state ${{F}_{1}}$ with opposite circular polarization ${{\vec{\sigma }}_{-1}}$ as well as with the quadrupole-active magnetoexciton states ${{F}_{3}}$ and ${{F}_{5}}$ with circular polarizations ${{\vec{\sigma }}_{1}}$ and ${{\vec{\sigma }}_{-1}}$ correspondingly.   The dispersion equation looks as
\begin{eqnarray}
&& \left( \varepsilon -\frac{{{x}^{2}}}{2} \right)={{f}_{osc}}\left\{ \frac{\left( 1-\frac{{{x}^{2}}}{2}+\frac{7}{16}{{x}^{4}} \right)}{\varepsilon +{{\delta }_{4}}-\frac{n_{c}^{2}\hbar {{\omega }_{c}}}{M\left( {{F}_{4}},B \right){{c}^{2}}}\cdot \frac{{{x}^{2}}}{2}} \right.+ \nonumber\\
&& +\frac{\frac{{{x}^{4}}}{16}{{\left| a_{0}^{-} \right|}^{2}}{{\left| d_{0}^{-} \right|}^{2}}}{\left( \varepsilon +1-\frac{{{E}_{ex}}\left( {{F}_{1}},B,0 \right)}{{{E}_{ex}}\left( {{F}_{4}},B,0 \right)}\left( 1-{{\delta }_{4}} \right)-\frac{n_{c}^{2}\hbar {{\omega }_{c}}}{M\left( {{F}_{1}},B \right){{c}^{2}}}\frac{{{x}^{2}}}{2} \right)}+ \nonumber\\
&& +\frac{{{\left| b_{1}^{-} \right|}^{2}}{{\left( \frac{\pi {{l}_{0}}}{{{L}_{c}}} \right)}^{2}}\frac{{{x}^{2}}}{2}\left( 1-\frac{{{x}^{2}}}{2}+\frac{7}{16}{{x}^{4}} \right)}{\left( \varepsilon +1-\frac{{{E}_{ex}}\left( {{F}_{3}},B,0 \right)}{{{E}_{ex}}\left( {{F}_{4}},B,0 \right)}\left( 1-{{\delta }_{4}} \right)-\frac{n_{c}^{2}\hbar {{\omega }_{c}}}{M\left( {{F}_{3}},B \right){{c}^{2}}}\cdot \frac{{{x}^{2}}}{2} \right)}+ \nonumber\\
&& \left. +\frac{{{\left| a_{0}^{-} \right|}^{2}}{{\left| d_{0}^{-} \right|}^{2}}{{\left( \frac{\pi {{l}_{0}}}{{{L}_{c}}} \right)}^{2}}\frac{{{x}^{6}}}{32}}{\left( \varepsilon +1-\frac{{{E}_{ex}}\left( {{F}_{5}},B,0 \right)}{{{E}_{ex}}\left( {{F}_{4}},B,0 \right)}\left( 1-{{\delta }_{4}} \right)-\frac{n_{c}^{2}\hbar {{\omega }_{c}}}{M\left( {{F}_{5}},B \right){{c}^{2}}}\cdot \frac{{{x}^{2}}}{2} \right)} \right\}
\end{eqnarray}
In both equations (86) and (87) the factor $\left( 1-\frac{{{x}^{2}}}{2}+\frac{7}{16}{{x}^{4}} \right)$ expresses the geometric selection rules (76) concerning the quantum transitions from the ground state of the crystal to the magnetoexcitons states ${{F}_{1}}$ and ${{F}_{4}}$ with their circular polarizations ${{\vec{\sigma }}_{-1}}$  and ${{\vec{\sigma }}_{1}}$ correspondingly excited under the influence of the photons with circular polarizations $\vec{\sigma }_{{{{\vec{k}}}_{\uparrow }}}^{-}$ or  $\vec{\sigma }_{{{{\vec{k}}}_{\downarrow }}}^{+}$in one case and $\vec{\sigma }_{{{{\vec{k}}}_{\uparrow }}}^{+}$ or $\vec{\sigma }_{{{{\vec{k}}}_{\downarrow }}}^{-}$in another one. The factor $\left( 1-\frac{{{x}^{2}}}{2}+\frac{7}{16}{{x}^{4}} \right)$ differs from 1 due to the nonzero in-plane components ${{\vec{k}}_{\parallel }}$ of the photon wave vectors $\vec{k}=\pm {{\vec{a}}_{3}}\frac{\pi }{{{L}_{c}}}+{{\vec{k}}_{\parallel }}$. The factor $\frac{{{x}^{4}}}{16}$ appears when the dipole-active states ${{F}_{1}}$ and ${{F}_{4}}$ with their circular polarizations ${{\vec{\sigma }}_{-1}}$ and ${{\vec{\sigma }}_{1}}$ are  excited by the photons with the opposite circular polarizations such as $\vec{\sigma }_{{{{\vec{k}}}_{\uparrow }}}^{+}$ or $\vec{\sigma }_{{{{\vec{k}}}_{\downarrow }}}^{-}$in the case of state ${{F}_{1}}$ as well such as $\vec{\sigma }_{{{{\vec{k}}}_{\downarrow }}}^{+}$ or $\vec{\sigma }_{{{{\vec{k}}}_{\uparrow }}}^{-}$ in the case of state ${{F}_{4}}$. The different from zero quantum transition probabilities are possible in this case only at the oblique incidence of the photons with ${{\vec{k}}_{\parallel }}\ne 0$ on the surface of the $QW$ embedded into the microcavity.
In the case of the strongly perpendicular incident light these quantum transitions are forbidden. The factor $\frac{{{x}^{2}}}{2}$ has another origin related with the difference by one between the quantum numbers ${{n}_{e}}$ and ${{n}_{h}}$ of the Landau quantization levels for electrons and holes what is reflected in the formulas (59) and (60).
The combinations of the terms $\left( 1-\frac{{{x}^{2}}}{2}+\frac{7}{16}{{x}^{4}} \right)$ and $\frac{{{x}^{4}}}{16}$ with the factor $\frac{{{x}^{2}}}{2}$ give rise to the factors $\frac{{{x}^{2}}}{2}\left( 1-\frac{{{x}^{2}}}{2}+\frac{7}{16}{{x}^{4}} \right)$  and $\frac{{{x}^{6}}}{32}$. All of them are present in the dispersion equations (86) and (87).

The energy spectrum of the cavity magnetoexciton-polaritons when the energy of the cavity mode is tuned to the magnetoexciton energy level ${{E}_{ex}}\left( {{F}_{1}},B,0 \right)$ was obtained solving the fifth order dispersion equation (85) and is depicted in the figures 9-11. They concern to the case of the cavity photons with circular polarizations $\vec{\sigma }_{{{{\vec{k}}}_{\uparrow }}}^{-}$ or $\vec{\sigma }_{{{{\vec{k}}}_{\downarrow }}}^{+}$, when the cavity mode energy $\hbar {{\omega }_{c}}$ was tuned to the magnetoexciton energy level ${{E}_{ex}}\left( {{F}_{1}},B,0 \right)$ in the way $\hbar {{\omega }_{c}}={{E}_{ex}}\left( {{F}_{1}},B,0 \right)/\left( 1-{{\delta }_{1}} \right)$ with dimensionless detuning ${{\delta }_{1}}=-0.01$.
Two values of the magnetic field strength B: 20T and 40T were considered. The energies $\hbar {{\omega }_{c}}$ in these cases are different. The obtained pictures essentially depend on the RSOC, arising under the influence of the external electric field applied perpendicularly to the surface of the QW being parallel to the external magnetic field. In the case of the 2D heavy holes the RSOC is characterized by the chirality terms of the third order in the wave vector $k$ components, wich inevitably lead to the collapse of the semiconductor energy gap. To avoid it in the heavy hole dispersion law the fourth order term of the type ${{\left| {\vec{k}} \right|}^{4}}$ was introduced being proportional to electron field strength with the parameter of the nonparabolicity C. Fig. 9 corresponds to the absence of the RSOC and to the parameters ${{E}_{z}}=0$ and $C=0$. Figs. 10 and 11 correspond to the presence of RSOC with the electric field strength ${{E}_{z}}=30kV/cm$ and the parameter C equal to 20 and 30 correspondingly. In the frame of each of these three cases the four variants were calculated. They arised due to two values of the magnetic field strength 20T and 40T mentioned above, and due to two values of the heavy hole unknown g-factors ${{g}_{h}}=\pm 5$. The electron $g$ factor was supposed to be equal ${{g}_{e}}=1$. Looking at the figures 9-11 one can observe that the bare cavity photon branch is intersected by the bare magnetoexciton state $F_{1}$, $F_{3}$, $F_{4}$, $F_{5}$ branches. Only the interaction of the magnetoexciton $F_{1}$ with the cavity photons with the selected circular polarization $\vec{\sigma }_{\uparrow }^{-}$ or $\vec{\sigma }_{\downarrow }^{+}$ gives rise to the well distinguishable polariton pictures. In another intersection points denoted as $(n-c)$ with $n=3,4,5$ the magnetoexciton-photon interactions are very small. The reconstructed polariton curves in the vicinities of these points are shown only qualitatively so as to help the eye. The real values of the splittings arising due to the anticrossings in these points are denoted as $\Delta(n-c)$. The coordinates of these intersection points $X( n-c)$ and $Y(n-c)$ together with the corresponding splittings are mentioned in the captions.

The forbidden bare magnetoexciton branches ${{F}_{2}}$ and ${{F}_{6}}$ are also drawn in the figures because their presence will influence on the kinetic and quantum statistic properties of the high density cavity magnetoexciton-polaritons. The description of another energy spectra in general outline is the same. But the positions of the different magnetoexciton energy branches on the energy scale essentially depend on the full set of the parameters ${B,E_{z},g_{e},g_{h},C}$ which is demonstrated in the figure 9-11.

Our theoretical model taking into account the electron structure of the 2D magnetoexcitons arising under the influence of the perpendicular strong magnetic and electric fields is applicable for the GaAs-type QWs only in the range of magnetic field strength $B\ge 10$T. A similar variant but without RSOC was described in the Refs.[2, 3]. The experimental investigations of the cavity exciton-polaritons in a range of small and stronger magnetic fields up to 14T were reported in the Ref.[15]. The more suitable theoretical model in this case was elaborated in Ref.[36], where the energy spectrum of the hydrogenic potential $V(r)={-{{e}^{2}}}/{r}\;$ in two dimensions was studied for perpendicular magnetic field in two weak and strong regimes. Using a two-point Pad\'{e} approximant a reliable interpolation between two limiting situations was proposed. This model was enlarged taking into account the dimension perpendicular to the QW [37]. The existence of many theoretical models will stimulate the further experimental investigations in the regime of high magnetic fields $B\ge 10$T. A short review of the comprehensive experimental findings found out in the Ref.[15] is needed to compare and to discuss them in two main models mentioned above. The experiments were performed using the 8nm-thick In$_{0.04}$Ga$_{0.96}$As QW embedded into the resonator with the distributed Bragg reflectors. The Wannier-Mott excitons have a resonance energy 1.484ev, an exciton binding energy equal approx. 7mev, and effective mass 0.046m0. The polariton linewidth was equal to 0.3mev. The interest to these objects is due to the observation of the BEC phenomenon with the participation of the cavity polaritons accompanied by the superfluidity formation with zero viscosity. Another stimulus is the elaboration of the polariton-based devices [38]. The effect of the magnetic field on the polaritons revealed in Ref.[15] is diverse already in the linear regime. The magnetic field induces the changes in the linewidth. The interpretation of these results on the base of the hydrogen interpolation model [36] takes into account the exciton energy shift including the Zeeman splitting, the modification of the exciton-photon coupling strength and the change of the Rabi splitting. The magnetic field is considered as a parameter changing the tuning of the exciton-photon resonances. The changes of the emission intensity and of the polariton linewidth were explained in Ref.[15] by the magnetically modified scattering processes with participation of the acoustical phonons. The increase of the exciton oscillator strength ${{f}_{osc}}$ was directly determined in Ref.[15] measuring the Rabi splitting $\Omega $ of a polariton branches. Taking into account the relation $\Omega \sim \sqrt{{N{{f}_{osc}}}/{{{L}_{eff}}}\;}$, where N is the number of QWs embedded into the resonator and ${{L}_{eff}}$ is the effective length, the ${{f}_{osc}}$ was calculated. These results are represented in the fig. 6 of the Ref.[15]. One can see that the $\hbar \Omega $ value increases from 3.4mev to 4.8mev at $B=14$T, whereas the ${{f}_{osc}}$ increases by a factor of two.

In our model the dependences of the Rabi frequency $\Omega $ on the magnetic field strength is determined by the relation $\Omega \sim l_{0}^{-1}\sim \sqrt{B}$, where ${{l}_{0}}$ is the magnetic length. It leads to the dependence for the oscillator strength ${{f}_{osc}}\sim |\Omega |^{2}\sim B$.
Another property of the cavity polariton spectra is the Zeeman splitting of the lower polariton branch. Two components of the bright exciton level are decoupled and couple independently with the photon of corresponding circular polarization. The observed experimentally in Ref.[15] the Zeeman splitting of the lower polariton branch reveals a decreasing in dependence on the magnetic field strength up to 14T, instead of the expected increasing.
In the frame of our model the magnetoexciton states F1-F6 are represented by the non-degenerate branches with the being evantail dependence on the magnetic field strength with increasing splitting between them in the absence of the RSOC.
Only in the presence of the RSOC, as well as taking into account the nonparabolicity of the heavy-hole dispersion law, the intersections and overlapping between the magnetoexciton branches appear. The supprising behavior of the Zeeman splitting revealed in the Ref.[15] possibly is related with the dependences of the electron and hole g-factors ${{g}_{e}}(B)$ and ${{g}_{h}}(B)$ on the magnetic field strength B [39-41].
\begin{figure}%[h]
\resizebox{0.48\textwidth}{!}{%
  \includegraphics{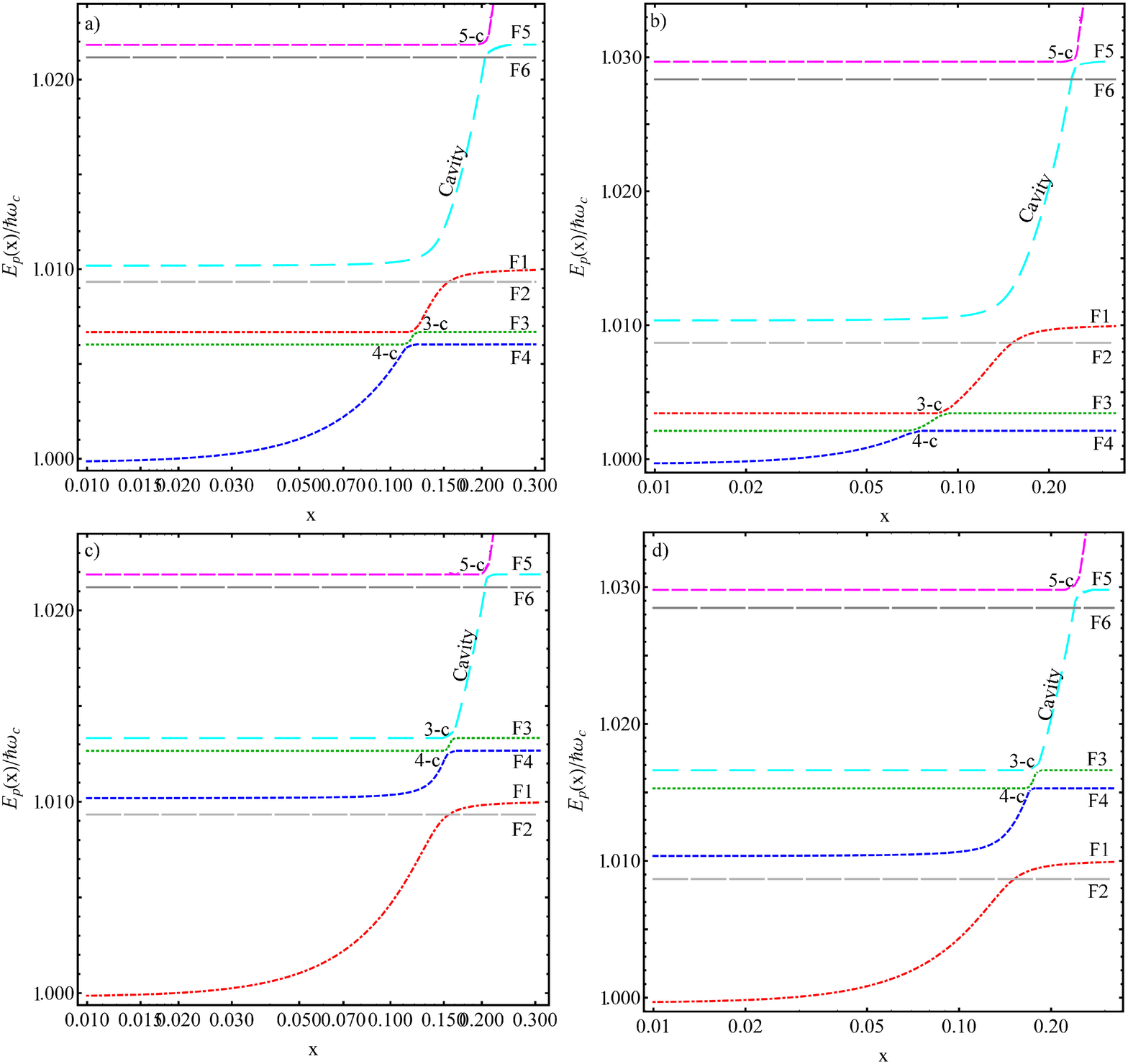}
}
\caption{Five cavity magnetoexciton-polariton energy branches in dependence on the dimensionless in-plane wave number $x$ arising as a result of the interaction of the cavity photons with circular polarizations $\vec{\sigma }_{{{{\vec{k}}}_{\uparrow }}}^{-}$ or $\vec{\sigma }_{{{{\vec{k}}}_{\downarrow }}}^{+}$ with four magnetoexciton energy branches ${{F}_{1}}$, ${{F}_{4}}$, ${{F}_{3}}$ and ${{F}_{5}}$ characterized by the circular polarizations ${{\vec{\sigma }}_{\pm 1}}$ , two branches being dipole-active and two quadrupole active, when the cavity mode energy is tuned to the magnetoexciton energy level ${{E}_{ex}}\left( {{F}_{1}},B,0 \right)$. The case of the absence of the external electric field and of the RSOC with the parameters ${{E}_{z}}=0$ and $C=0$ is represented in four variants. They appear from the combinations of two values of the magnetic field strength $B$ and of two values of the hh $g-$ factor  ${{g}_{h}}$ as follows:  a) $B=20$T, ${{g}_{h}}=5;$ b) $B=40$T, ${{g}_{h}}=5;$ c) $B=20$T, ${{g}_{h}}=-5;$ d) $B=40$T, ${{g}_{h}}=-5$. The electron $g-$ factor was selected ${{g}_{e}}=1$. For the completeness two forbidden magnetoexciton energy branches ${{F}_{2}}$ and ${{F}_{6}}$ are added. The intersection points and the splittings of the polariton curves in these points: $a){{X}_{a}}\left( 4-c \right)=0,113;$ ${{Y}_{a}}\left( 4-c \right)=1,00601;$ ${{\Delta }_{a}}\left( 4-c \right)=8,44\cdot {{10}^{-6}}$. ${{X}_{a}}\left( 3-c \right)=0,120;$ ${{Y}_{a}}\left( 3-c \right)=1,00670;$ ${{\Delta }_{a}}\left( 3-c \right)=8,46\cdot {{10}^{-6}}.$ ${{X}_{a}}\left( 5-c \right)=0,209;$ ${{Y}_{a}}\left( 5-c \right)=1,02184;$ ${{\Delta }_{a}}\left( 5-c \right)=8,44\cdot {{10}^{-6}}$, $b){{X}_{b}}\left( 4-c \right)=0,071;$ ${{Y}_{b}}\left( 4-c \right)=1,00206;$ ${{\Delta }_{b}}\left( 4-c \right)=7,26\cdot {{10}^{-6}}$. ${{X}_{b}}\left( 3-c \right)=0,089;$ ${{Y}_{b}}\left( 3-c \right)=1,00338;$ ${{\Delta }_{b}}\left( 3-c \right)=7,31\cdot {{10}^{-6}}$. ${{X}_{b}}\left( 5-c \right)=0,243;$ ${{Y}_{b}}\left( 5-c \right)=1,02969;$ ${{\Delta }_{b}}\left( 5-c \right)=7,28\cdot {{10}^{-6}}.$ $c){{X}_{c}}\left( 4-c \right)=0,156;$ ${{Y}_{c}}\left( 4-c \right)=1,01265;$ ${{\Delta }_{c}}\left( 4-c \right)=8,41\cdot {{10}^{-6}}$. ${{X}_{c}}\left( 3-c \right)=0,160;$ ${{Y}_{c}}\left( 3-c \right)=1,01333;$ ${{\Delta }_{c}}\left( 3-c \right)=8,44\cdot {{10}^{-6}}.{{X}_{c}}\left( 5-c \right)=0,208; {{Y}_{c}}\left( 5-c \right)=1,02190; {{\Delta }_{c}}\left( 5-c \right)=8,41\cdot {{10}^{-6}}.$ $d){{X}_{d}}\left( 4-c \right)=0,171;$ ${{Y}_{d}}\left( 4-c \right)=1,01529;$ ${{\Delta }_{d}}\left( 4-c \right)=5,94\cdot {{10}^{-6}}$. ${{X}_{d}}\left( 3-c \right)=0,179;$ ${{Y}_{d}}\left( 3-c \right)=1,01660;$ ${{\Delta }_{d}}\left( 3-c \right)=6,00\cdot {{10}^{-6}}$. ${{X}_{d}}\left( 5-c \right)=0,243;$ ${{Y}_{d}}\left( 5-c \right)=1,02976;$ ${{\Delta }_{d}}\left( 5-c \right)=5,96\cdot {{10}^{-6}}$}
\end{figure}
\begin{figure}%[h]
\resizebox{0.48\textwidth}{!}{%
  \includegraphics{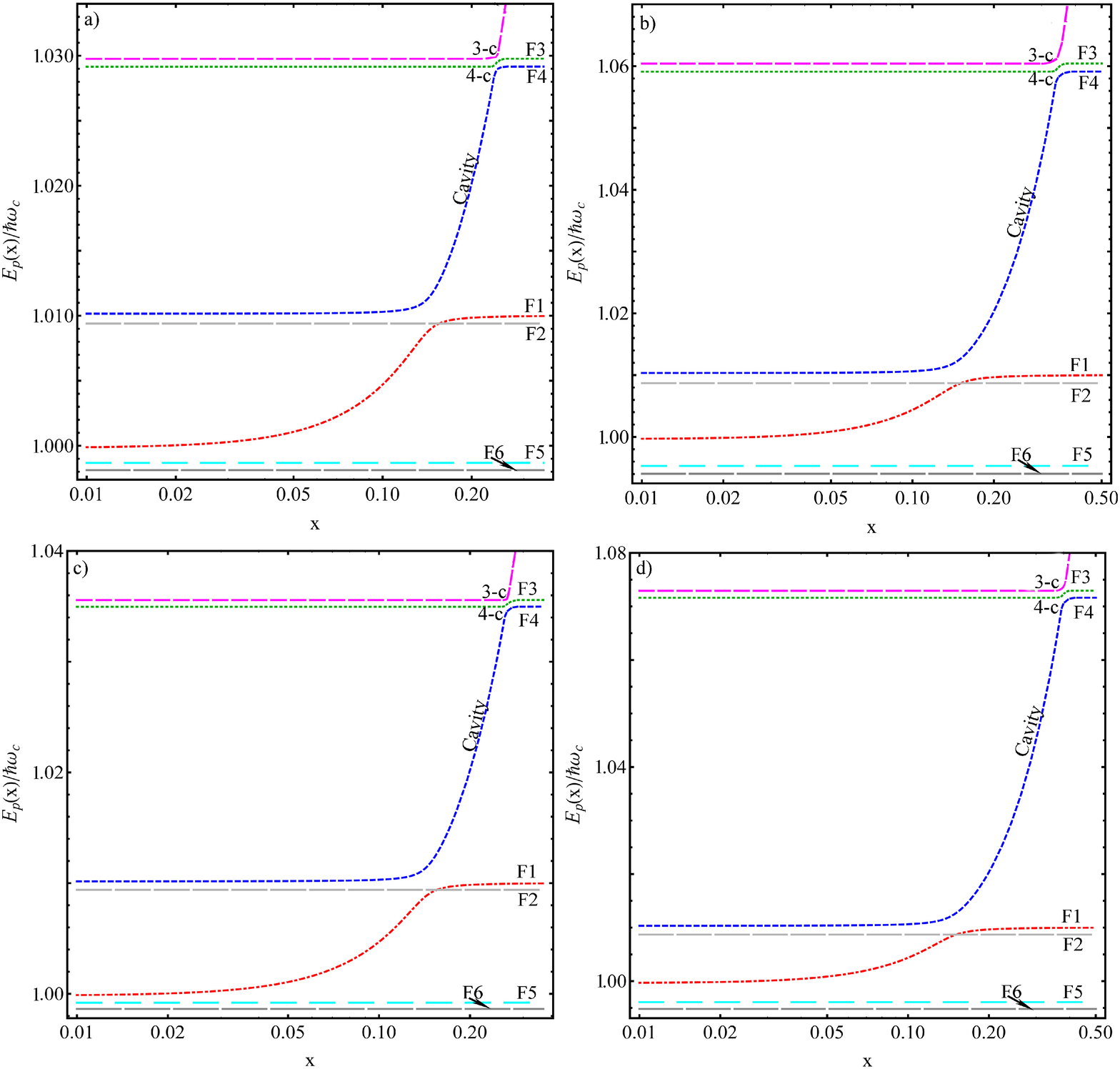}
}
\caption{Five cavity magnetoexciton-polariton energy branches in dependence on the dimensionless in-plane wave number $x$ arising as result of the interaction of the cavity photons with circular polarizations $\vec{\sigma }_{{{{\vec{k}}}_{\uparrow }}}^{-}$ or $\vec{\sigma }_{{{{\vec{k}}}_{\downarrow }}}^{+}$ with four magnetoexciton energy branches ${{F}_{1}}$, ${{F}_{4}}$, ${{F}_{3}}$ and ${{F}_{5}}$ characterized by the circular polarizations ${{\vec{\sigma }}_{\pm 1}}$ , two branches being dipole-active and two quadrupole active, when the cavity mode energy is tuned to the magnetoexciton energy level ${{E}_{ex}}\left( {{F}_{1}},B,0 \right)$. The influence of the external electric field and of the RSOC with the parameters ${{E}_{z}}=30$kV/cm and $C=20$ is represented in four variants. They appear from the combinations of two values of the magnetic field strength $B$ and of two values of the hh $g-$factor  ${{g}_{h}}$ as follows:  a) $B=20$T, ${{g}_{h}}=5;$ b) $B=40$T, ${{g}_{h}}=5;$ c) $B=20$T, ${{g}_{h}}=-5;$ d) $B=40$T, ${{g}_{h}}=-5$. The electron $g-$factor was selected ${{g}_{e}}=1$. For the completeness two forbidden magnetoexciton energy branches ${{F}_{2}}$ and ${{F}_{6}}$ are added. The intersection points and the splittings of the polariton curves in these points: $a)X\left( 1-c \right)=0,145;$ $Y\left( 1-c \right)=1,01015$. $ {{X}_{a}}\left( 4-c \right)=0,241;$ ${{Y}_{a}}\left( 4-c \right)=1,02919;$ ${{\Delta }_{a}}\left( 4-c \right)=11,23\cdot {{10}^{-6}}$. $ {{X}_{a}}\left( 3-c \right)=0,245;$ ${{Y}_{a}}\left( 3-c \right)=1,02980;$ ${{\Delta }_{a}}\left( 3-c \right)=11,29\cdot {{10}^{-6}}.$ $b){{X}_{b}}\left( 4-c \right)=0,343;$ ${{Y}_{b}}\left( 4-c \right)=1,05920;$ ${{\Delta }_{b}}\left( 4-c \right)=15,74\cdot {{10}^{-6}}$. ${{X}_{b}}\left( 3-c \right)=0,347;$ ${{Y}_{b}}\left( 3-c \right)=1,06040;$ ${{\Delta }_{b}}\left( 3-c \right)=16\cdot {{10}^{-6}}$. $c){{X}_{c}}\left( 4-c \right)=0,264;$ ${{Y}_{c}}\left( 4-c \right)=1,03492;$ ${{\Delta }_{c}}\left( 4-c \right)=11,19\cdot {{10}^{-6}}$. ${{X}_{c}}\left( 3-c \right)=0,268;$ ${{Y}_{c}}\left( 3-c \right)=1,03561;$ ${{\Delta }_{c}}\left( 3-c \right)=11,26\cdot {{10}^{-6}}.$ $d){{X}_{d}}\left( 4-c \right)=0,377;$ ${{Y}_{d}}\left( 4-c \right)=1,07161;$ ${{\Delta }_{d}}\left( 4-c \right)=15,60\cdot {{10}^{-6}}.$ $ {{X}_{d}}\left( 3-c \right)=0,380;$ ${{Y}_{d}}\left( 3-c \right)=1,07287;$ ${{\Delta }_{d}}\left( 3-c \right)=15,91\cdot {{10}^{-6}}$}
\end{figure}
\begin{figure}%[h]
\resizebox{0.48\textwidth}{!}{%
  \includegraphics{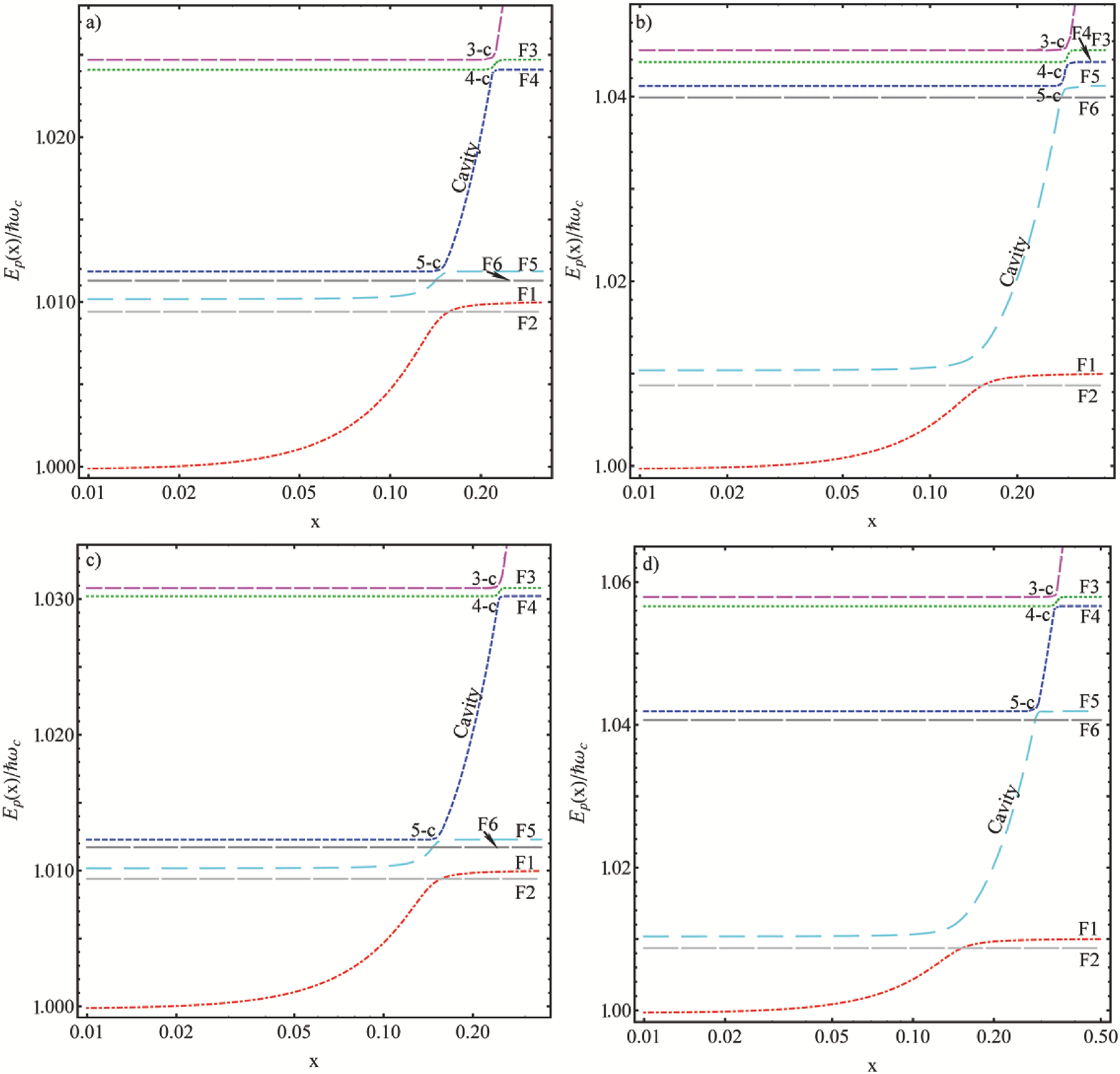}
}
\caption{Five cavity magnetoexciton-polariton energy branches in dependence on the dimensionless in-plane wave number $x$ arising as result of the interaction of the cavity photons with circular polarizations $\vec{\sigma }_{{{{\vec{k}}}_{\uparrow }}}^{-}$ or $\vec{\sigma }_{{{{\vec{k}}}_{\downarrow }}}^{+}$ with four magnetoexciton energy branches ${{F}_{1}}$, ${{F}_{4}}$, ${{F}_{3}}$ and ${{F}_{5}}$ characterized by the circular polarizations ${{\vec{\sigma }}_{\pm 1}}$ , two branches being dipole-active and two quadrupole active, when the cavity mode energy is tuned to the magnetoexciton energy level ${{E}_{ex}}( {{F}_{1}},B,0 )$. The influence of the external electric field and of the RSOC with the parameters ${{E}_{z}}=30$kV/cm and $C=30$ is represented in four variants. They appear from the combinations of two values of the magnetic field strength $B$ and of two values of the hh $g-$ factor  ${{g}_{h}}$ as follows:  a) $B=20$T, ${{g}_{h}}=5;$ b) $B=40$T, ${{g}_{h}}=5;$ c) $B=20$T, ${{g}_{h}}=-5;$ d) $B=40$T, ${{g}_{h}}=-5$. The electron $g-$ factor was selected ${{g}_{e}}=1$. For the completeness two forbidden magnetoexciton energy branches ${{F}_{2}}$ and ${{F}_{6}}$ are added. The intersection points and the splittings of the polariton curves in these points:$ a){{X}_{a}}(5-c)=0,148;$ ${{Y}_{a}}(5-c)=1,01182;$ ${{\Delta }_{a}}(5-c)=9,47\cdot {{10}^{-6}}.$ ${{X}_{a}}(4-c)=0,218;$ ${{Y}_{a}}(4-c)=1,02409;$ ${{\Delta }_{a}}(4-c)=9,44\cdot {{10}^{-6}}.$ $ {{X}_{a}}(3-c)=0,222;$ ${{Y}_{a}}(3-c)=1,02470;$ ${{\Delta }_{a}}(3-c)=9,49\cdot {{10}^{-6}}.$ $b){{X}_{b}}(5-c)=0,285;$ ${{Y}_{b}}(5-c)=1,04116;$ ${{\Delta }_{b}}(5-c)=10,39\cdot {{10}^{-6}}.$ ${{X}_{b}}(4-c)=0,295;$ ${{Y}_{b}}(4-c)=1,04371;$ ${{\Delta }_{b}}(4-c)=10,24\cdot {{10}^{-6}}.$ ${{X}_{b}}(3-c)=0.3;$ ${{Y}_{b}}(3-c)=1,04502;$ ${{\Delta }_{b}}(3-c)=10,41\cdot {{10}^{-6}}.$ $c){{X}_{c}}(5-c)=0,151;$ ${{Y}_{c}}5-c)=1,01227;$ ${{\Delta }_{c}}(5-c)=10,67\cdot {{10}^{-6}}.$ ${{X}_{c}}(4-c)=0,247;$ ${{Y}_{c}}(4-c)=1,03021;$ ${{\Delta }_{c}}(4-c)=10,62\cdot {{10}^{-6}}.$ ${{X}_{c}}(3-c)=0,249;$ ${{Y}_{c}}(3-c)=1,03081;$ ${{\Delta }_{c}}( 3-c)=10,68\cdot {{10}^{-6}}.$ $d){{X}_{d}}(5-c)=0,288;$ ${{Y}_{d}}(5-c)=1,04190;$ ${{\Delta }_{d}}(5-c)=16,16\cdot {{10}^{-6}}.$ ${{X}_{d}}(4-c)=0,336;$ ${{Y}_{d}}( 4-c)=1,05672;$ ${{\Delta }_{d}}(4-c)=15,96\cdot {{10}^{-6}}.$ ${{X}_{d}}(3-c)=0,341;$ ${{Y}_{d}}(3-c)=1,05799;$ ${{\Delta }_{d}}(3-c)=16,18\cdot {{10}^{-6}}.$}
\end{figure}
\section{Conclusions:}
The properties of the 2D cavity polaritons subjected to the action of a strong perpendicular magnetic and electric fields were investigated. The exact solutions of the Landau quantization of the 2D heavy-holes accompanied by the RSOC with third order chirality terms, by the ZS effects and by the non-parabolicity of their dispersion low were obtained following the method proposed by Rashba [1]. The results of Ref.[1] concerning the conduction electrons were supplemented taking into account the effects of Zeeman splitting. Using the wave functions for the 2D electrons and holes, the Hamiltonians describing in the second quantization representation the Coulomb electron-electron and the electron-radiation interactions were derived. The electron-radiation interaction Hamiltonian makes possible to construct another Hamiltonian describing the magnetoexciton-photon interaction and to develop the theory of the magnetoexciton-polaritons using the wave functions of the 2D magnetoexcitons. The six magnetoexciton states arising due to the composition of two lowest Landau levels for conduction electrons with three LLLs for heavy-holes were taken into consideration. Between them two states ${{F}_{1}}$ and ${{F}_{4}}$ are dipole-active, another two, ${{F}_{3}}$ and ${{F}_{5}}$, are quadrupole-active and the last two, ${{F}_{2}}$ and ${{F}_{6}}$ are forbidden in the inter-band optical quantum transitions as well as from the ground state of the crystal to the magnetoexciton states in the GaAs-type QWs. The dispersion equation describing the magnetoexciton-polariton energy spectrum includes the first four states ${{F}_{1}},{{F}_{3}},{{F}_{4}}$ and ${{F}_{5}}$ interacting with the cavity photons in two approximations. These four magnetoexciton degrees of freedom together with the branch of the cavity photons give rise to fifth order dispersion equation with five polariton-type renormalized energy branches.
The obtained picture was complemented by two magnetoexciton energy-branches ${{F}_{2}}$ and ${{F}_{6}}$ forbidden in the optical quantum transitions. The cavity polariton energy spectrum essentially depends on the detuning between the energy of the cavity mode $\hbar {{\omega }_{c}}$ and the energy of the magnetoexciton level. One variant of such detuning concerns the energy level ${{E}_{ex}}\left( {{F}_{1}},B,0 \right)$ of the dipole-active state ${{F}_{1}}$ with circular polarization ${{\vec{\sigma }}_{-1}}$ interacting with the cavity photons with circular polarization $\vec{\sigma }_{{{{\vec{k}}}_{\uparrow }}}^{-}$ or $\vec{\sigma }_{{{{\vec{k}}}_{\downarrow }}}^{+}$ propagating inside the microcavity with wave vectors $\vec{k}=\pm {{\vec{a}}_{3}}\frac{\pi }{{{L}_{c}}}+{{\vec{k}}_{||}}$ correspondingly. The cavity mode energy was chosen to be equal $\hbar {{\omega }_{c}}\left( B \right)=\frac{{{E}_{ex}}\left( {{F}_{1}},B,0 \right)}{1-{{\delta }_{1}}}$ with dimensionless detuning ${{\delta }_{1}}=-0,01$. The value $\hbar {{\omega }_{c}}\left( B \right)$ changes in dependence on the magnetic field strength B and on other parameters of the theory simultaneously with the change of the magnetoexciton level  energy ${{E}_{ex}}\left( {{F}_{1}},B,0 \right)$. Two different sign $\pm $ of the quantized component ${{k}_{z}}$ of the wave vector $\vec{k}$ happen to be very important, giving rise to the gyrotropy effect. It consists in the dependence of the polariton energy spectrum on the direction of the photon propagation. In one case it is determined by the wave vector ${{\vec{k}}_{\uparrow }}$ with ${{k}_{z}}=\pi /{{L}_{c}}$ whereas in another case by the wave vector ${{\vec{k}}_{\downarrow }}$ with ${{k}_{z}}=-\pi /{{L}_{c}}$.

\begin{acknowledgments}
I.V.P. and E.V.D. thanks the Foundation for Young Scientists of the Academy of Sciences of Moldova for financial support (14.819.02.18F).
\end{acknowledgments}

\end{document}